\documentclass[useAMS,usenatbib]{mn2e}

\usepackage{amsmath}
\usepackage{amsfonts}
\usepackage{amssymb}
\usepackage{graphicx}
\usepackage{mathrsfs}
\usepackage{latexsym}
\usepackage{graphics}
\usepackage{hyperref}
\usepackage{mathtools}
\usepackage[english]{babel}
\usepackage{bm}
\usepackage{color}
\usepackage{xspace}
\usepackage{listings}

\def\be{\begin{equation}}
\def\ee{\end{equation}}
\def\ben{\begin{equation*}}
\def\een{\end{equation*}}
\def\na{\begin{align}}
\def\ea{\end{align}}

\newcommand\hf{\textsc{hyper-fit}\xspace}

\newcommand\R{{\textsc R}\xspace}
\newcommand\Python{{\textsc PYTHON}\xspace}

\newcommand\profit{{\textsc ProFit}\xspace}
\newcommand\pyprofit{{\textsc PyProFit}\xspace}
\newcommand\libprofit{\texttt{libprofit}\xspace}
\newcommand\galfit{{\textsc GALFIT}\xspace}
\newcommand\sersic{{\textsc S\'{e}rsic}\xspace}

\date{\today}

\title[ProFit: profile fitting]{ProFit: Bayesian Profile Fitting of Galaxy Images}
\author[A.~S.~G. Robotham, D.~S. Taranu, R. Tobar, A. Moffett, S.~P. Driver]{%\small%
A.~S.~G. Robotham$^1$, D.~S. Taranu$^{1,2}$, R. Tobar$^1$, A. Moffett$^1$, S.~P. Driver$^1$\\
$^1$ICRAR, M468, University of Western Australia, Crawley, WA 6009, Australia\\
$^2$CAASTRO: ARC Centre of Excellence for All-sky Astrophysics, Building A28, School of Physics, The University of Sydney,\\NSW 2006, Australia\\
}

\begin{document}

\date{\today}

\pagerange{\pageref{firstpage}--\pageref{lastpage}} \pubyear{2016}

\maketitle

\label{firstpage}

\begin{abstract}

We present \profit, a new code for Bayesian two-dimensional photometric galaxy profile modelling. \profit consists of a low-level {\textsc c++} library (\libprofit), accessible via a command-line interface and documented API, along with high-level \R (\profit) and \Python (\pyprofit) interfaces (available at \url{github.com/ICRAR/libprofit}, \url{github.com/ICRAR/ProFit}, and \url{github.com/ICRAR/pyprofit} respectively). \R{} \profit is also available pre-built from {\textsc CRAN}, however this version will be slightly behind the latest GitHub version. \libprofit offers fast and accurate two-dimensional integration for a useful number of profiles, including \sersic, Core-\sersic, broken-exponential, Ferrer, Moffat, empirical King, point-source and sky, with a simple mechanism for adding new profiles. We show detailed comparisons between \libprofit{} and \galfit. \libprofit is both faster and more accurate than \galfit at integrating the ubiquitous \sersic profile for the most common values of the \sersic index $n$ ($0.5 < n < 8$).

%We present our new 2D galaxy profiling code \profit. The core functionality is contained in a C++ library (\libprofit) that can accessed through the command-line or via a low-level API. High level interfaces written in \R and \Python are provided in this initial release (\profit and \pyprofit respectively), with other community forks strongly encouraged. The low-level library and higher level interfaces are already public \footnote{\libprofit, \profit and \pyprofit are available at \url{https://github.com/ICRAR/libprofit}, \url{https://github.com/ICRAR/ProFit}, and \url{https://github.com/ICRAR/pyprofit respectively}}. The basic premise is that the core library of \libprofit offers fast and accurate 2D image evaluation for a useful number of profiles. Initially \libprofit comes with \sersic, Core-\sersic, Ferrer, Moffat, empirical King, point-source and sky profile types. It has a simple well documented mechanism for adding new user-defined profiles. Detailed comparisons are made between \libprofit and \galfit (the latter being a popular tool for 2D image fitting). They perform at a broadly similar level, where \libprofit is typically faster and more accurate at generating a given \sersic profile (the profile type of most interest in the extra-galactic community).

The high-level fitting code \profit is tested on a sample of galaxies with both SDSS and deeper KiDS imaging. We find good agreement in the fit parameters, with larger scatter in best-fitting parameters from fitting images from different sources (SDSS versus KiDS) than from using different codes (ProFit versus GALFIT). A large suite of Monte Carlo-simulated images are used to assess prospects for automated bulge-disc decomposition with ProFit on SDSS, KiDS and future LSST imaging. We find that the biggest increases in fit quality come from moving from SDSS- to KiDS-quality data, with less significant gains moving from KiDS to LSST.

%\profit is used to fit a sample of galaxies where we have access to both SDSS and KiDS imaging. We find good agreement in the fit parameters and evidence of more scatter being caused by differing data sources as opposed to different fitting codes (\profit and \galfit in this case). A large suite of simulations are made to determine future prospects for the automated running of \profit on a mixture of current and future data sources. We find that the biggest increases in data quality (from the point-of-view of galaxy decomposition) will come from moving from SDSS to KiDS quality data. A smaller, but still notable, improvement should occur when transitioning to full LSST quality data.

\end{abstract}

\begin{keywords}
galaxies: statistics -- galaxies: structure -- methods: statistical -- techniques: photometric -- galaxies: fundamental parameters -- methods: data analysis
\end{keywords}

\section{Introduction}\label{section_introduction}

% Need to separately discuss single- vs multi-component fitting

Galaxy modelling can be broadly separated into two categories: light profile fitting and kinematic dynamical modelling. In dynamical modelling, part or all of the six-dimensional phase space structure is constrained by kinematic data and dynamical constraints such as the Poisson and/or Jeans' equations. By contrast, profile fitting (or profiling) quantifies the \emph{projected} structure of a galaxy on the sky, usually by fitting radially-varying density profiles to one-dimensional, azimuthally-averaged profiles or two-dimensional images. Both methods have been used to infer the presence of distinct structural components in galaxies, such as discs, bulges, bars and spiral arms (so-called decomposition).

% Who (if anyone) should be credited with identifying distinct components (bulge+disk) or suggesting distinct evolutionary pathways?

\citet{hubb26} is credited as the first to develop a galaxy classification scheme including distinct features or components, including spiral arms and a concentrated central component. As early as \citet{de-v58} it had been quantitatively demonstrated, using photoelectric measurements along the major and minor axes, that M31 comprised a distinct bulge and disc component. In the modern era, large samples of thousands of galaxies have been decomposed quantitatively into distinct components \citep{alle06, gado09, sima11, lang16, kenn16}. Such data sets have been used to infer distinct evolutionary pathways and mechanisms \citep{korm04} for galaxies of different types and compositions and to develop galaxy formation models \citep{zava08, driv13, lace16}. Improving these theories requires both better data (by volume, depth and quality) and analysis methods. This paper offers a new route to reliably divide galaxies into their constituent parts using a new publicly released library and interface (\libprofit and \profit respectively) for galaxy profile modelling and decomposition. These will be used by the authors for near future large-scale studies, and are being made available to the wider community along with the prospect of practical support and future updates.

% And the exponential profile - de Vaucouleurs '59 noted that M31 has a bulge + disc. Who was first?

The earliest efforts of galaxy structural characterization concentrated on simple one-dimensional intensity profile fitting, leading to the early discovery of the de-Vaucouleurs profile \citep{de-v48} for describing early-type galaxies and the realization that disc galaxies had close to exponential profiles \citep{de-v59}. \citet{sers63} generalized these profiles into the $r^{1/n}$ law that remains widely popular today, commonly referred to as the \sersic profile \citep{grah05}. Increasingly complicated one-dimensional component fitting came in work by \citet{korm77} and \citet{kent85}, where galaxies were decomposed into distinct components rather than treated as having a single light profile.

The earliest efforts at two-dimensional galaxy profiling came with the work of \citet{andr95}, \citet{byun95}, \citet{de-j96}, \citet{scha95} and \citet{wada99}. The latter four approaches are broadly similar in application, and are recognizably similar to modern two-dimensional galaxy fitting efforts. The basic philosophy was the same as for previous one-dimensional work, the aim being to find the distinct light components of galaxies, but here the analysis was made using image pixels directly rather than by fitting azimuthally averaged ellipses of light.

A large number of key insights into galaxy properties have been uncovered through both one-dimensional and two-dimensional structural analysis. Early work found that the disc component of galaxies is very well represented by an exponential drop-off in light \citep[e.g.][]{free70, korm77}. Of recent interest in the astronomy literature is the relationship between galaxy mass and size, both globally \citep{shen03, lang15} and for individual components \citep{lang16}. There is also a large body of work investigating the luminosity-surface brightness relation for different classes of galaxy \citep[see reviews by][]{ferg94, grah13}, as well as investigating the distribution of mass for different types of structure \citep[e.g.][]{dres87, bens07, driv07, kelv14, kenn16, moff16a, moff16b}.

Such global morphological properties of galaxies are finally being utilized and predicted by the newest theoretical work, both in the regime of semi-analytic models \citep{lace16} and cosmological hydrodynamic simulations \citep{voge14, scha15}. Given the constraints that galaxy morphology can offer theory, it will be increasingly important that well quantified structural measurements for galaxy properties are extracted from current and future large-area, deep, high-resolution photometric surveys \citep[e.g. Kilo Degree Survey (KiDS), Dark Energy Survey (DES), Hyper Suprime Camera (HSC) and Large Synoptic Survey Telescope (LSST):][respectively]{kuij15, des05, miya13, abel09}.

There remains some debate regarding the relative merits of one-dimensional and two-dimensional image analysis \citep[e.g.][]{savo16}. This introduction will not serve as a thorough discussion of the various issues, but we will discuss a few key points. One-dimensional fitting relies heavily on an additional step in the analysis: the galaxy must be collapsed into a one-dimensional profile in some manner, usually using software such as IRAF's ELLIPSE task. It is impossible to do this in an entirely model independent manner when the galaxy is anything other than smoothly changing isophotes; however, such an approach can easily cope with a smoothly varying orientation in the isophotes. A caveat to this is that it is unclear formally how the atmospheric point-spread function (PSF) should be treated in one-dimensional fitting \citep[note that there are analytic solutions for the specific case of face-on convolution, see][]{prit81}. Even a galaxy constructed from perfect concentric ellipses will become artificially circular towards the profile centre when convolved with a circular PSF. Whilst this effect is captured from the one-dimensional profile information, it is non-trivial to properly propagate this information such that you measure the intrinsic two-dimensional profile properties of the galaxy. Collapsing two-dimensional information into a one-dimensional form is almost always a lossy process (it can be no better than lossless), and real azimuthal profile fluctuations may be lost entirely. Correctly propagating errors is also non-trivial, and in practice these are often not used at all \citep[an approach advocated in][]{savo16}.

There are two other critical aspects of this collapsing process that can have a dramatic effect on galaxy profiling: the centre of the galaxy must be fixed exactly a-priori and the components must share the same centre. The freedom to fit for the galaxy centre is often important for capturing the steepest inner parts of a galaxy properly (especially if the centre is sub-pixel) and galaxy components sometimes do not share exactly the same centres \citep{lang15}.

Two-dimensional fitting is closer in spirit to Bayesian generative model fitting, e.g.\ creating a two-dimensional distribution of flux that is then propagated through the atmospheric PSF. As such it is an attractive approach to quantifying image components. However, the isophotes of popular galaxy profile functions are poorly behaved when integrating over two-dimensional pixels, requiring relatively expensive numerical estimation, while complex geometry (like rotating isophotes) is more difficult to capture fully, with most codes ignoring this issue entirely. In the regime where the true galaxy is well represented by the two-dimensional model being used a two-dimensional fit {\it should} be preferable. In principle, effects such as twisting isophotes can be incorporated into a two-dimensional generative model technique \citep[see][]{peng10}, but this is rarely attempted in practice.

Because of these relative advantages and disadvantages between one-dimensional and two-dimensional approaches, their application to galaxy profiling has pragmatically diverged. One-dimensional codes are more popular in the regime of highly resolved galaxies where complex geometric effects (such as twisting isophotes) are clearly visible and the issue of PSF convolution is less significant. Because of the more automated nature of two-dimensional codes (they do not require a carefully controlled collapse of the data to one-dimensional), they have tended to be more popular for large samples of galaxies where the fitting is made with little user interaction \citep[e.g.][]{sima02, alle06, haus07, kelv12,  haus13}. Our long-term ambition is to decompose galaxy structures from a large number of imaging data sources for many millions of galaxies, so our use-case is squarely in the highly-automated regime. For this reason in particular, we focused our efforts on two-dimensional decomposition codes.

Current publicly-available two-dimensional galaxy profile fitting codes include {\textsc BUDDA} \citep{de-s04}, \galfit \citep{peng10}, {\textsc GALMORPH} \citep{hydephd}, {\textsc GIM2D} \citep{sima02}, and {\textsc IMFIT} \citep{erwi15} --- all of which support the commonly-used \sersic profile. These codes share their origins in the early two-dimensional profiling efforts of \citet{wada99} and \citet{bens02} and have been utilized in a number of notable large-scale galaxy studies: \citet{alle06, gado09, hyde09, sima11, kelv12, haus13, meer15, lang16, kenn15, kenn16}. Notable alternatives include {\textsc GASPHOT} \citep{pign06}, which is a one-dimensional fitting code, and {\textsc MGE} \citep{capp02}, which abandons traditional \sersic profile bulge-disc decomposition for more flexible (but arguably less physical) concentric Gaussians. In all of the other codes, the model is treated generatively by numerically integrating a radially-varying profile over a rectangular pixel grid; in practice, most codes improve integration accuracy by oversampling the native pixel grid, particularly near the centre of the galaxy. Once this idealized model is integrated, it can be convolved with a target PSF, either via brute-force or fast Fourier transform (FFT) techniques. These codes all have their positive features and weaknesses, and differ in fine aspects such as how the likelihood is computed and maximized. However, no existing code achieves the desired design requirements that our new code, \profit, aims to address, namely that it:

\begin{itemize}
\item should be legally open source under a GPL, BSD, MIT license, or similar;
\item should be practically collaborative and hosted in a publicly available location such
      as GitHub or BitBucket;
\item should be written in a low-level modern object-oriented language, aiding modularity, extensions and robustness for complex software,
\item has the core image generation code clearly separated from the
      optimization code, available as a shared library,
      and accessible via an easy-to-use API;
\item offers a useful range of built-in models and have a well documented mechanism for adding more;
\item supports sub $\sim$1\% flux weighted error model image generation for built-in models, and be testable in this regard;
\item should be fast at generating these model images;
\item has a range of options to define likelihoods;
      i.e.\ Normal, Poisson, Chi-Squared and Student-T statistics;
\item should be able to fit a large number of components in a flexible manner,
      e.g.\ components being fixed, free or locked together with other components;
%\item be able to fit physical properties to multiple bands, ASGR: don't have this feature yet, so best not to suggest it
%      e.g. stellar mass,
\item should be untied to a specific optimizer,
      i.e.\ able to use a range of (user-selectable) optimization engines;
\item should be able to accept prior distributions of an arbitrary type for model parameters (vital for Bayesian model evaluation);
\item has the ability to fit parameters in linear or log space
      as deemed appropriate by the user for the problem (vital for Bayesian model evaluation);
\item should be able to handle parameter limits and constraints in flexible manner (e.g.\ prevent some parameters getting larger than other etc);
\item has a core interface that does not require the use of text file inputs,
      i.e.\ there is a completely functional interactive interface to the software (though scripting with files should be allowable);
\item is simple to parallelize.
\end{itemize}

The reasons for desiring the above features are multifaceted, but broadly it is to aid the structural analysis of large amounts of data obtained from new and future imaging surveys such as KiDS, DES, HSC and LSST. Also, the addition of optional priors and log or linear parameter scaling is paramount for proper Bayesian model evaluation. Automating a historic code is a complicated affair \citep{kelv12, haus13}, and there are in practice many subtle issues with trying to bootstrap a new code from legacy software. Starting from scratch with a new code offers full flexibility in language choice, design flexibility and low-level choice regarding pixel integration and likelihood calculations. Our estimation was it would be at least as much effort to adapt any available open-source code to meet the above requirements that \profit delivers.

This paper discusses the development and application of the core image generation code (\libprofit, Section \ref{sec:imgen}) and the fitting code (\profit, Section \ref{sec:fit}). The \profit code is then applied to various examples (Section \ref{sec:examples}), spanning a detailed case study (Section \ref{sec:example1}) and the analysis of 10 exemplar bulge-disc galaxies that are included with the \profit package (Section \ref{sec:example2}). Estimations of decomposition fidelity is then made for current and future data sets (Section \ref{sec:future}). Finally, we discuss and summarize the results of this paper (Section \ref{sec:discuss}).

\section{Image Generation with \libprofit{}}
\label{sec:imgen}

\begin{figure*}
	\centering
	\includegraphics[width=\columnwidth*2]{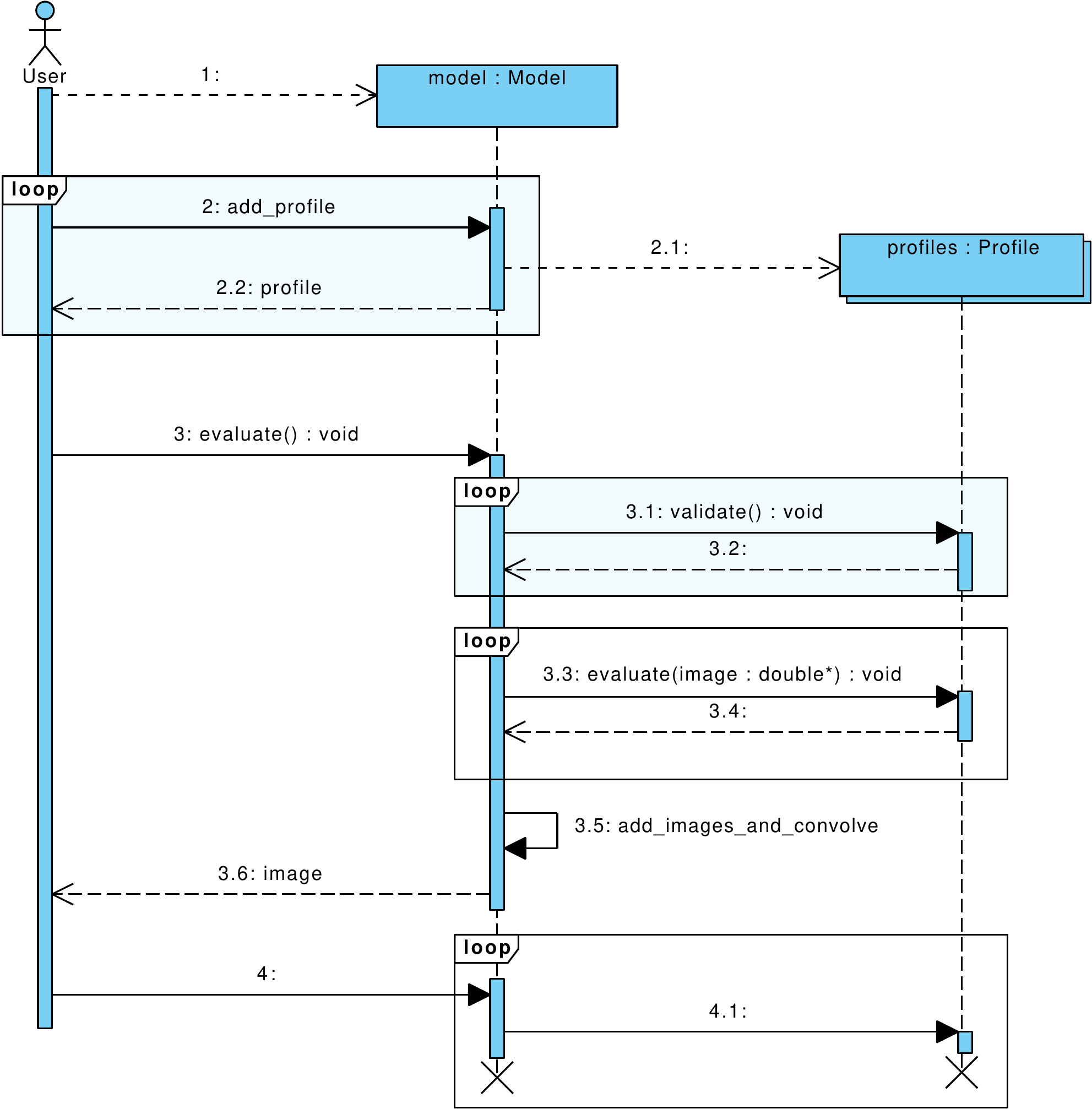}
	\caption{Unified Modelling Language sequence diagram showing how to interact with \libprofit (UML is a standard language commonly used when designing software). Users create a model, append profiles to it and finally evaluate it.}
	\label{fig:libprofit}
\end{figure*}

The image generation code is contained in a shared library written in {\textsc c++} called \libprofit{}. \libprofit{} enables the user to construct a \textit{model} to which \textit{profiles} are appended. Each profile can be fine-tuned, and fully describes a component of the resulting image. After profiles are appended the model is evaluated, and the resulting image can be retrieved. \libprofit{} cycles through each of the profiles generating individual images for each of them, which are optionally convolved and finally added. Hence, it is trivial to describe a one-, two-, or multi-component system. This external interface, internal organization and execution of the \libprofit library are depicted in Fig.~\ref{fig:libprofit} using the Unified Modelling Language.

\subsection{Galaxy Profiles}
\label{sec:profiles}

\begin{figure*}
	\centering
	\includegraphics[width=12cm]{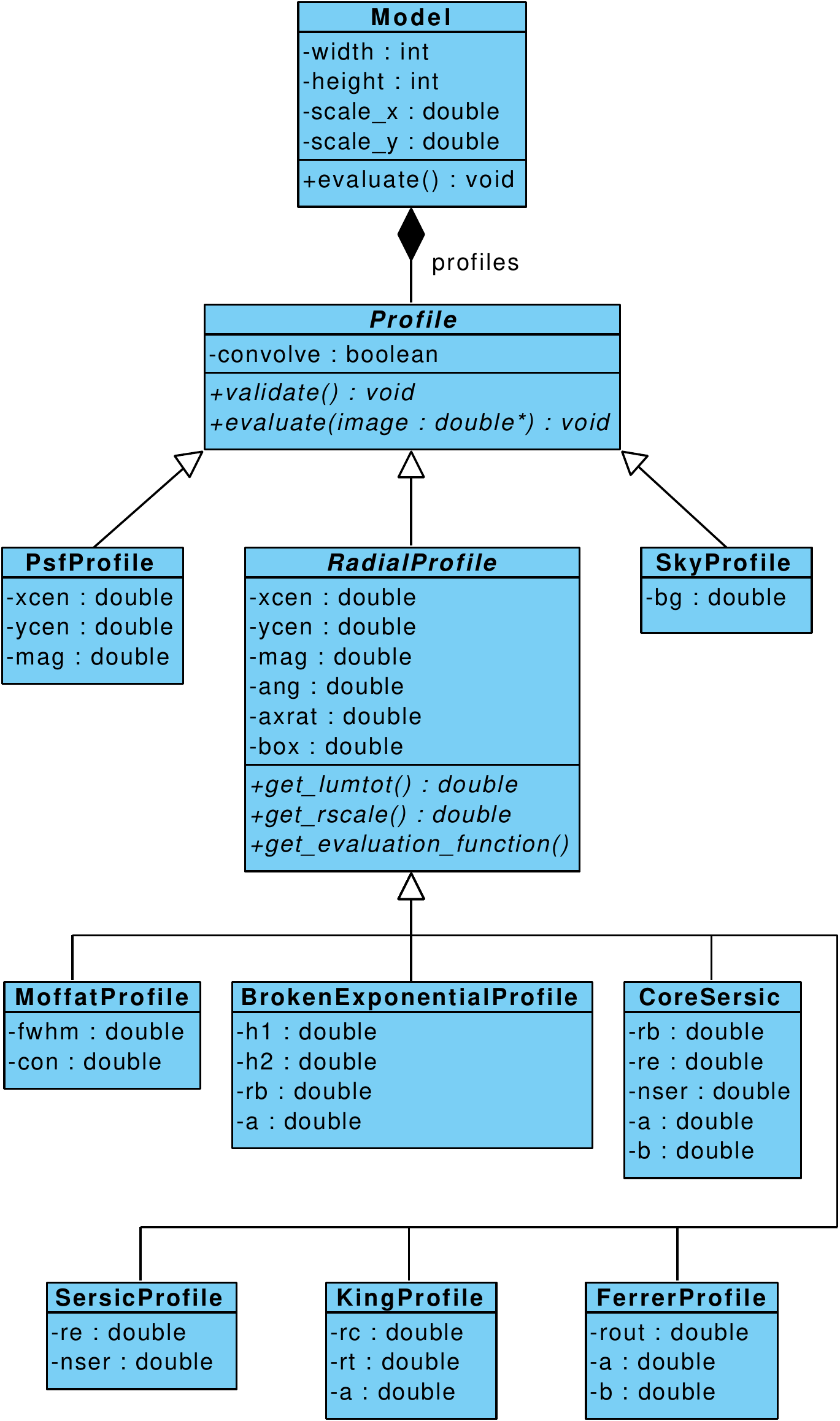}
	\caption{UML class diagram showing the eight profile types available in v1.0 of 
\libprofit{} (from left to right, top to bottom): PSF, sky, Moffat, broken-exponential, Core-\sersic{}, \sersic{}, King, and Ferrer. The bottom six
derive from a common type, the radial profile (centre), a generic
form of profile that takes an analytic one-dimensional profile as
its input. It is easy to add more profiles of this type if they can
be described with a one-dimensional form and the total integral
is calculable and finite (even if not analytic). All profile types
inherit from a common class and form part of a model.}
	\label{fig:profiles}
\end{figure*}

\libprofit{} includes a default set of popular profiles: \sersic, Core-\sersic, broken-exponential, Moffat, Ferrer, modified empirical King, PSF and sky (see Fig.~\ref{fig:profiles}).
The first six share some key similarities: they can be fully described by an analytic radial profile that is then evaluated in two-dimensions over pseudo-elliptical isophotes. They also share similarities in terms of how they must be evaluated in a given pixel: if the gradient of the model image is varying rapidly over the scale of a pixel then using a simple trapezoidal integration scheme (effectively taking the flux value predicted at the centre and assuming this to reflect the average for the pixel) can be error prone.

For this reason the \sersic, Core-\sersic, broken-exponential, Moffat, Ferrer and King profiles all share the same core c++ code that evaluates the flux in a pixel and determines whether more accuracy is required. This means new profiles can be defined quite simply within \libprofit{}, effectively reducing to the one-dimensional form of the radial profile and a scheme to calculate the total flux within the profile. For \sersic, Moffat and Ferrer, the total integration is known (or has been calculated by the authors) analytically, so this makes them particularly simple cases.

All eight profiles are discussed in more detail below, with the caveat that the detail of the integration scheme discussed with respect to the \sersic profile also applies to the Core-\sersic, broken-exponential, Moffat, Ferrer, and King.

\subsubsection{\sersic Profile}

\begin{figure*}
	\centering
	\includegraphics[width=5.8cm]{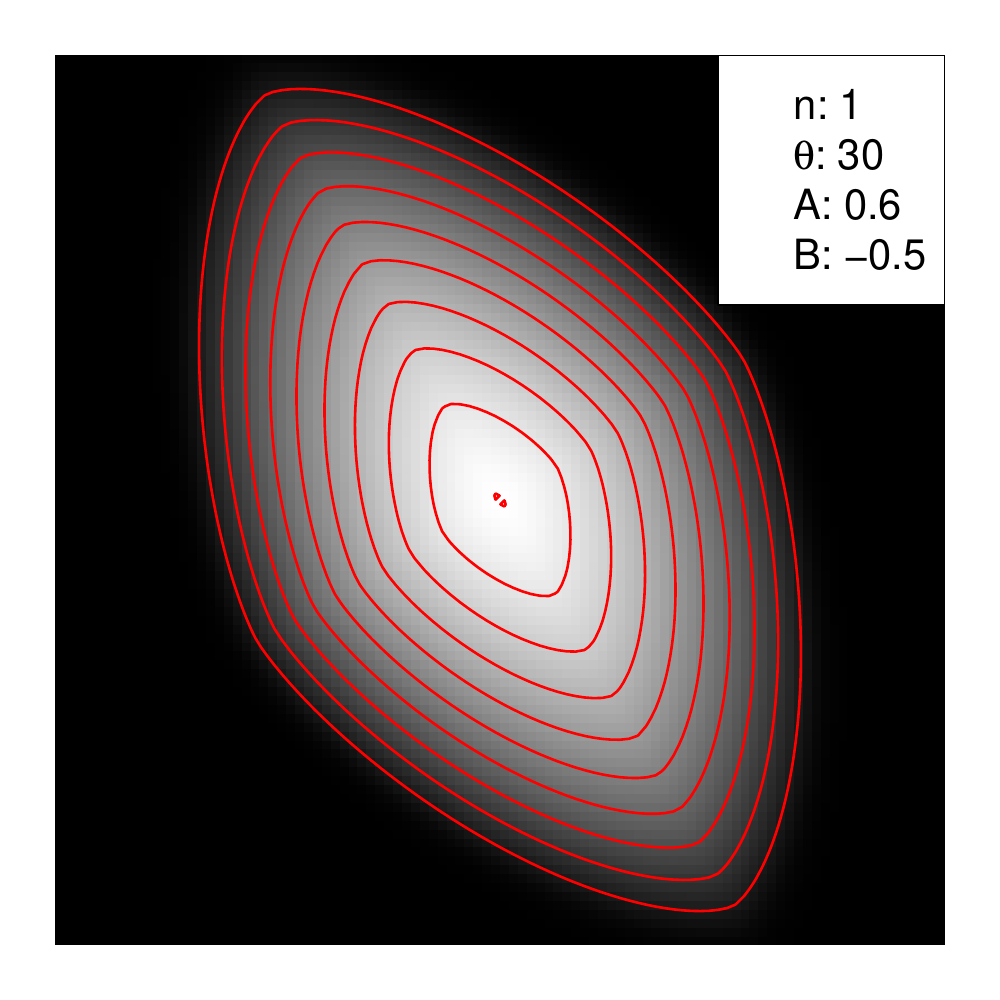}
	\includegraphics[width=5.8cm]{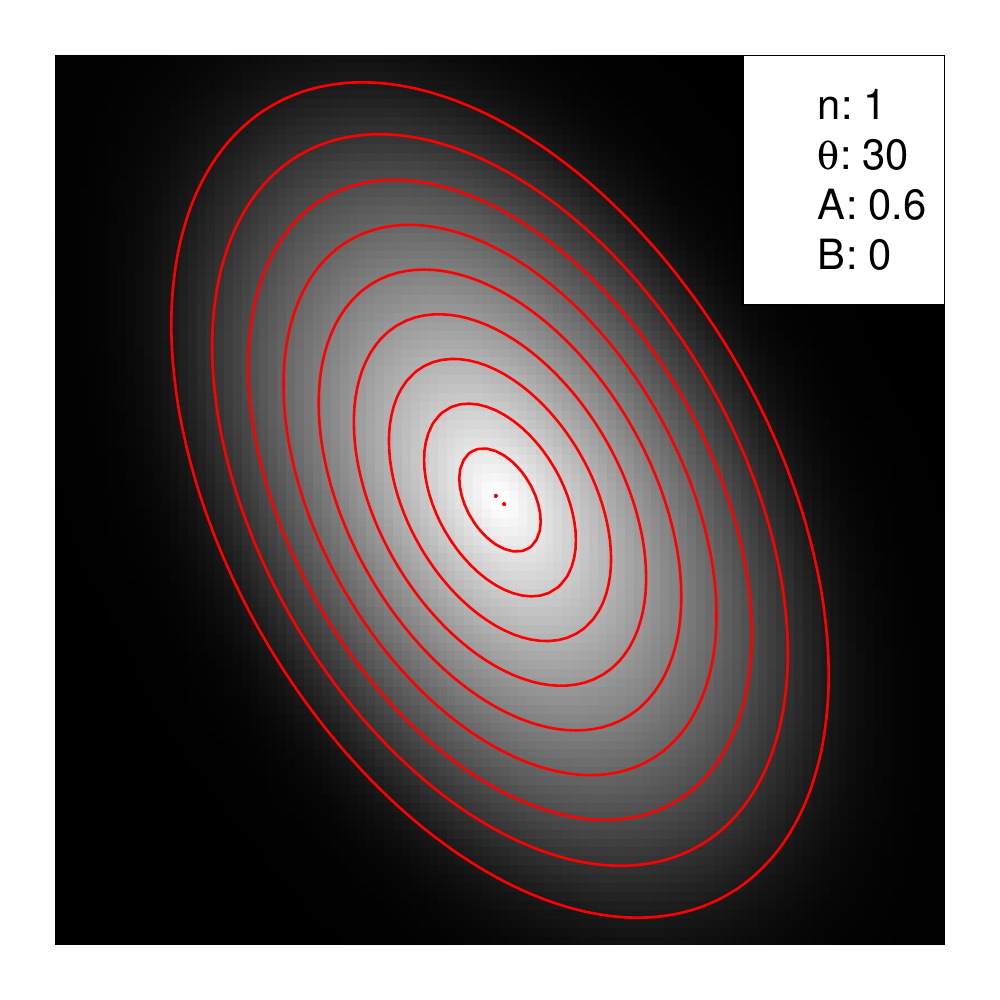}
	\includegraphics[width=5.8cm]{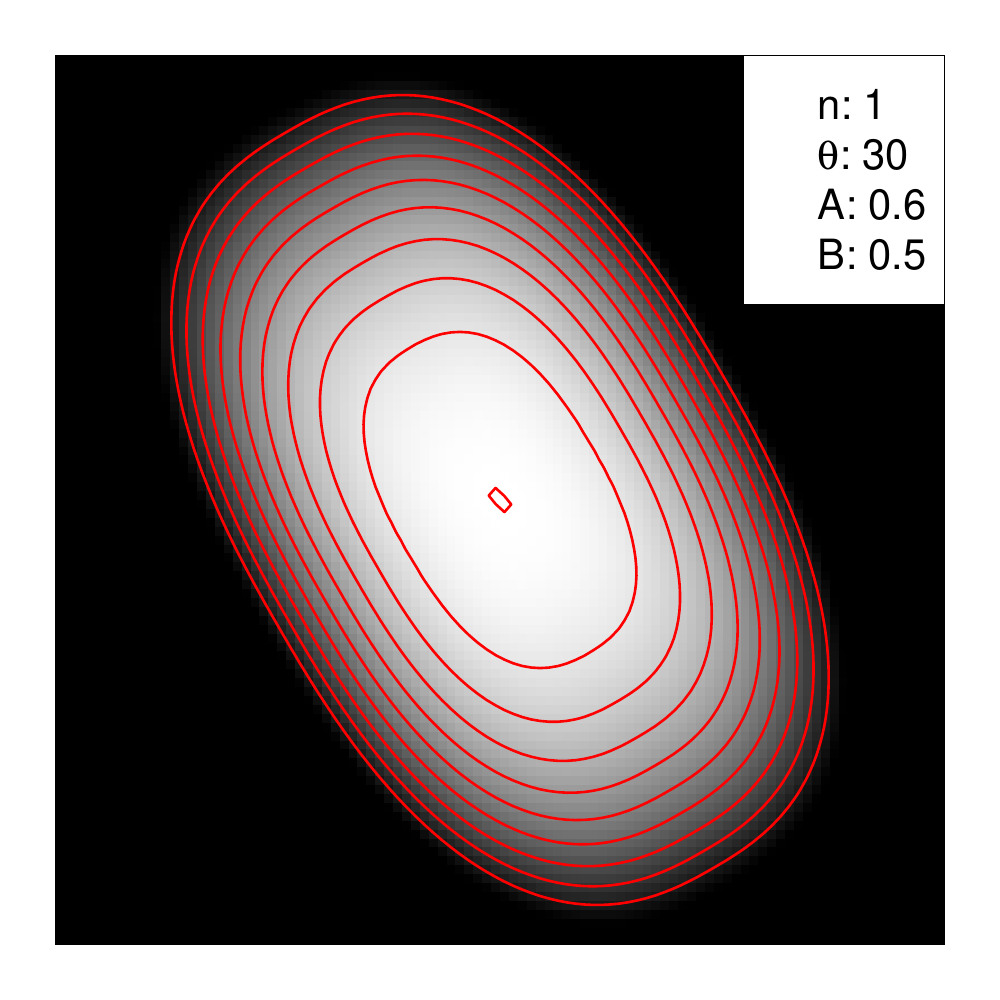}
	\caption{Comparison of kite-like negative boxiness (left), normal ellipses with no boxiness (middle) and box-like positive boxiness (right).}
	\label{fig:boxiness}
\end{figure*}

For galaxy fitting purposes the \sersic profile \citep{sers63} is a modern day work-horse, having enough flexibility in profile characteristics to fit a large variety of galaxy types \citep[see e.g.][]{kelv12}. The intensity profile is usually parameterized as:

\begin{equation}
I(R)=I_e \exp \left\{ -b_n \left[ \left( \frac{R_m}{R_e} \right)^{1/n}-1 \right] \right\},
\end{equation}

\noindent where $I_e$ is the intensity of the profile at $R_e$ (the radius containing half of the flux), $n$ is the \sersic index (which has special cases of Normal / Gaussian with $n=0.5$, exponential with $n=1$ and de-Vaucouleurs with $n=4$) and $b_n$ is a derived parameter that ensures the correct integration properties at $R_e$ and is the quantile at which the Gamma probability distribution function integrates to 0.5 given a shape parameter of $2n$. This can be computed directly using high-level interfaces to statical distribution libraries (e.g.\ in \R $b_n=$ {\tt qgamma(0.5, 2*n)}). $R_m$ is the modified radius where we want to make the evaluation, which in one dimension is simply the radius from the profile centre. In two dimensions $R_m$ has a more complex form

\begin{equation}
R_m=\sqrt{(x-x_{cen})^2 + (y-y_{cen})^2},
\end{equation}

\noindent where $x,y$ is the two-dimensional location where we wish to make the evaluation and $x_{cen},y_{cen}$ is the location of the profile centre. When the two-dimensional profile is non-circular, the isophotal contours become elliptical annuli, so calculating $R_m$ becomes a series of computations that rotate and circularize the annuli. When the major-axis angle $\theta$ is defined such that it is 0$^{\circ}$ vertically and increases positively as the galaxy is rotated counter-clockwise (as per \galfit) these steps become

\begin{eqnarray}
R_t&=&\sqrt{(x-x_{cen})^2 + (y-y_{cen})^2},\\
\theta_t&=&\arctan{\frac{x-x_{cen}}{y-y_{cen}}}+\theta,\\
R_m&=&\sqrt{(R_t\sin(\theta_t) A_{rat})^2 + (R_t\cos(\theta_t))^2},
\end{eqnarray}

\noindent where $A_{rat}$ is the minor to major axis ratio (so always a number between 0 and 1, where 0 is an infinitely thin line and 1 is a circle or disc). The final modifier to the standard one-dimensional \sersic profile intensity is to allow for apparent `boxiness'. Boxiness is often used to create the more rectangular visual appearance of galaxy bulges or elliptical galaxies. It effectively modifies the unit circle (or ellipse) away from the standard $L^2$ norm and either towards a diamond/kite appearance ($L^1$ norm) or a boxy appearance ($L^\infty$ norm). This is achieved though a small manipulation of the final $R_m$ calculation above

\begin{equation}
\label{eq:Rmod}
R_m=\left[(R_t\sin(\theta_{t})A)^{(2+B)} + (R_t\cos(\theta_{t}))^{(2+B)}\right]^\frac{1}{2+B},
\end{equation}

\noindent where $B$ is the boxiness, which for practical (as opposed to mathematical) purposes is usually allowed to vary between $-1$ and 1 (thus covering the full range of apparent visual boxiness). Fig.~\ref{fig:boxiness} shows the effect of different negative and positive values of boxiness. A consideration for the user is the relative expense of computing the boxiness. Powers that differ from the $L^2$ norm are much more expensive to compute, in particular non-integer values might be a factor of a couple slower when computing a model image. In most standard C math libraries ({\tt libm}) it is faster to make successive calls to {\tt sqrt} and {\tt csqrt} rather than a single call to {\tt pow}. \libprofit{} acknowledges this fact and calls the most efficient set of functions as appropriate for certain exponents.

Whilst the above is an accurate description of the computations that need to be made to calculate an arbitrary two-dimensional \sersic profile, they are not optimal. \libprofit{} uses rotation matrix arithmetic internally to achieve the above steps efficiently, projecting pixel coordinates on to the major and minor axes of the ellipse. This can be a factor of a few faster than explicitly computing rotations with expensive calls to trigonometric functions.

It is not necessary to parameterize the profile using $I_e$ to define the normalization, as this is rarely intuitive. Instead, the user can specify the total flux contained within a profile in magnitudes ($m$) given a magnitude zero-point ($m_{zero}$), and calculations are made internally to convert this to the correct value of $I_e$. These include appropriate modifications for the axis ratio ($A_{rat}$) and boxiness ($B$):

\begin{eqnarray}
\label{eq:sersnorm}
R_{box}&=&\frac{\pi(2+B)}{2\beta\left(\frac{1}{2+B},\frac{1}{2+B}\right)},\\
L_{tot} &=& \frac{2 \pi A_{rat} R_e^2 n \Gamma(2n) \exp(b_n)}{R_{box}b_n^{2n}},\\
I_e&=&\frac{10^{-0.4(m-m_{zero})}}{L_{tot}},
\end{eqnarray}

\noindent where $\beta$ is the Beta function defined as $\beta(a,b)=\Gamma(a)\Gamma(b)/\Gamma(a+b)$, and $\Gamma$ is the standard Gamma function. If the profile has no boxiness, then the above is simplified by noting that $R_{box}(B=0)=1$ and therefore the denominator term for $L_{tot}$ disappears. To aid flexibility, it is also possible in \libprofit{} to define the profile normalisation using the mean surface brightness within $R_e$ ($<\mu>$), where:

\begin{equation}
<\mu>=m+2.5\log_{10}(\pi Re^2 A_{rat})-2.5\log_{10}(0.5).
\end{equation}

\noindent \libprofit{} follows the convention that $<\mu>$ is determined for an elliptical radius containing half of the objects flux, i.e.\ it is not modified for boxiness. This can be useful when parameterizing a galaxy where the galaxy magnitude is entirely unknown since galaxies span a much smaller dynamic range in surface brightness than total magnitude \citep{driv05}.

\begin{figure}
	\includegraphics[width=\columnwidth]{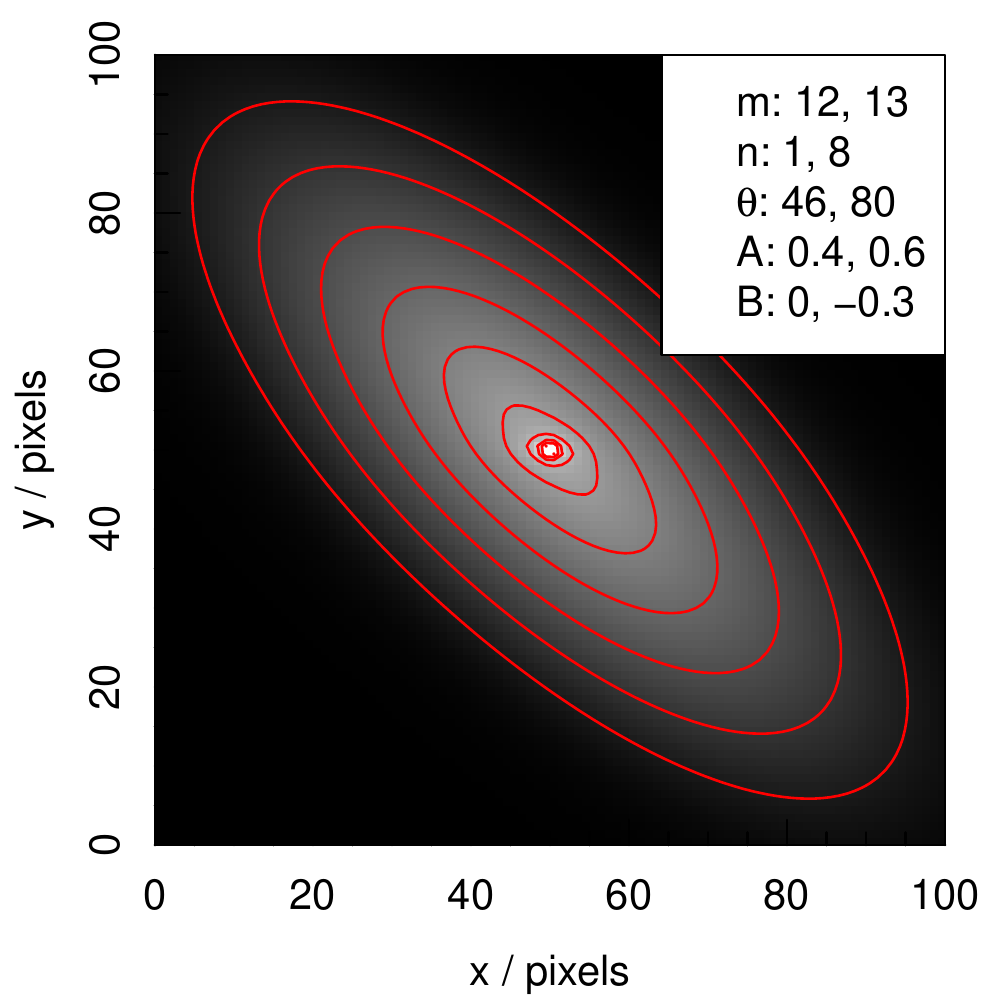}
	\caption{Example \libprofit{} model image with no convolution by the PSF.}
	\label{fig:examplesersicmodel}
\end{figure}

In conclusion, the \sersic profile is specified fully through the provision of eight parameters that are parsed into the \libprofit{} image building routine as a list structure of equal length vectors. Below is an example specification of a two-component \sersic model in the \R implementation of \profit that directly calls the \libprofit API to generate model images, where $x_{cen}=$\hspace{1mm}xcen, $y_{cen}=$\hspace{1mm}ycen, $m=$\hspace{1mm}mag, $R_e=$\hspace{1mm}re, $n=$\hspace{1mm}nser, $\theta=$\hspace{1mm}ang, $A_{rat}=$\hspace{1mm}axrat, and $B=$\hspace{1mm}box:

\begin{lstlisting}
modellist = list(
  sersic = list(
    xcen   = c(50, 50),
    ycen   = c(50, 50),
    mag = c(12, 13),
    re  = c(14, 5),
    nser  = c(1, 8),
    ang  = c(46, 80),
    axrat  = c(0.4, 0.6),
    box = c(0,-0.3)
  )
)
\end{lstlisting}

This model is shown in Fig.~\ref{fig:examplesersicmodel}. As expected, the high \sersic index bulge component dominates the flux near the galaxy centre as well as at very large radii, due to the shallow slope of an $n=8$ \sersic profile. This is visible as side lobes in the surface brightness contours jutting out from the visually dominant disc component running diagonally across the image.

\subsubsection{Core-\sersic}

Closely related to the \sersic profile is the Core-\sersic profile introduced by \citet{grah03}. This was designed to parameterize the strong inner core (i.e.\ suppression compared to pure \sersic) profile of well-resolved elliptical galaxies. The inner components of such galaxies had previously been parameterized well with the Nuker profile \citep{laue95}, at least in the inner parts of the profile. However, the Nuker profile fits poorly at large radii and does not have a well behaved integral to infinite radius for typical parameters. The Core-\sersic addresses these issue by allowing the inner component to behave in a Nuker-like manner, with the outer parts being described by a standard \sersic profile. The intensity profile can be parameterised as

\begin{equation}
I(R_m)=I_r\left[1+\left(\frac{R_m}{R_b}\right)^{-a}\right]^\frac{b}{a} \exp \left[-b_n \left(\frac{R_m^a+R_b^a}{R_e^a}\right)^\frac{1}{n a}\right],
\end{equation}

\noindent where $I_r$ is the intensity at a parameter dependent reference point (the centre if $R_b=0$), $R_b$ is the break radius that controls the point where the outer \sersic-like profile transitions to become an inner Nuker-like profile, $R_e$ is the standard \sersic half light radius (and $b_n$ has the same meaning as for the pure \sersic profile) and $a$ controls the strength of transition (large value is a sharper transition) and $b$ controls the inner power law. $R_m$ is the modified radius where we want to make the evaluation, which in one dimension is simply the radius from the profile centre. In two dimensions $R_m$ has a more complex form where, as with the \sersic profile, we can use this basic form and compute annuli that are modified from pure circular ellipses by introducing geometric transformations that compute an effective $R_m$, as per Equation \ref{eq:Rmod}.

Accounting for the allowed geometric distortions of axial ratio and boxiness, the value of $I_r$ for a given magnitude can be calculated numerically with

\begin{eqnarray}
L_{tot} &=& \frac{2 \pi A_{rat} \int_0^\infty R_m I(R_m; I_r=1) dR_m}{R_{box}},\\
I_r&=&\frac{10^{-0.4(m-m_{zero})}}{L_{tot}},
\end{eqnarray}

The integral above has no simple analytic solution, so it is instead calculated numerically inside \libprofit{} using either the \R or GSL implementation of the {\textsc QUADPACK} library (specifically adaptive Gauss-Kronrod quadrature integration). This is a much better solution than attempting to re-normalize the generated image since much of the flux could exist beyond the confines of the image pixels, especially for large values of $n$.

Below is an example of a Core-\sersic model in the \R implementation of \profit that directly calls the \libprofit API to generate model images, where $x_{cen}=$\hspace{1mm}xcen, $y_{cen}=$\hspace{1mm}ycen, $m=$\hspace{1mm}mag, $R_{b}=$\hspace{1mm}rb, $R_{e}=$\hspace{1mm}re, $n=$\hspace{1mm}nser, $a=$\hspace{1mm}a, $b=$\hspace{1mm}b, $\theta=$\hspace{1mm}ang, $A_{rat}=$\hspace{1mm}axrat, and $B=$\hspace{1mm}box:

\begin{lstlisting}
modellist = list(
  coresersic = list(
    xcen   = 50,
    ycen   = 50,
    mag = 15,
    rb = 5,
    re = 10,
    nser=4,
    a = 2,
    b=1.3,
    ang  = 30,
    axrat  = 0.4,
    box = 0
  )
)
\end{lstlisting}

\subsubsection{Broken-Exponential Profile}

The broken-exponential profile is a popular means to parameterise rolling or sharp truncations in the surface profiles of discs, which are usually close to exponential in profile until they enter the moderate- to low-surface brightness regime. \citet{erwi08} introduced a simple modification to the standard exponential profile (which is itself a subset of possible \sersic profiles) in order to capture the range of truncations observed. The intensity profile can be parameterized as

\begin{eqnarray}
S&=&(1 + e^{ - a R_b } )^{-\frac{1}{a}\left(\frac{1}{h_1}-\frac{1}{h_2}\right)},\\
I(R_m)&=&I_0 S e^{ \left(\frac{-R_m}{h_1}\right)} \left[ 1+ e^{a (R_m - R_b)} \right]^{\frac{1}{a}\left(\frac{1}{h_1}-\frac{1}{h_2}\right)},
\end{eqnarray}

\noindent where $I_0$ is the intensity at the centre of the profile, $h_1$ is the scale length of the inner exponential profile, $h_2$ is the scale length of the outer exponential profile, $R_b$ is the radius at which the profile transitions from the inner to outer part and $a$ controls the sharpness of this transition. $R_m$ is the modified radius where we want to make the evaluation, which in one dimension is simply the radius from the profile centre. In two dimensions $R_m$ has a more complex form where, as with the \sersic profile, we can use this basic form and compute annuli that are modified from pure circular ellipses by introducing geometric transformations that compute an effective $R_m$, as per Equation \ref{eq:Rmod}.

Accounting for the allowed geometric distortions of axial ratio and boxiness, the value of $I_0$ for a given magnitude can be calculated analytically with

\begin{eqnarray}
L_{tot} &=& \frac{2 \pi A_{rat} \int_0^{R_t} R_m I(R_m; I_0=1) dR_m}{R_{box}},\\
I_0&=&\frac{10^{-0.4(m-m_{zero})}}{L_{tot}},
\end{eqnarray}

\noindent where $m$, $m_{zero}$ and $R_{box}$ have the same meanings as per Equation \ref{eq:sersnorm}.

An example specification of a broken exponential model in the \R implementation of \profit that directly calls the \libprofit API to generate model images, where $x_{cen}=$\hspace{1mm}xcen, $y_{cen}=$\hspace{1mm}ycen, $m=$\hspace{1mm}mag, $h_1=$\hspace{1mm}h1, $h_2=$\hspace{1mm}h2, $\R_b=$\hspace{1mm}rb, $a=$\hspace{1mm}a, $\theta=$\hspace{1mm}ang, $A_{rat}=$\hspace{1mm}axrat, and $B=$\hspace{1mm}box, is

\begin{lstlisting}
modellist = list(
  brokenexp = list(
    xcen   = 40,
    ycen   = 60,
    h1 = 10,
    h2 = 5,
    rb = 8,
    ang = 30,
    a = 0.2,
    axrat  = 0.6,
    box = -0.1
  )
)
\end{lstlisting}

\subsubsection{Moffat Profile}
\label{subsubsec:moffat}

The Moffat function was designed as a means to parameterize a telescope PSF \citep{moff69}, and is often used on occasions when the PSF is predicable and when you might not have many point sources within the field of view to estimate it empirically (e.g.\ for space telescopes with small fields-of-view like the {\it Hubble Space Telescope}). Whilst this was the purpose behind its design, in practice in modern applications, it is rare to see analytic approximations of the PSF (such as the Moffat) used, the modern preference being for empirically derived estimates. Regardless, for reasons of flexibility, it is included as a profile option in \libprofit{}. The Moffat is a re-parameterization of the bivariate Student-T distribution, meaning it creates a Normal-like core with Lorentzian-like wings. The intensity profile can be parameterized as

\begin{equation}
I(R_m)=I_0 \left[1+\left(\frac{R_m}{R_d}\right)^2 \right]^{-c},
\end{equation}

\noindent where

\begin{equation}
R_d=\frac{FWHM}{2\sqrt{2^\frac{1}{c}-1}},
\end{equation}

\noindent where $I_0$ is the intensity at the centre of the profile, $c$ is the profile concentration ($c=1$ is pure Lorentzian, $c=\infty$ is pure Normal), and $FWHM$ is the full-width half max of the profile across the major-axis of the intensity profile. $R_m$ is the modified radius where we want to make the evaluation, which in one dimension is simply the radius from the profile centre. In two dimensions, $R_m$ has a more complex form where, as with the \sersic profile, we can use this basic form and compute annuli that are modified from pure circular ellipses by introducing geometric transformations that compute an effective $R_m$, as per Equation \ref{eq:Rmod}. Thus in \libprofit{} the Moffat function need not be circular, which is often the case for the PSF of a wide-field telescope where corner distortions tend to be radial or tangentially aligned with respect to the field centre.

Accounting for the allowed geometric distortions of axial ratio and boxiness, the value of $I_0$ for a given magnitude can be calculated analytically with:

\begin{eqnarray}
L_{tot} &=& \frac{\pi A_{rat} {R_d}^2}{(c-1) R_{box}},\\
I_0&=&\frac{10^{-0.4(m-m_{zero})}}{L_{tot}},
\end{eqnarray}

\noindent where $m$, $m_{zero}$ and $R_{box}$ have the same meanings as per Equation \ref{eq:sersnorm}.

An example specification of three Moffat models in the \R implementation of \profit that directly calls the \libprofit API to generate model images, where $x_{cen}=$\hspace{1mm}xcen, $y_{cen}=$\hspace{1mm}ycen, $m=$\hspace{1mm}mag, $FWHM=$\hspace{1mm}fwhm, $c=$\hspace{1mm}con, $\theta=$\hspace{1mm}ang, $A_{rat}=$\hspace{1mm}axrat, and $B=$\hspace{1mm}box, is

\begin{lstlisting}
modellist = list(
  moffat = list(
    xcen   = c(34,10,85),
    ycen   = c(74,64,13),
    mag = c(10,13,16),
    fwhm  = c(3, 3, 3),
    con  = c(5, 5, 5),
    ang  = c(45, 45, 45),
    axrat  = c(0.95, 0.95, 0.95),
    box = c(0.1, 0.1, 0.1)
  )
)
\end{lstlisting}

Since the Moffat profile is almost exclusively used to model the image PSF, \libprofit{} is hard-coded to never convolve the profile with another PSF, even if this is provided. This is to help eliminate potential user error.

\subsubsection{Modified Ferrer/s Profile}

The modified Ferrer profile (a type of projected potential derived from Ferrers' functions, as discussed in \citet{laur05}) is a useful parameterization of galaxy features that have very strong drop-offs, most typically bar structures. Whilst it should formally be called the Ferrers profile, it is very common for it to be written as Ferrer (the assumption is that authors have erroneously attributed the final `s' as possessive, but the name derives from Norman Macleod Ferrers). For convenience, the profile can be named as Ferrer or Ferrers in \libprofit and in higher level interfaces. The intensity can be parameterized as

\begin{equation}
I(R_m)=I_0 \left[1-\left(\frac{R_m}{R_{out}}\right)^{(2-b)}\right]^{a},
\end{equation}

\noindent where $I_0$ is the intensity at the centre of the profile, $R_{out}$ is the outer truncation radius (the profile is 0 beyond this radius), $a$ controls the global power-law slope to the profile centre, and $b$ controls the strength of truncation as $R_M$ approaches $R_{out}$. $R_m$ is the modified radius where we want to make the evaluation, which in one dimension is simply the radius from the profile centre. In two dimensions $R_m$ has a more complex form where, as with the \sersic profile, we can use this basic form and compute annuli that are modified from pure circular ellipses by introducing geometric transformations that compute an effective $R_m$, as per Equation \ref{eq:Rmod}. For the Ferrer profile it is common to introduce some positive boxiness to reflect the visually rectangular isophotes of galaxy bars.

Accounting for the allowed geometric distortions of axial ratio and boxiness, the value of $I_0$ for a given magnitude can be calculated analytically with

\begin{eqnarray}
L_{tot} &=& \frac{\pi A_{rat} {R_{out}}^2 a \beta\left(a, 1+\frac{2}{2-b}\right)}{R_{box}},\\
I_0&=&\frac{10^{-0.4(m-m_{zero})}}{L_{tot}},
\end{eqnarray}

Below is a sample specification of a Ferrer model in the \R implementation of \profit that directly calls the \libprofit API to generate model images, where $x_{cen}=$\hspace{1mm}xcen, $y_{cen}=$\hspace{1mm}ycen, $m=$\hspace{1mm}mag, $R_{out}=$\hspace{1mm}rout, $a=$\hspace{1mm}a, $b=$\hspace{1mm}b, $\theta=$\hspace{1mm}ang, $A_{rat}=$\hspace{1mm}axrat and $B=$\hspace{1mm}box:

\begin{lstlisting}
modellist = list(
  ferrer = list(
    xcen   = 50,
    ycen   = 50,
    mag = 15,
    rout = 20,
    a = 0.5,
    b=0.7,
    ang  = 30,
    axrat  = 0.3,
    box = 0.6
  )
)
\end{lstlisting}

\subsubsection{Modified Empirical King profile}

The modified King profile in \libprofit{} shares its origins with the empirically motivated two-dimensional profile presented in \citet{king62}, and remains a popular function for parameterizing globular cluster light profiles. It should be emphasized that it is {\it not} the same as the equally popular profile presented in \citet{king66}, the latter being parameterized in three dimensions, theoretical in origin and non-trivial to project. \libprofit{} uses the same basic modification to the \citet{king62} as used in \galfit, where the intensity can be parameterized as

\begin{equation}
I(R_m)=I_r \left\{\frac{1}{\left[1+\left(\frac{R_m}{R_c}\right)^2\right]^\frac{1}{a}}-\frac{1}{\left[1+\left(\frac{R_t}{R_c}\right)^2\right]^\frac{1}{a}}\right\}^a,
\end{equation}

\noindent where $I_r$ is the intensity at a parameter-dependent reference point (the centre if $R_t=\infty$), $R_c$ is the core radius, $R_t$ is the outer truncation radius (the profile is 0 beyond this radius) and $a$ controls the global power-law. When $a=2$, this parameterisation is identical to the one presented originally in \citet{king62}. $R_m$ is the modified radius where we want to make the evaluation, which in one dimension is simply the radius from the profile centre. In two dimensions $R_m$ has a more complex form where, as with the \sersic profile, we can use this basic form and compute annuli that are modified from pure circular ellipses by introducing geometric transformations that compute an effective $R_m$, as per Equation \ref{eq:Rmod}.

Accounting for the allowed geometric distortions of axial ratio and boxiness, the value of $I_0$ for a given magnitude can be calculated analytically with:

\begin{eqnarray}
L_{tot} &=& \frac{2 \pi A_{rat} \int_0^{R_t} R_m I(R_m; I_r=1) dR_m}{R_{box}},\\
I_r&=&\frac{10^{-0.4(m-m_{zero})}}{L_{tot}},
\end{eqnarray}

An example specification of a modified King profile in the \R implementation of \profit that directly calls the \libprofit API to generate model images, where $x_{cen}=$\hspace{1mm}xcen, $y_{cen}=$\hspace{1mm}ycen, $m=$\hspace{1mm}mag, $R_{c}=$\hspace{1mm}rc, $R_{t}=$\hspace{1mm}rt $a=$\hspace{1mm}a, $\theta=$\hspace{1mm}ang, $A_{rat}=$\hspace{1mm}axrat and $B=$\hspace{1mm}box, is

\begin{lstlisting}
modellist = list(
  king = list(
    xcen   = 50,
    ycen   = 50,
    mag = 15,
    rc = 5,
    rt = 30,
    a = 2,
    ang  = 30,
    axrat  = 0.8,
    box = 0
  )
)
\end{lstlisting}

\subsubsection{Point Spread Function Profile}
\label{subsubsec:psf}

The PSF profile for an unresolved point source (referred to as a {\tt pointsource} within \libprofit and \profit to distinguish it from the PSF) is a simple prescription requiring the $x_{cen},y_{cen}$ location of the desired point source, the total flux contained in magnitudes $m$ and either an empirical PSF image or analytic PSF model. In the former case, the PSF image is linearly interpolated on to the specified location in the image (thus fractional pixel values can be provided) and is renormalized to ensure the flux contained is correct given the magnitude zero-point ($m_{zero}$); otherwise, the point source is accurately integrated with the specified PSF model using an analytic profile (e.g.\ the Moffat, as discussed in Section \ref{subsubsec:moffat}). An example of a model containing three point sources in the \R implementation of \profit, where $x_{cen}=$\hspace{1mm}xcen, $y_{cen}=$\hspace{1mm}ycen and $m=$\hspace{1mm}mag, is

\begin{lstlisting}
modellist = list(
  pointsource = list(
    xcen   = c(34,10,85),
    ycen   = c(74,64,13),
    mag = c(10,13,16)
  )
)
\end{lstlisting}

Numerical integration of an analytic PSF is considerably more accurate than interpolation of an empirical PSF image, so we generally recommend using the former approach if it is at all possible. As mentioned in Section \ref{subsubsec:moffat}, it is more typical to see astronomers using empirical PSFs. Since the intrinsic PSF is unknown, using such a description will result in a simple linear interpolation to fractional pixel locations. To accommodate this fact, \libprofit{} and \profit allow for the specification of an oversampled empirical PSF. If PSF modelling is required, this offers both a pragmatic and accurate solution.

\subsubsection{Sky Profile}

The simplest profile of all is the sky. In \libprofit{}, this is specified as the flux per pixel without any adjustment made for the magnitude zero-point ($m_{zero}$), i.e. it is the flat sky level pedestal directly measured from the image in its native flux units (any subsequent conversions must be made explicitly by the user). An example sky model in the \R implementation of \profit requires only the background level bg:

\begin{lstlisting}
modellist = list(
  sky = list(
    bg = 3e-12
  )
)
\end{lstlisting}

Currently only flat sky profiles are available in \libprofit{} and \profit, with the assumption that in typical \profit use cases the sky is a pedestal term, with no gradient of curvature. Depending on community feedback additional complexity for the sky profile could be added in future. However, it is clear that the majority of modern survey image processing involves an explicit sky-subtraction stage that reduces the data to a pedestal-only sky profile \citep[e.g.\ see][]{kuij15}. Like any other profile type, it is possible to fit for the sky profile background within \libprofit; however, in many use cases, the background will be known a-priori and should not be fit. A common issue with the leaving the sky as a fitted parameter is how degenerately it behaves with \sersic-index outer profiles, which also appear close to flat at many multiples of $R_e$. To guard against this effect, it is wise to fit using pixels that are dominated by galaxy flux.

\subsubsection{A Combined Model}

Combining models in \libprofit{} is simple given the nested list nature of object specification. A structure containing the \sersic, Ferrer bar, point-source and sky models is given by

\begin{lstlisting}
modellist = list(
  sersic = list(
    xcen   = c(50, 50),
    ycen   = c(50, 50),
    mag = c(12, 13),
    re  = c(14, 5),
    nser  = c(1, 8),
    ang  = c(46, 80),
    axrat  = c(0.4, 0.6),
    box = c(0,-0.3)
  ),
  ferrer = list(
    xcen = 50,
    ycen = 50,
    mag = 14,
    rout = 12,
    a = 0.3,
    b = 1.5,
    ang = 130,
    axrat = 0.2,
    box = 0.5
  ),
  pointsource = list(
    xcen   = c(34,10,85),
    ycen   = c(74,64,13),
    mag = c(18,15,16)
  ),
  sky = list(
    bg = 3e-12
  )
)
\end{lstlisting}

This is the list structure used in \R{} \profit to specify models, and in principle an unlimited number of profiles can be provided to create complex model images. In practice, for galaxy image fitting simple models are made of the region of direct interest. This structure is also the mechanism for providing initial conditions for galaxy image fitting. This structure can be easily manipulated from outside of the \R environment, or directly within \R.

\begin{figure}
	\includegraphics[width=\columnwidth]{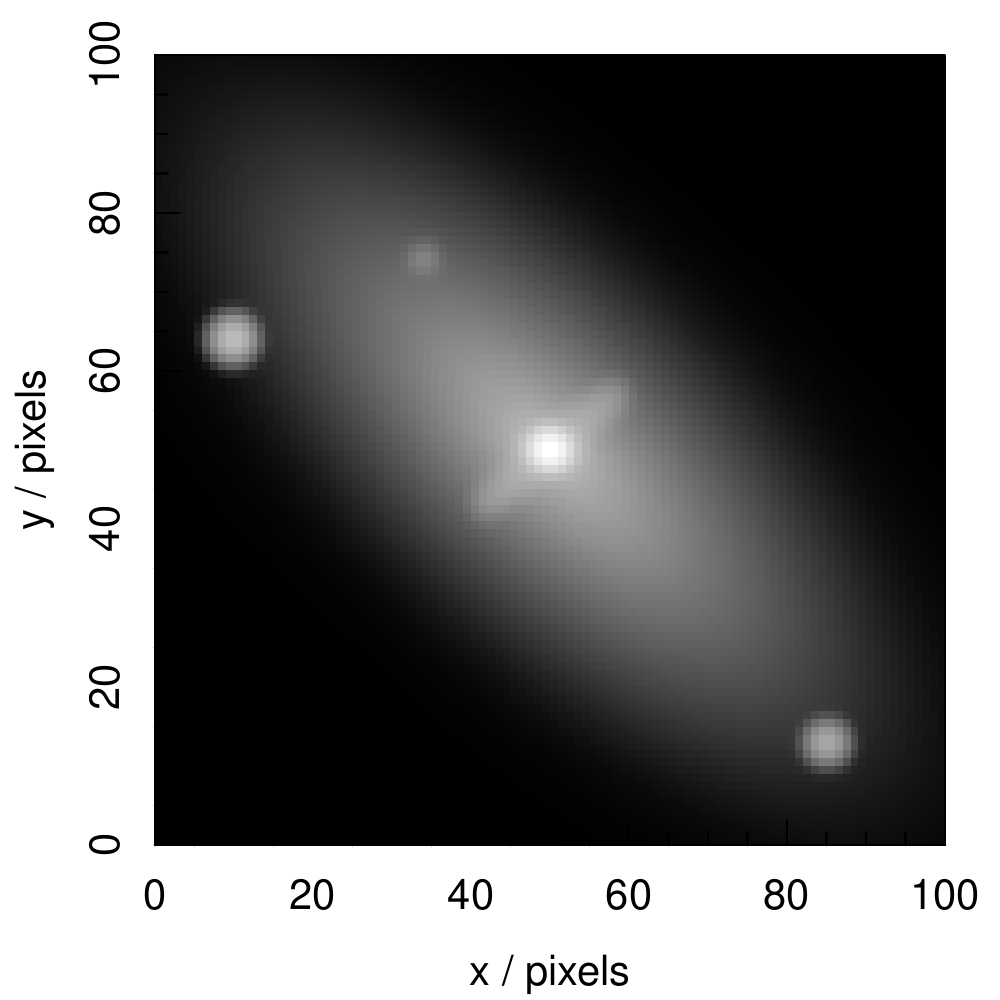}
	\caption{Example \libprofit{} model image with convolution by the PSF, which is an idealised Gaussian PSF with full width half max (FWHM) of 1.3 pixels. For clarity no surface brightness contours are over-drawn. This image has the same two-component \sersic model as shown in Figure \ref{fig:examplesersicmodel}, but with the addition of a Ferrer bar profile, three PSFs and a sky background.}
	\label{fig:examplecombmodel}
\end{figure}

Fig.~\ref{fig:examplecombmodel} shows the image generated by the example model above. This is very similar to Fig.~\ref{fig:examplesersicmodel} but with the addition of a Ferrer bar profile, three bright point sources and a sky background (which is not visible due to the image scaling). More details on how \libprofit{} generates these images is provided in the following section.

\subsection{Accurate and Fast Pixel Integration}
\label{sec:profit-acc}

The main focus of development effort in \libprofit{} was optimizing model image generation for speed {\it and} accuracy, particularly for the \sersic profile (since this is by some margin the most popular general purpose profile for galaxy modelling). The first requirement to achieve rapid image generation was that the core code should be written in a fast low-level language. {\textsc c++} was chosen for these purposes, with the idea that \libprofit{} should provide a generic low-level library that higher level languages (e.g.\ \R and \Python) can access easily.

With {\textsc c++} chosen as the implementation language, the main considerations left were how to best approach the problem of model image generation algorithmically. Two-dimensional modelling routines are circumspect in how they achieve pixel integration, with most published references stating that inner pixels are `oversampled' in order to achieve reasonable accuracy in the pixel flux (note this is different to the type of oversampling required to ensure accurate image convolution, as discussed in Section \ref{subsec:convolution}). Depending on the profile being created the degree of oversampling can vary by many orders of magnitude. Getting accurate solutions for the flux in a pixel containing the peak of a steep (high $n$) \sersic profile is clearly a very different prospect to estimating the flux of a pixel a large distance from the centre, where the profile is very flat. To combat this problem, \libprofit{} uses a unified oversampling scheme (inspired by adaptive quadrature) for all profiles that have an analytic radial intensity profile --- initially the \sersic, Core-\sersic, broken-exponential, Moffat, Ferrer, and King profiles discussed above. Here, we concentrate on the evaluation of the \sersic profile, but the basic principles hold true for all radially varying profiles. In \libprofit{} all such profiles inherit from a generic class of radial type profiles (see Fig.~\ref{fig:profiles}), where the pixel integration scheme is shared in a common method. The profiles are specified by their radial intensity and total luminosity (as per Section \ref{sec:profiles}). This minimizes code duplication and simplifies the addition of new profiles to \libprofit{}, where the user only needs to be able to express the one-dimensional form of the profile shape and a method to integrate it to get the total luminosity (which can simply be a numerical integration if there is no convenient analytic solution).

\begin{figure*}
	\centering
	\includegraphics[width=13cm]{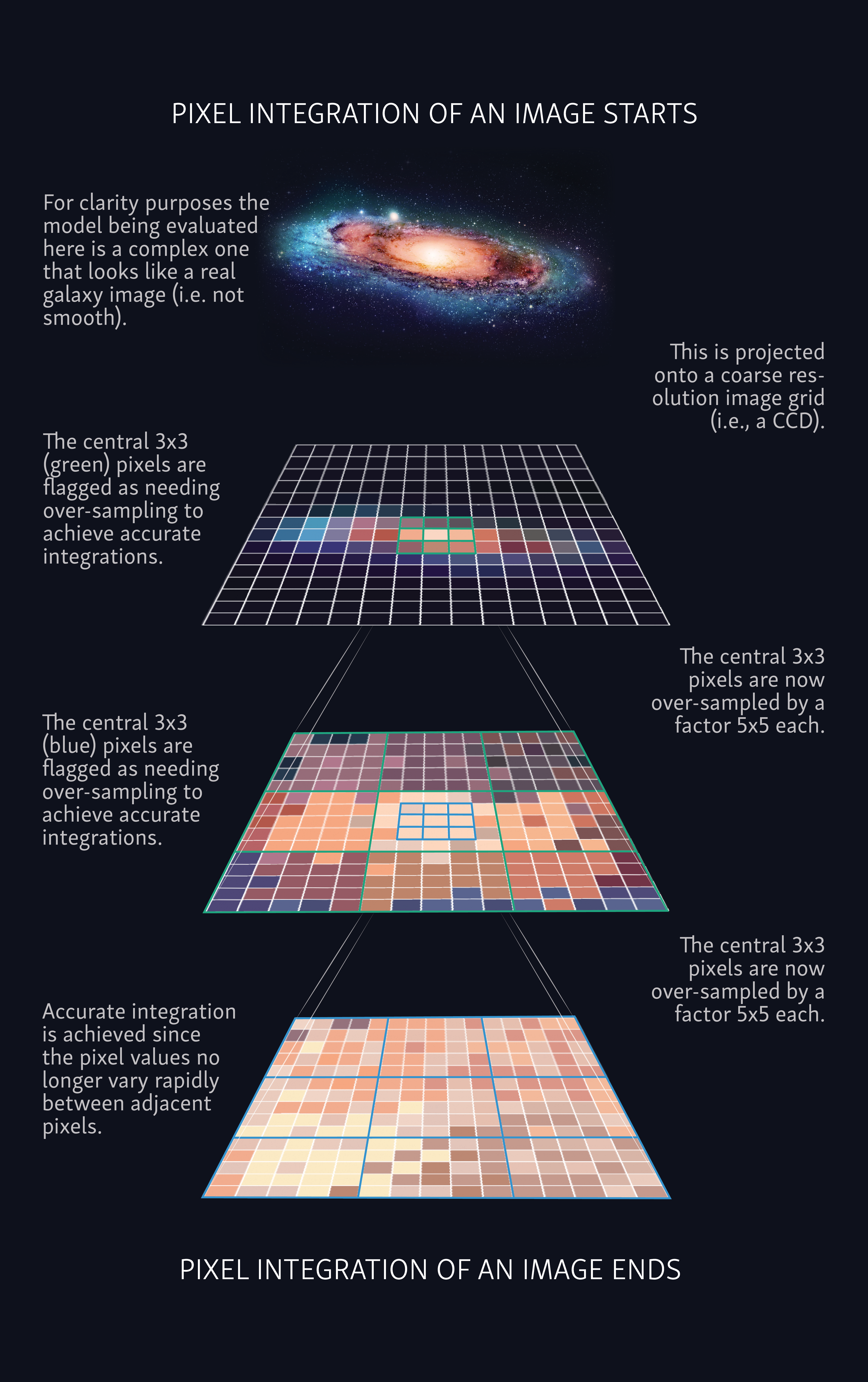}
	\caption{Simple schematic of \profit recursively up-sampling the inner most pixel in order to achieve a desired level of per-pixel integration accuracy. In this example the inner $3\times3$ pixels are up-sampled by $5\times5$ each. Of this new up-sampled grid the inner $3\times3$ pixels are up-sampled by $5\times5$ again, after which the recursion stops. In \profit this assessment is made by the rate-of-change of flux between pixels, i.e.\ if the change is rapid then pixels are identified as requiring up-sampling.}
	\label{fig:pixint}
\end{figure*}

The qualitative version of the pixel integration scheme is that an oversampling resolution, a recursion depth, and an accuracy level is specified for a given pixel. The grid is initially oversampled by the integer factor provided, for each sub-pixel the gradient along the minor axis (this being the steepest descent) is computed and when it is over a tolerance specified by the accuracy value this pixel is selected for recursive oversampling. In this situation, the sub-pixel becomes redefined as the parent pixel and the scheme above is repeated. Fig.~\ref{fig:pixint} is a simple schematic of the central $3\times3$ pixels being recursively oversampled by a factor $5\times5$ to achieve an accurate numerical integration in the regime where the profile is rising steeply in the core of the galaxy.

The default allowed maximum depth for sub-pixel recursion is 2, but it can be user-specified. The default oversampling factor is 8, meaning at most sub-regions of pixels would be oversampled by a factor $(8^2)^2=8^4$. A special case is made for the central pixel in the \sersic profile (containing the flux peak). To ensure accurate results even for extreme examples of very steep profiles of barely resolved galaxies, the maximum recursion depth is increased to 10, giving potential oversampling of $(8^{10})^2\sim10^{9}$. In extreme cases, where deep recursion is triggered on the very central pixel of each layer, the practical worst case scenario is that $(8^2) \times 10=640$ calculations are actually required because only the sub-pixels containing the profile peak require deeper recursion.

This combination of accuracy and speed means \libprofit{} performs well in model image generation bench-marking whilst maintaining user-definable accuracy with sensible defaults. Detailed comparisons were made with reference to \galfit, where the target was to achieve similar or better pixel level accuracy whilst ensuring the model image computation time was no longer for typical values of the \sersic index ($n$). For these comparisons, a modified build of \galfit was used that did not read or write to disk and instead took arguments and produced outputs in memory (ensuring this relatively large I/O bottleneck was not a factor).

The exact pixel fluxes were computed using the {\textsc R2CUBA} \R package that is an interface to the {\textsc CUBA} {\textsc c++} library that offers deterministic algorithms for multidimensional numerical integration. The algorithm used was {\textsc CUHRE}, which offers a user-definable level of integration accuracy. For these tests, it was set to be better than 0.1 per cent pixel accuracy (in practice achieving an order of magnitude better accuracy in most pixels), so at most this amount of inaccuracy may come from the reference {\it exact} pixel fluxes. It is worth highlighting why this approach cannot be used in general: to create a typical $20\times20$ pixel image (a very small image by galaxy fitting standards) with this level of accuracy takes $\sim$3 s (and much longer for some profiles). In comparison, \libprofit{} and \galfit can both create similar image in less than 0.01 s, i.e.\ they are both at least 300 times faster than using a multidimensional integration library, even a fast one written in {\textsc c++}. It is not clear why using this specialized library is so slow, but it ruled out using it within \libprofit for general application.

\begin{figure}
	\includegraphics[width=\columnwidth]{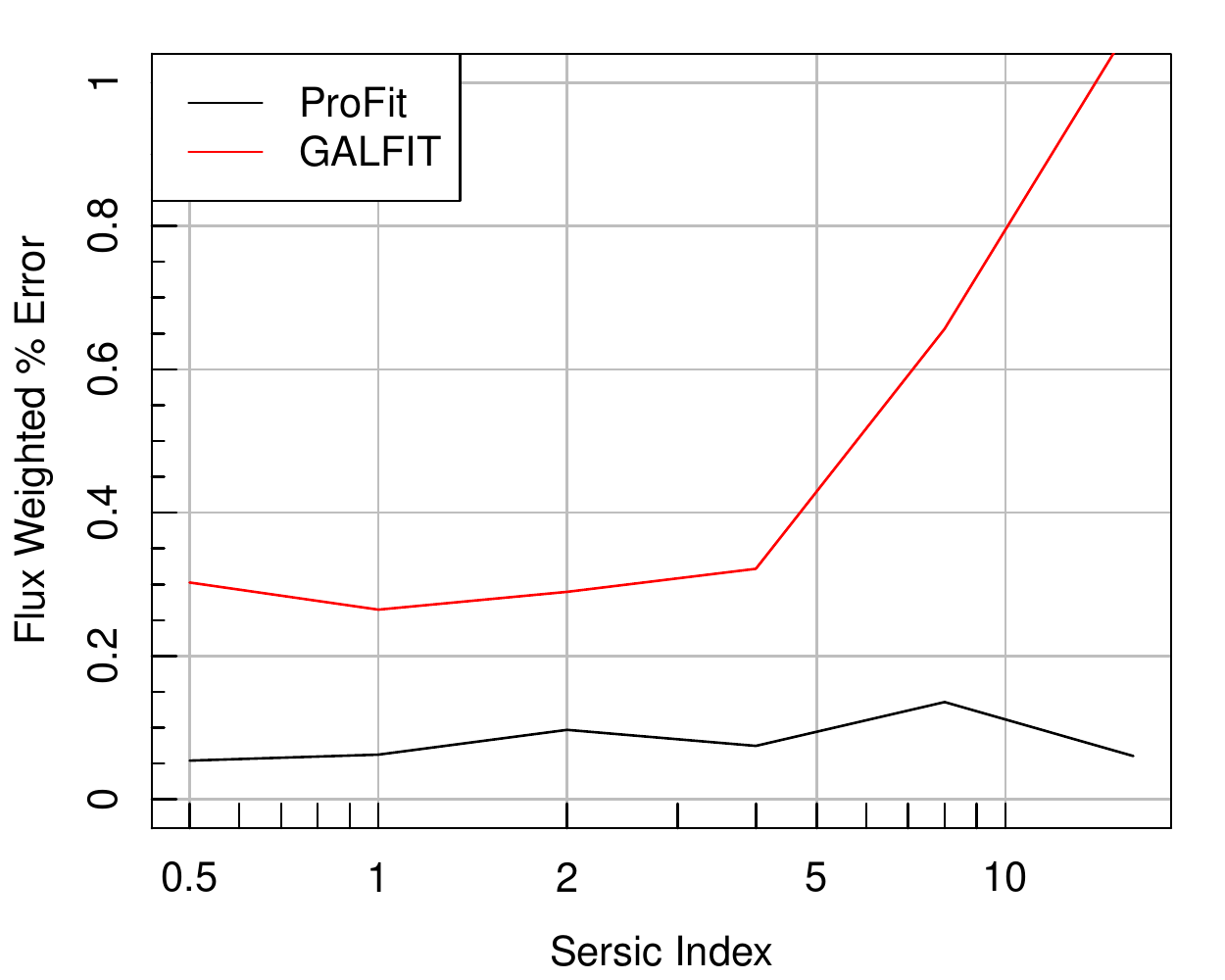}
	\caption{\libprofit{} versus \galfit flux weighted model image error for \sersic indices 0.5--16 for $R_e=2$pixels, $\theta=60^\circ$, and $A/B=0.3$ in a $20 \times 20$ pixel image. A subset of these results (for $n=1, 4, 8, 16$) are shown in detail in Fig.~\ref{fig:pixelaccuracy}.}
	\label{fig:fluxweighterror}
\end{figure}

Fig.~\ref{fig:fluxweighterror} compares the flux-weighted pixel error of \libprofit{} and \galfit as a function of \sersic index for \sersic profiles (again, \sersic profiles have been high-lighted due to their popularity in the galaxy profiling community) with $R_e=2$pixels, $\theta=60^\circ$ and $A/B=0.3$. Two things are immediately clear: \libprofit{} has a lower flux-weighted error per pixel (typically 0.1 per cent), and this error does not vary significantly with \sersic index. \galfit achieves a flux weighted error per pixel of 0.3 per cent for low values of the \sersic index, but this inflates to $>1$ per cent for larger values of $n$.

\begin{figure*}
	\centering
	Exact Image Generated with Cubic Quadrature Routine via Cuba\\
	\includegraphics[width=4.3cm]{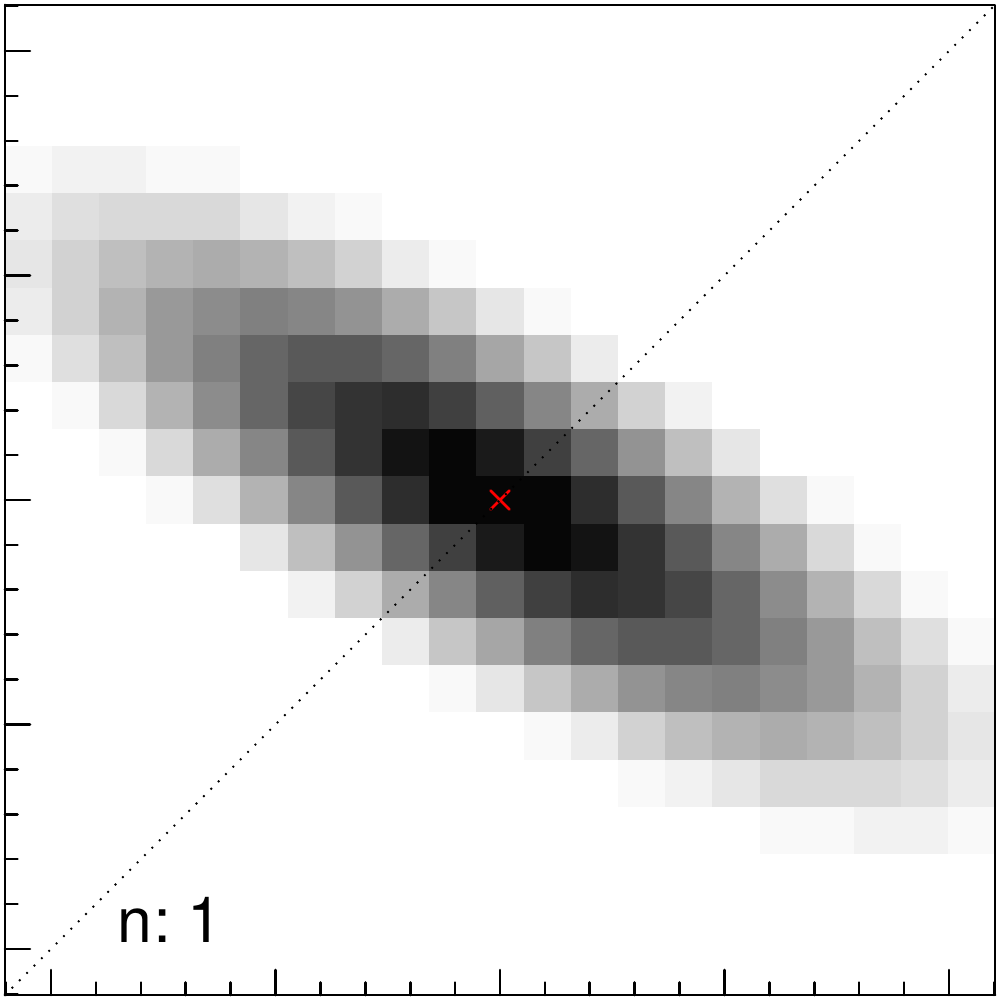}
	\includegraphics[width=4.3cm]{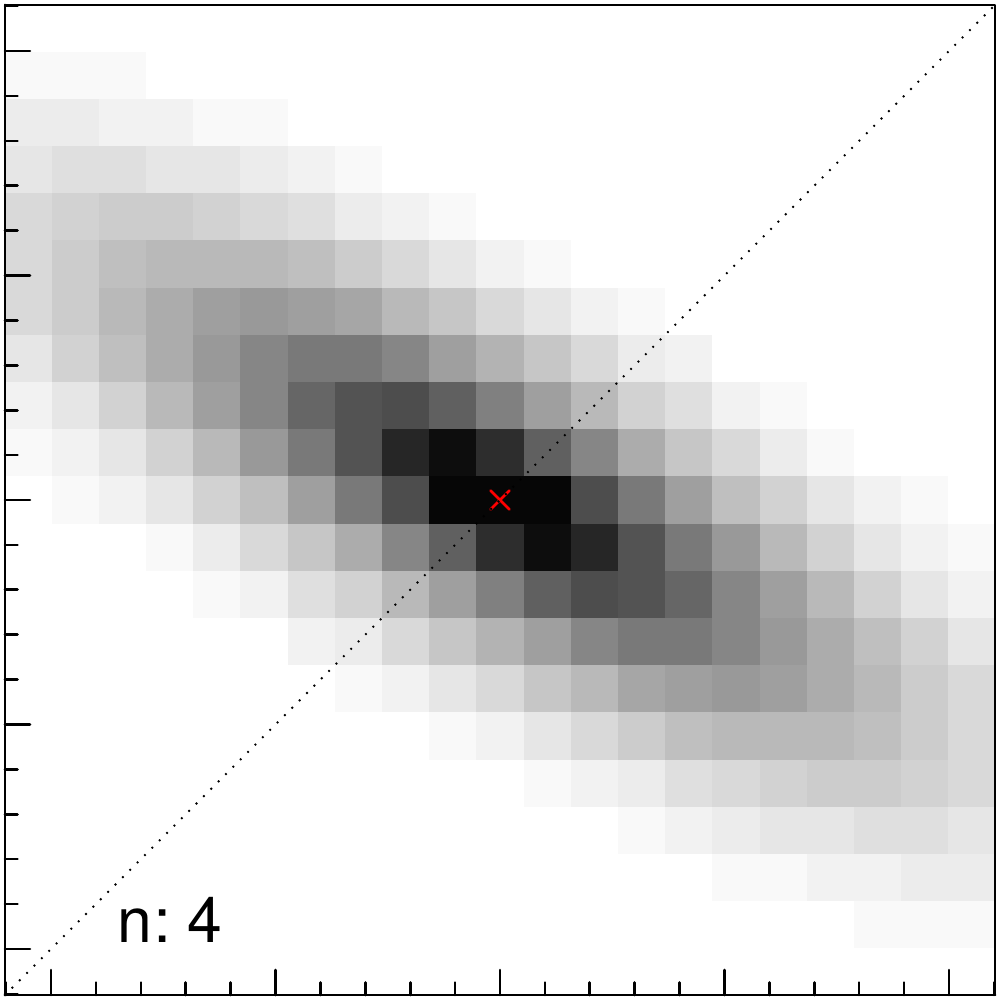}
	\includegraphics[width=4.3cm]{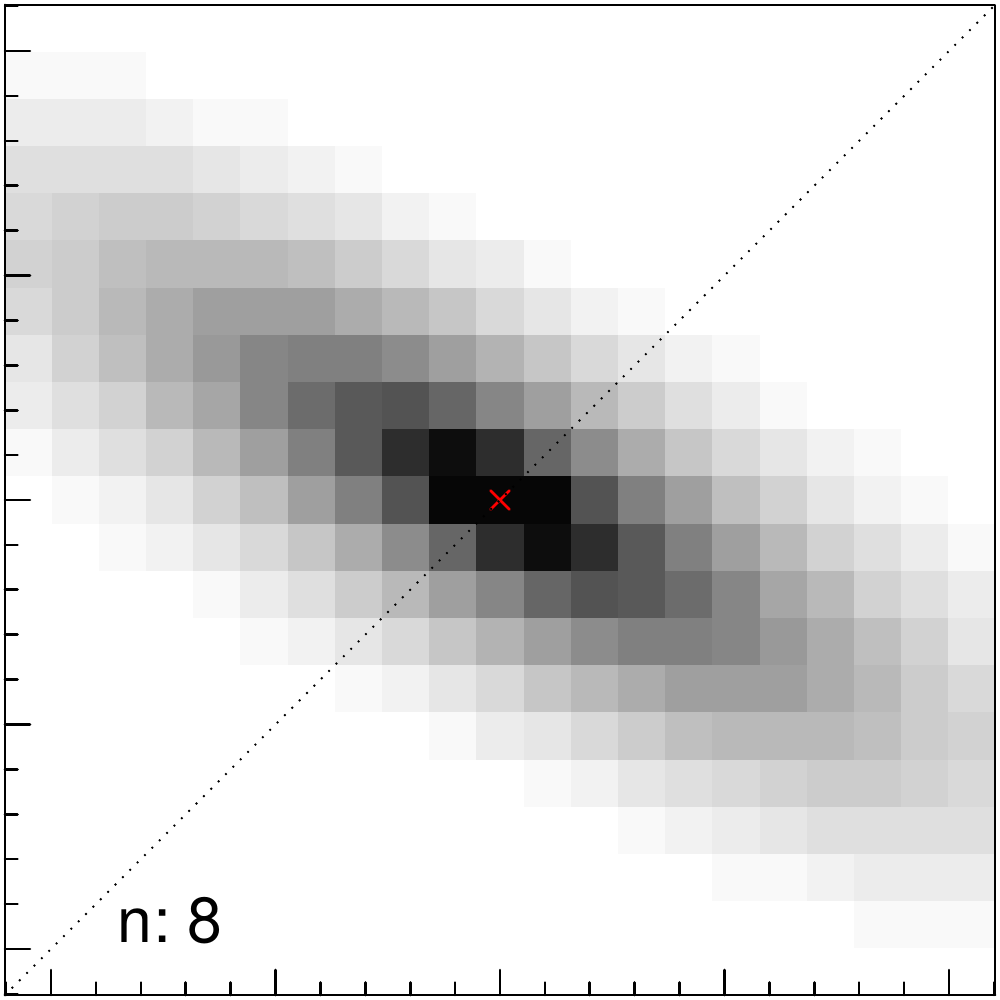}
	\includegraphics[width=4.3cm]{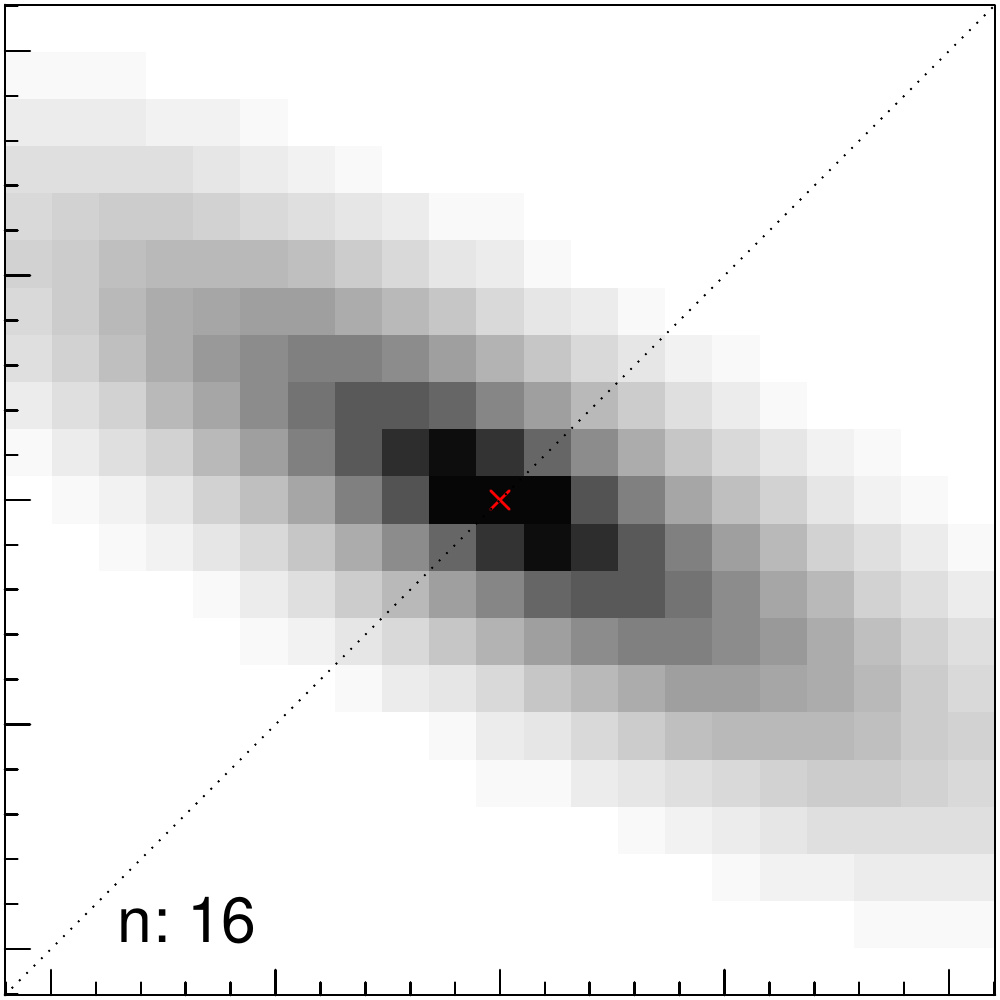}\\
	Error in Approximate \libprofit{} Image\\
	\includegraphics[width=4.3cm]{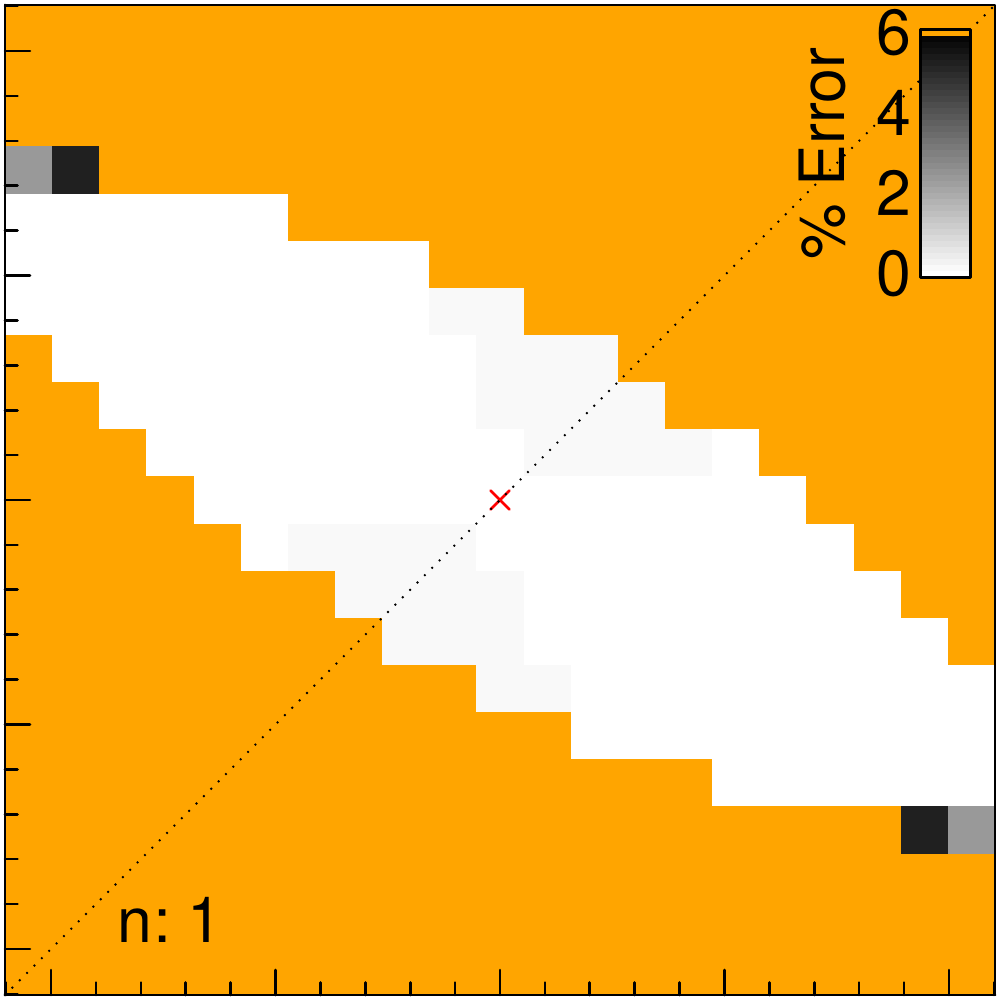}
	\includegraphics[width=4.3cm]{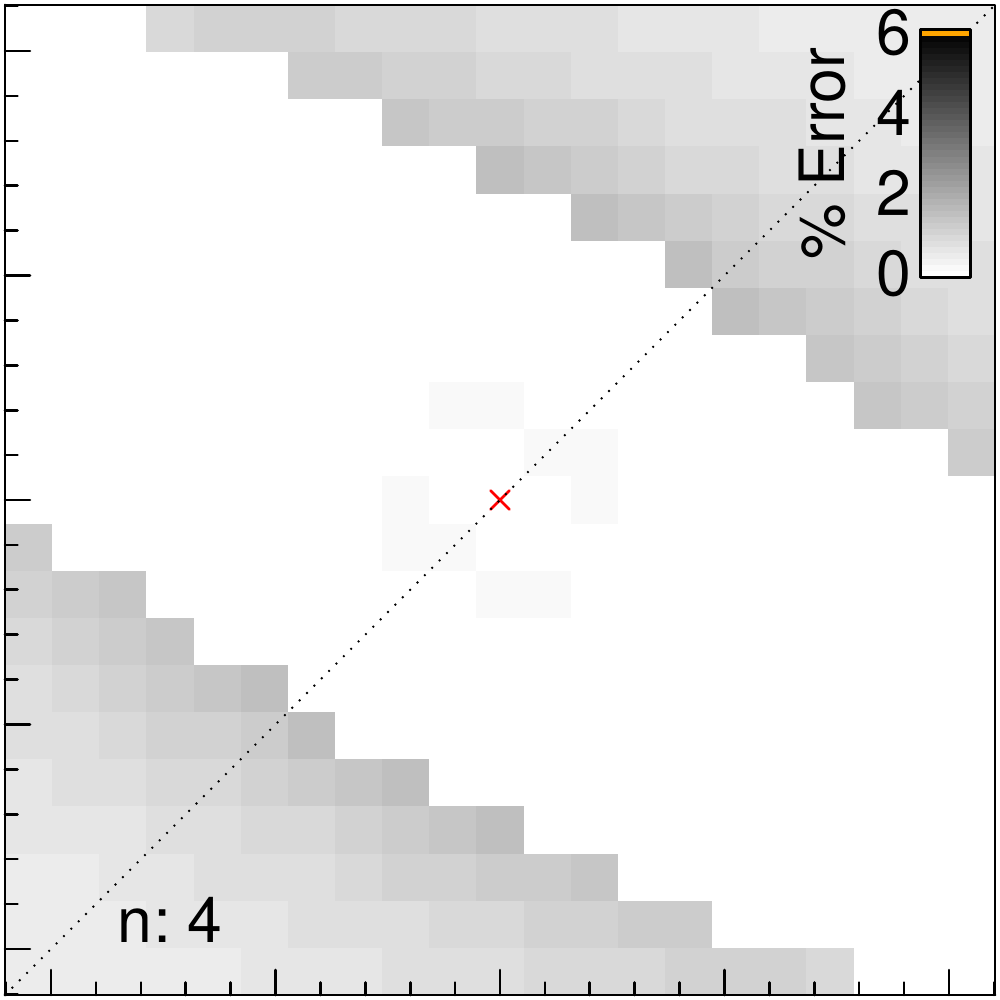}
	\includegraphics[width=4.3cm]{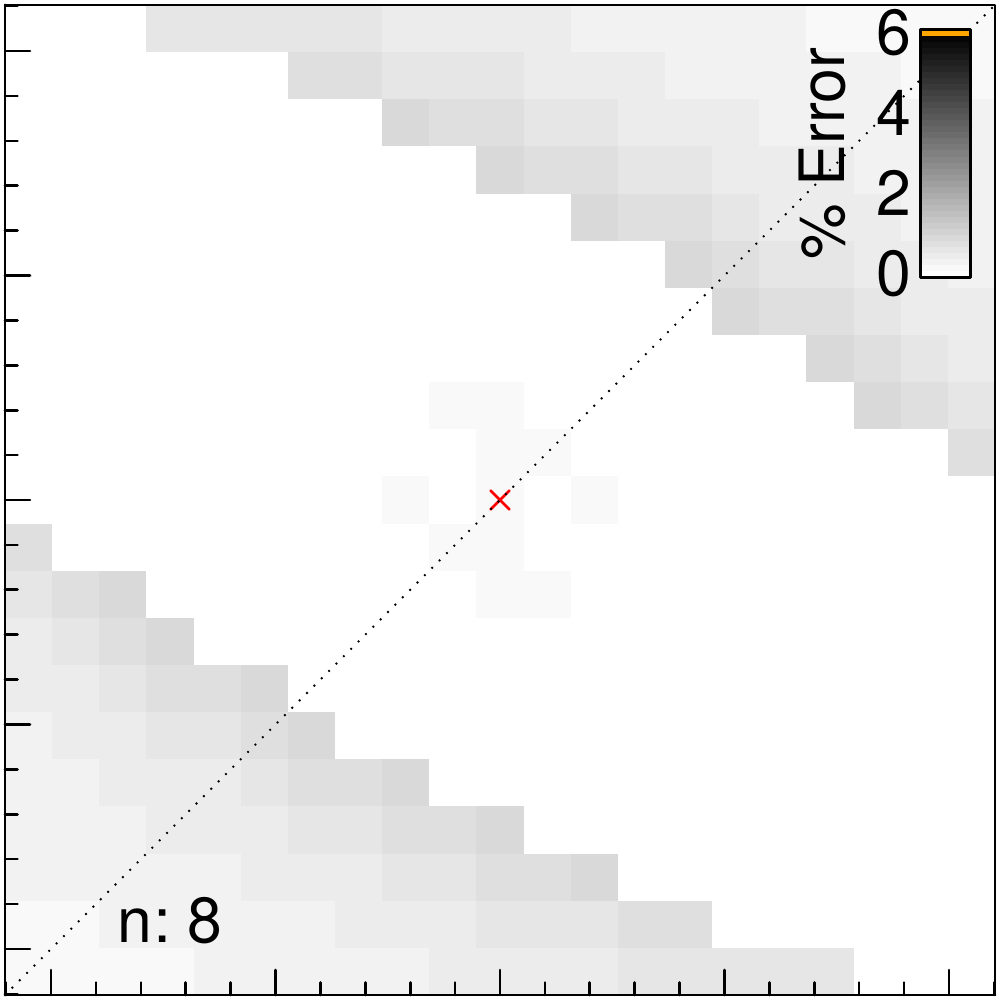}
	\includegraphics[width=4.3cm]{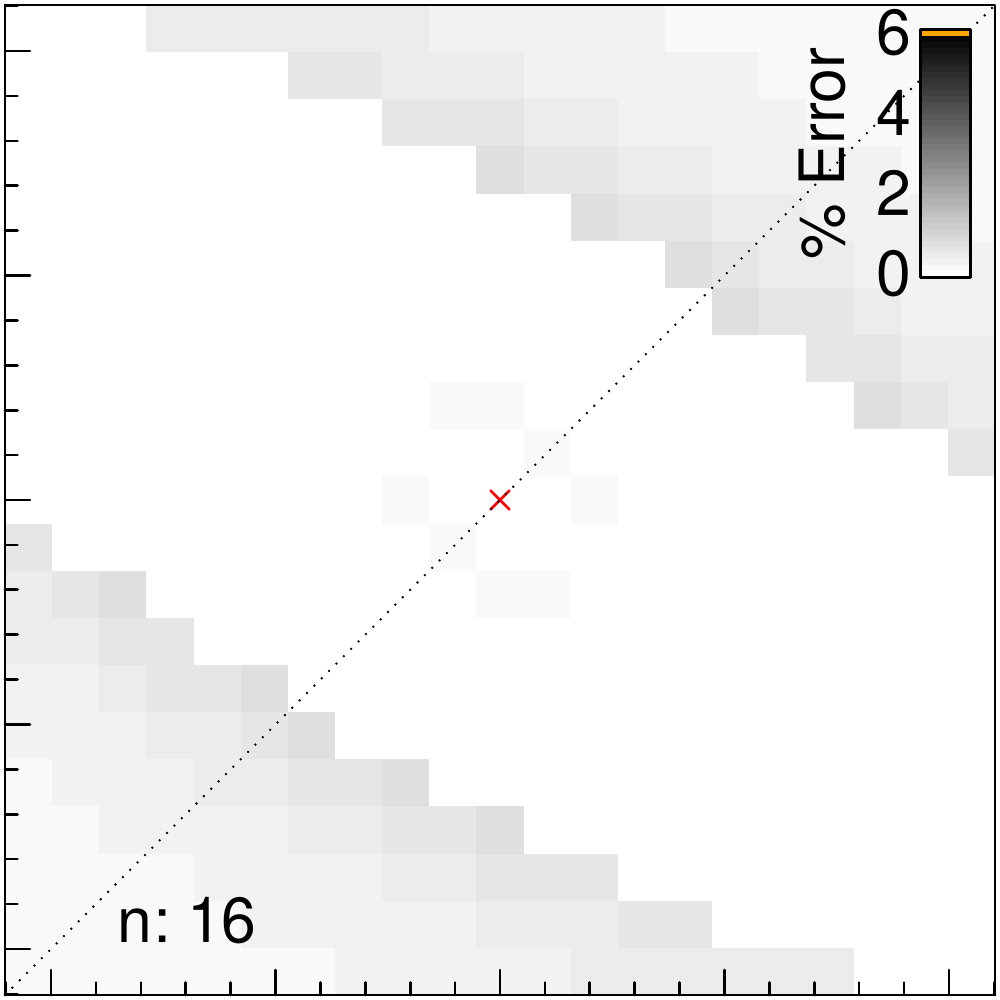}\\
	Error in Approximate \galfit Image\\
	\includegraphics[width=4.3cm]{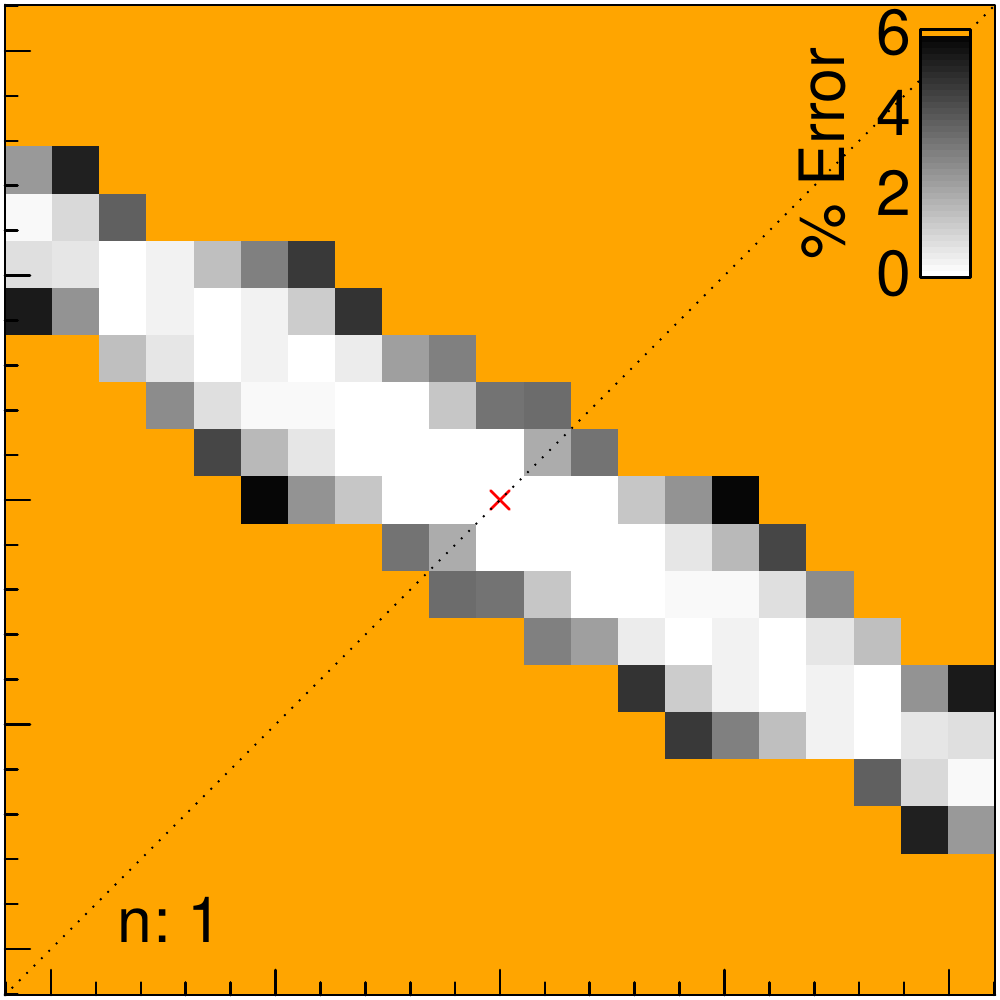}
	\includegraphics[width=4.3cm]{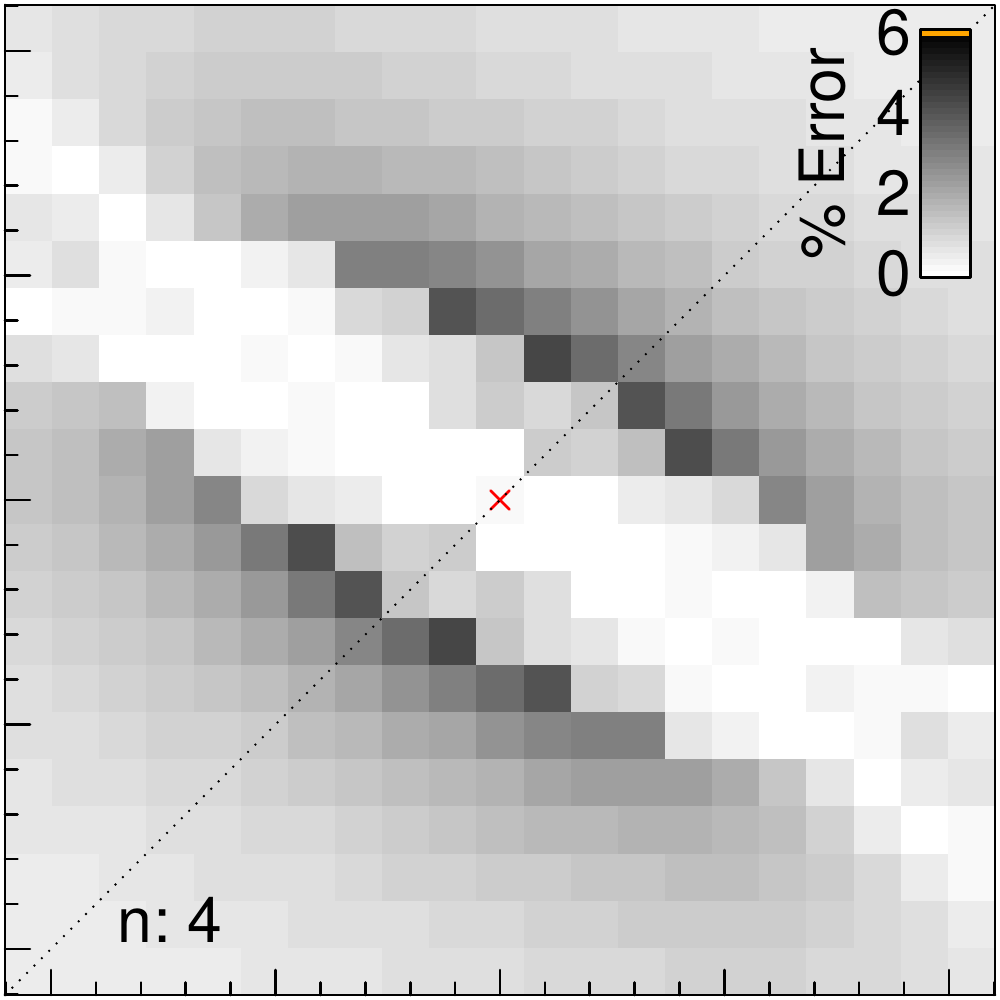}
	\includegraphics[width=4.3cm]{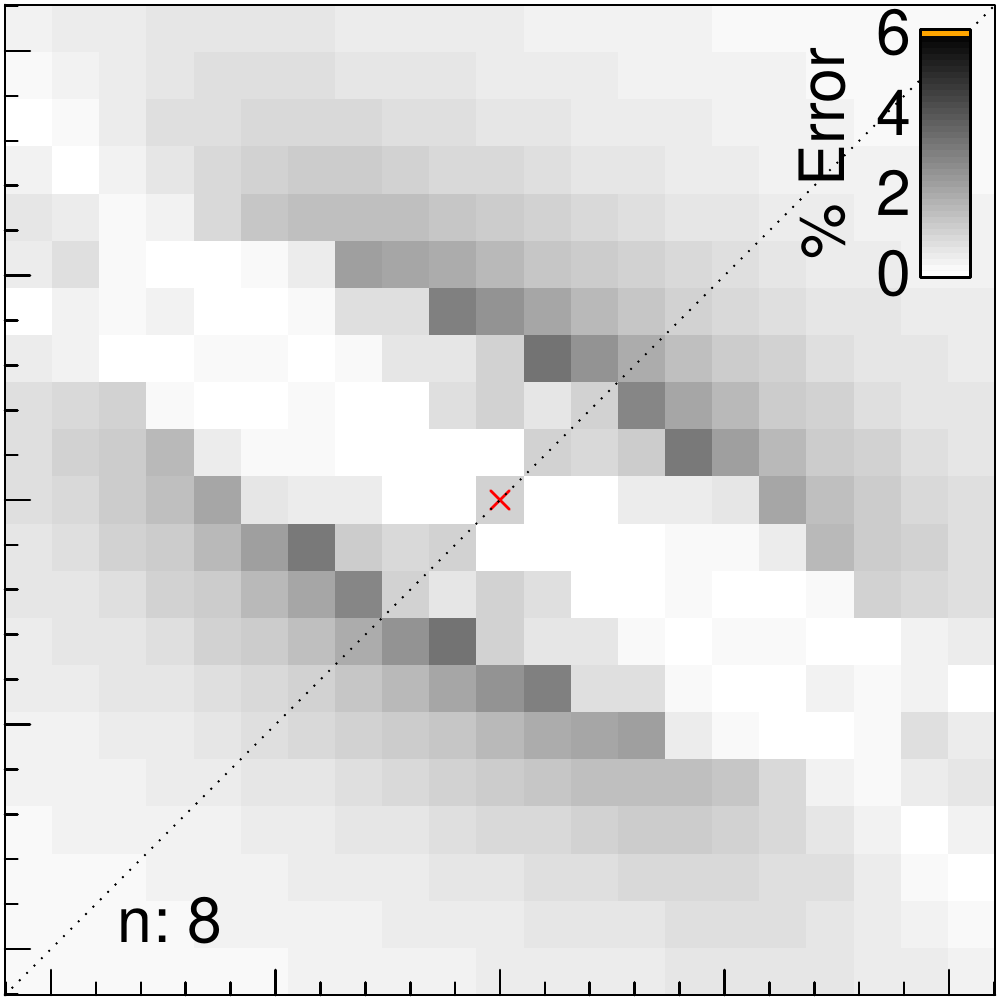}
	\includegraphics[width=4.3cm]{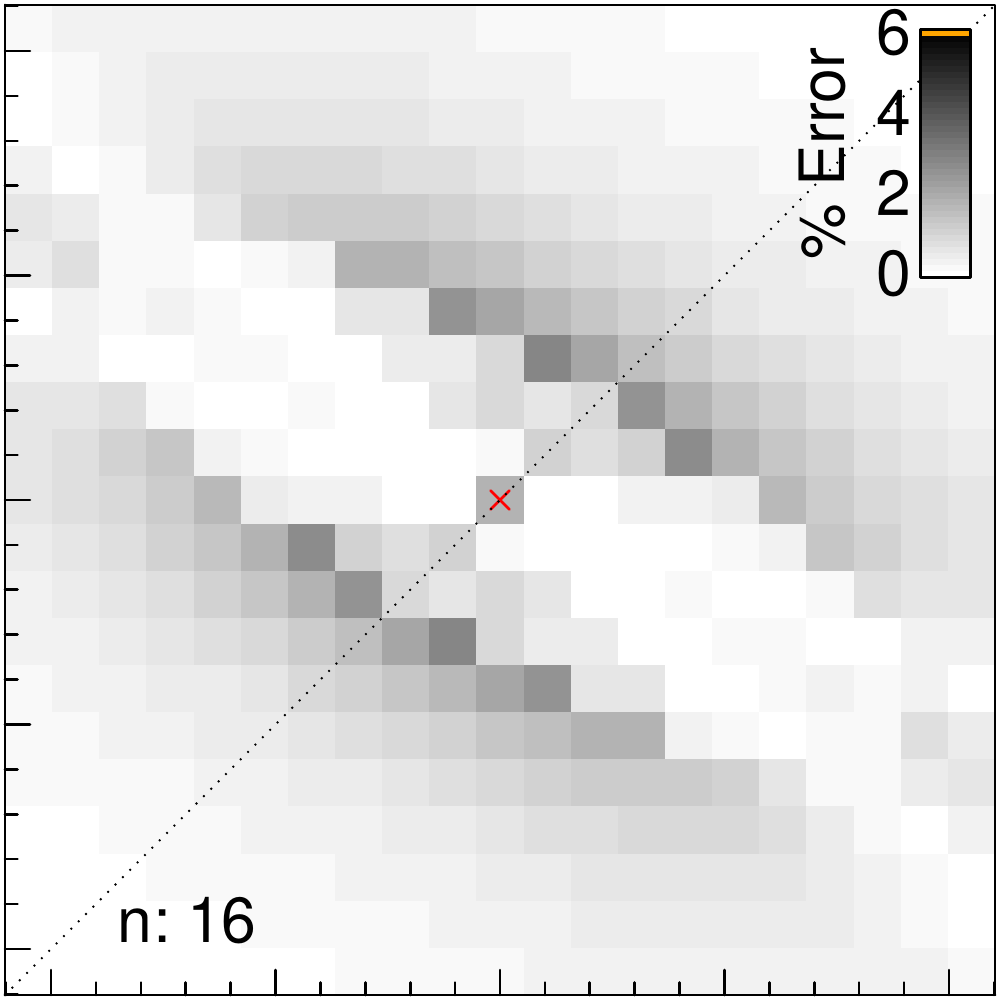}\\
	\caption{A detailed view of model error during image generation in \libprofit{} versus \galfit for \sersic indices 1, 2, 4 and 16 (as labelled), $R_e=2$pixels, $\theta=60$ degrees and $A/B=0.3$ in a $20x20$ pixel image. The examples shown here are a subset of the flux weighted errors shown in Figure \ref{fig:fluxweighterror}. The top row shows the exact Cubic Quadrature derived model being generated, using a sinh scaling for the greyscale. The second row down shows the \libprofit{} error residuals at each pixel location. The third row down shows the \galfit error residuals at each pixel location. Orange highlights regions that have errors greater than the fixed greyscale limits, i.e.\ above 6\%. The pixels along the diagonal dotted line are shown in Figure \ref{fig:pixelaccuracydiag}.}
	\label{fig:pixelaccuracy}
\end{figure*}

\begin{figure*}
	\centering
	Error Across Central Diagonal Pixels\\
	\includegraphics[width=\columnwidth]{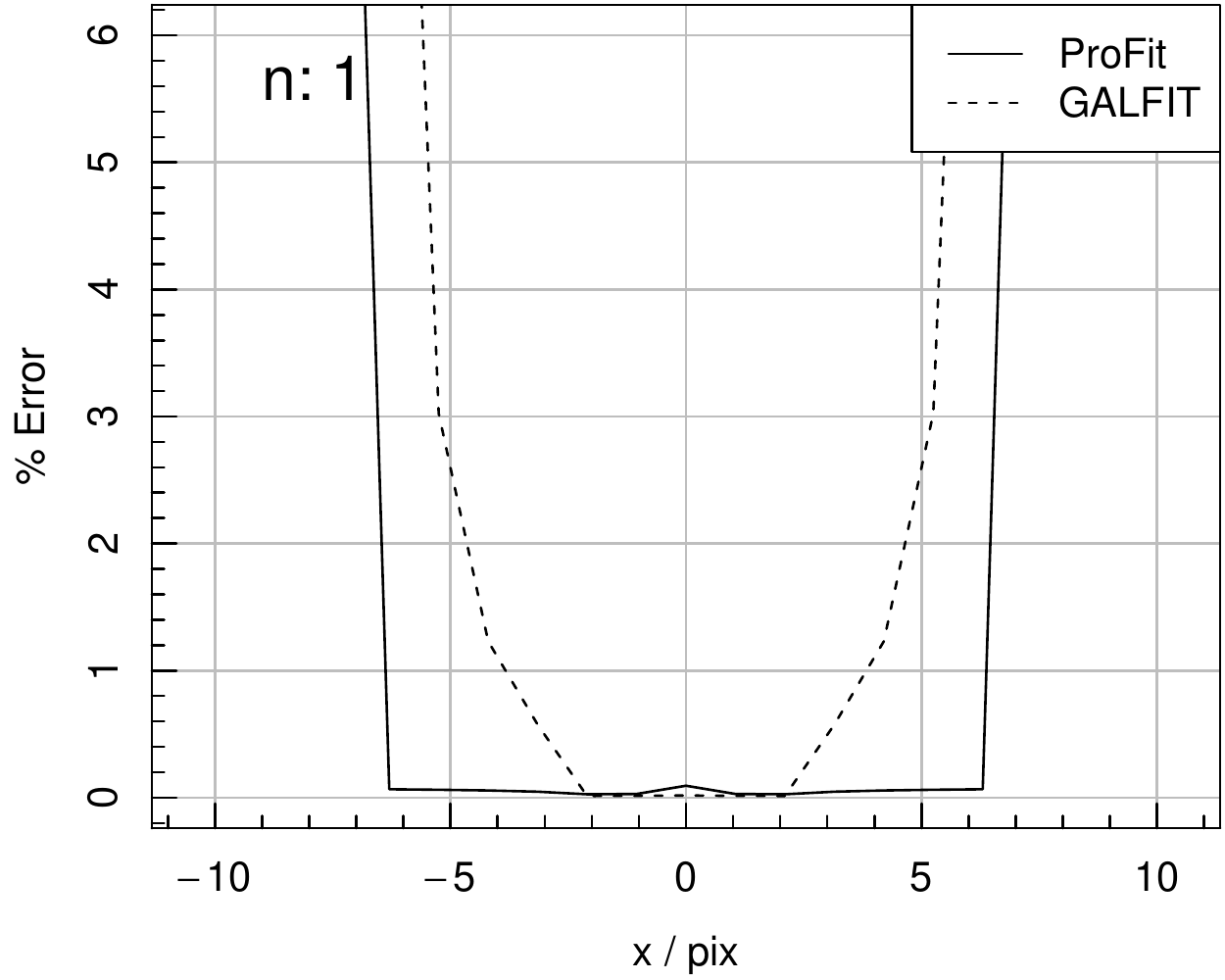}
	\includegraphics[width=\columnwidth]{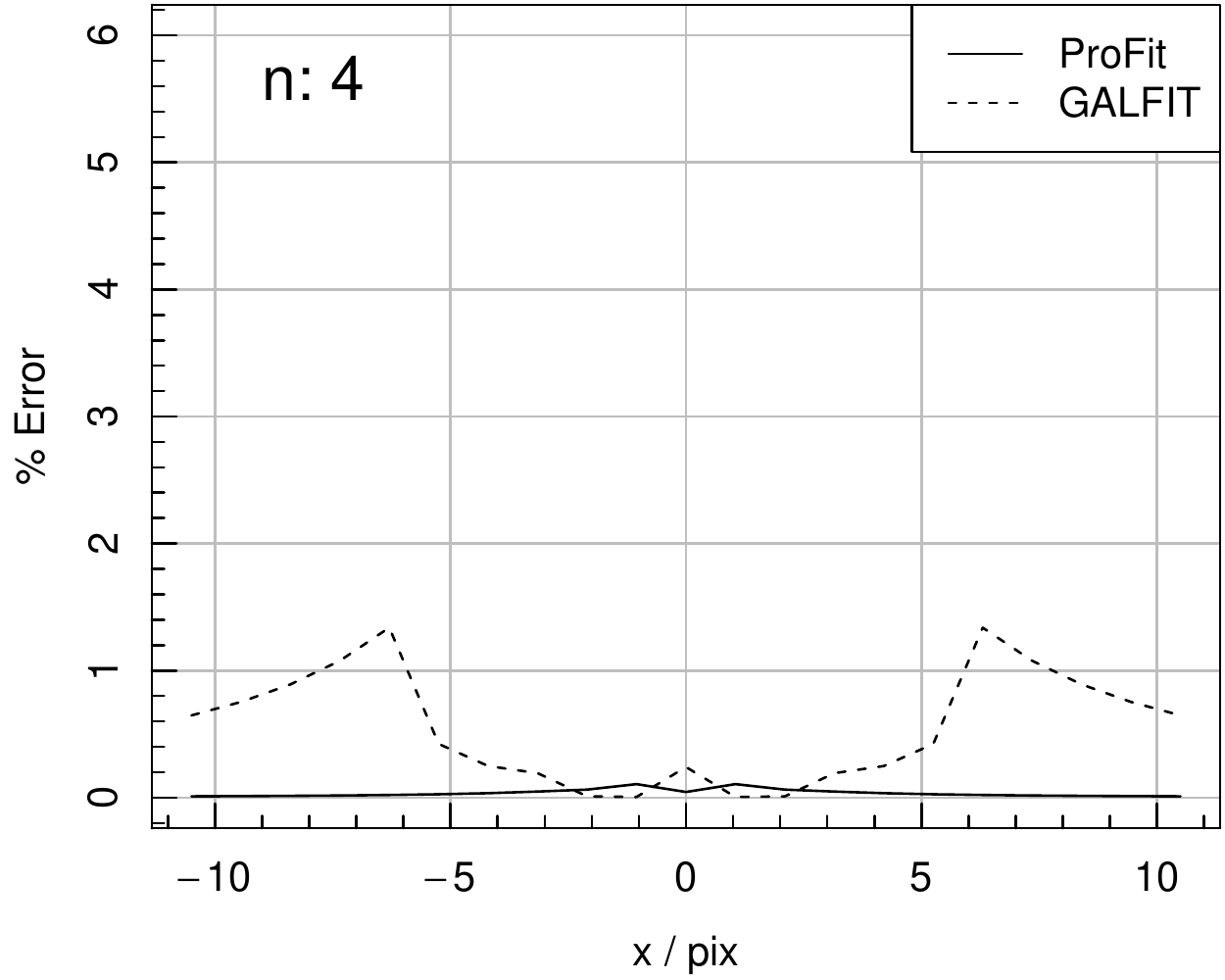}\\
	\includegraphics[width=\columnwidth]{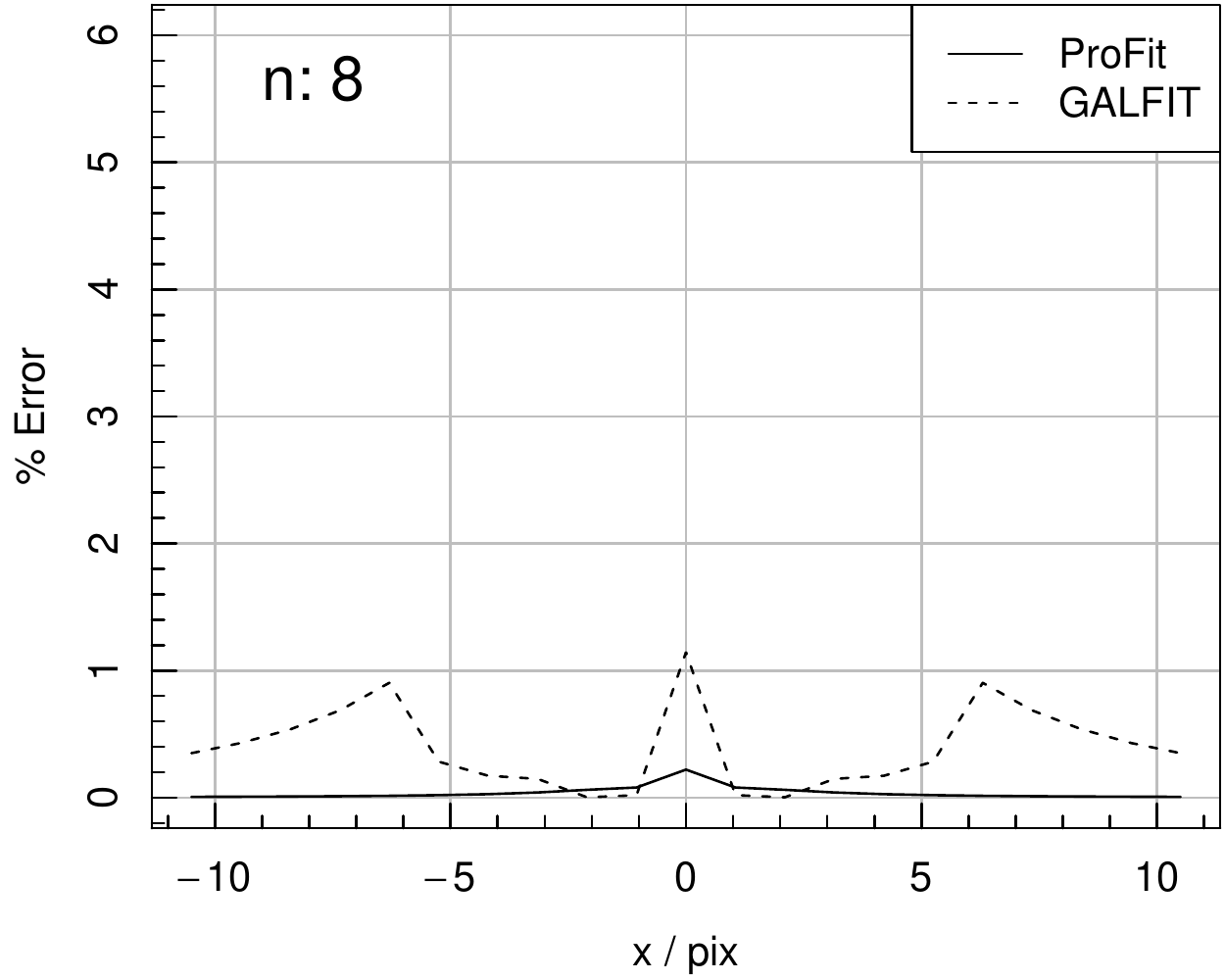}
	\includegraphics[width=\columnwidth]{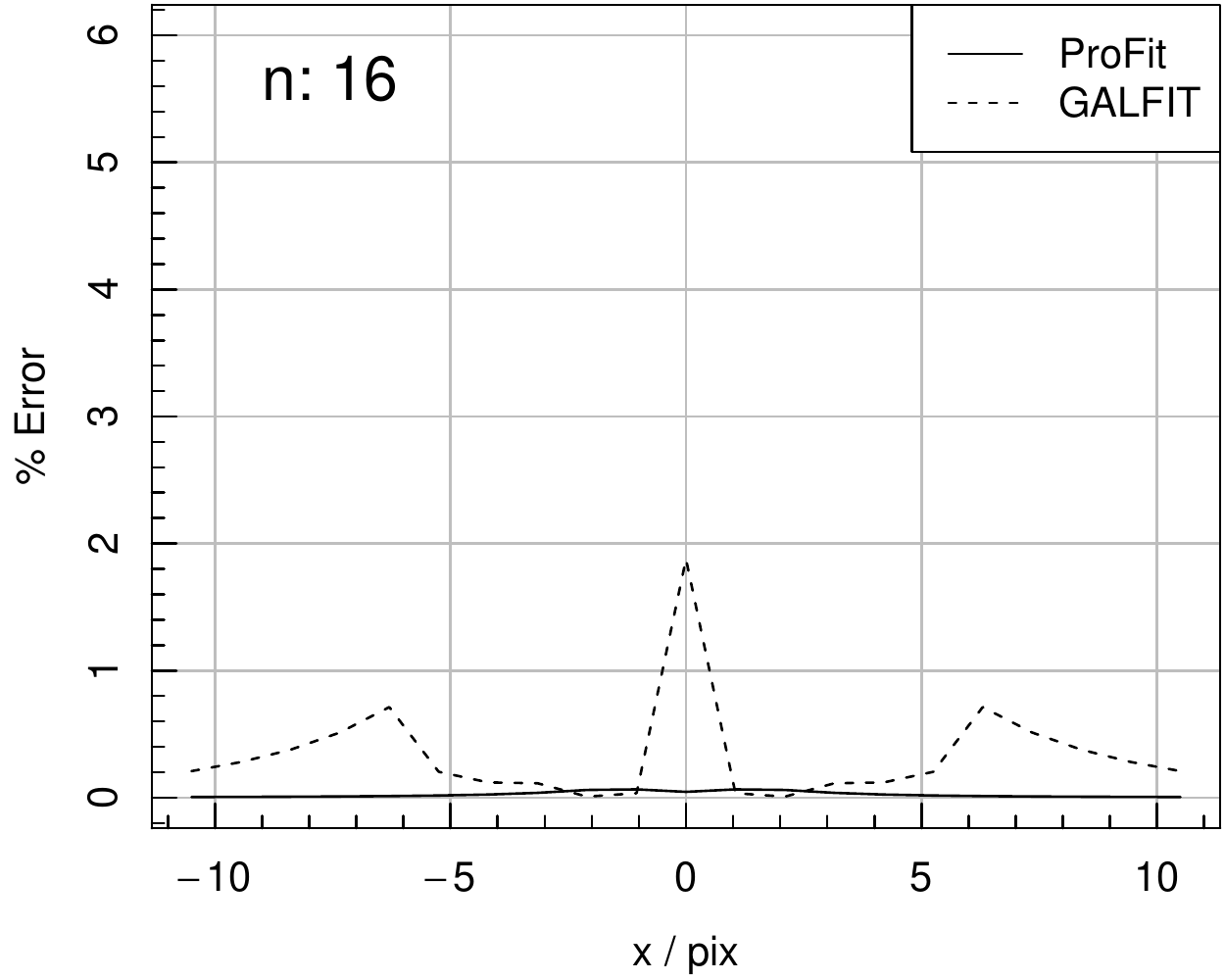}\\
	\caption{The pixel errors for pixels on the diagonal dotted lines in panel rows 1, 2 and 3 of Fig.~\ref{fig:pixelaccuracy} (i.e.\ close to the minor-axis centre of the image, where pixel errors tend to be largest).}
	\label{fig:pixelaccuracydiag}
\end{figure*}

The image generation is shown in more detail for these example profiles in Fig.~\ref{fig:pixelaccuracy}. This shows the actual error as a function of pixel location on the two-dimensional image for the target model for both \libprofit{} and \galfit. Fig.~\ref{fig:pixelaccuracydiag} shows the one-dimensional pixel error across the central portion of pixels that incorporate the peak pixel (which is usually the hardest to integrate accurately for the \sersic profile). There are specific regions of pixels where \galfit has smaller errors, but the general trend is that \libprofit{} is more accurate over a large range of pixels, and never experiences large integration error in pixels that contain significant fractions of the profile flux. In the case of both \libprofit{} and \galfit outer pixels might often have large relative error, but in these cases the pixels contain almost no flux and do not drive the overall model image to a large flux-weighted error.

\begin{figure}
	\includegraphics[width=\columnwidth]{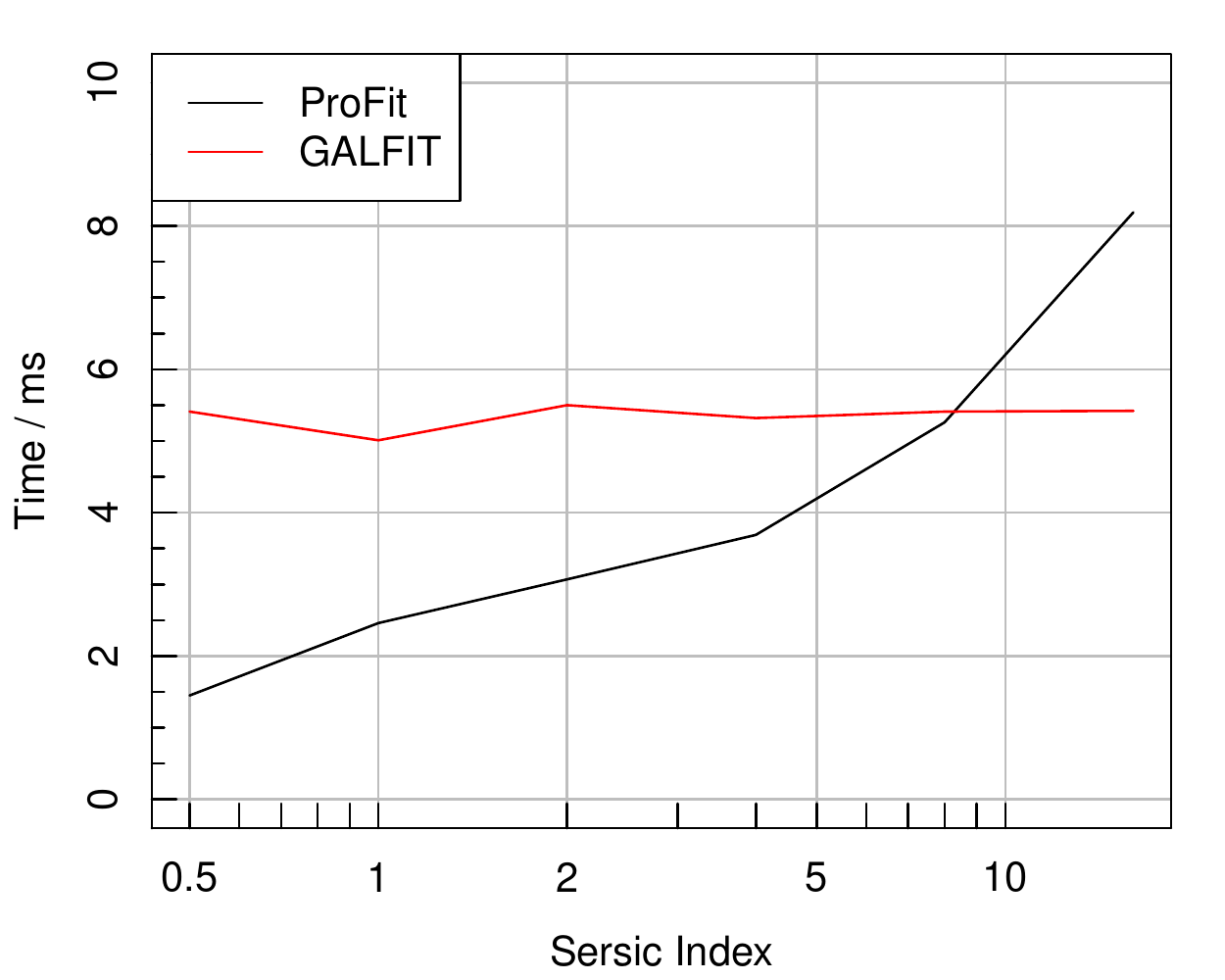}
	\caption{\libprofit{} versus \galfit model image generation time for \sersic indices 0.5 to 16 for $R_e=2$pixels, $\theta=60^\circ$ and $A/B=0.3$ in a $20 \times 20$ pixel image.}
	\label{fig:imagetime}
\end{figure}

As established in the design goals of \libprofit{}, generating images accurately is important, but there is a trade-off to be made with the required computation time. It is unavoidable that some pixels will be harder to evaluate due to the rapid change in flux within a pixel (especially true for the central pixel containing the peak of the \sersic profile flux), and more computing time will be needed for correctly evaluating the pixel flux. \libprofit{} uses adaptive recursion to adjust the integration resources as required in order to maintain a relatively constant level of pixel integration accuracy. This is evident from Figure \ref{fig:imagetime}, which compares the image generation times for \libprofit{} and \galfit for the same \sersic profiles as presented in Figure \ref{fig:fluxweighterror}. It is clear that \libprofit{} varies the integration time (which is akin to more resources spent evaluating {\it hard} pixels) more in order to maintain relatively constant flux weighted errors (see Figure \ref{fig:fluxweighterror}), whereas \galfit maintains nearly constant integration times.

% Isn't GALFIT hard-coded to oversample each pixel by 10x10 in the 10x10 central pixels? Therefore it wastes time oversampling tiny galaxies and doesn't oversample large galaxies enough, no?

\begin{figure}
	\includegraphics[width=\columnwidth]{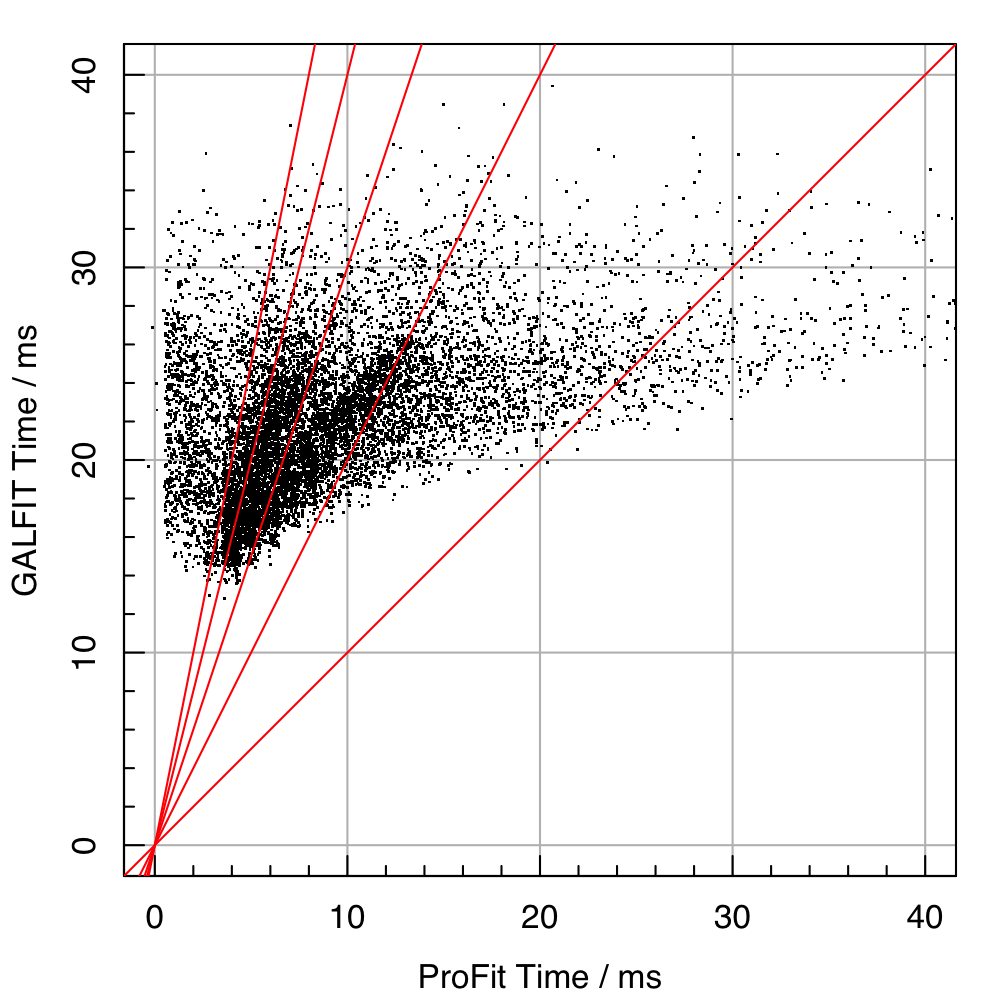}
	\caption{\libprofit{} versus \galfit model image generation times for a large grid of $200 \times 200$ pixel images. Galaxy size (10 $R_e$), \sersic index (11 $n$), rotation angle (10 $\theta$) and axial ratio (10 $A/B$) were sampled on a grid, with 11,000 model images generated in total. The red lines show 1:1, 1:2, 1:3, 1:4 and 1:5 relative speeds for the same target model image, \libprofit{} being substantially faster in almost all situations.}
	\label{fig:TimevTime}
\end{figure}

\begin{figure*}
	\centering
	Time to create $200\times200$ pixel images\\
	\includegraphics[width=\columnwidth]{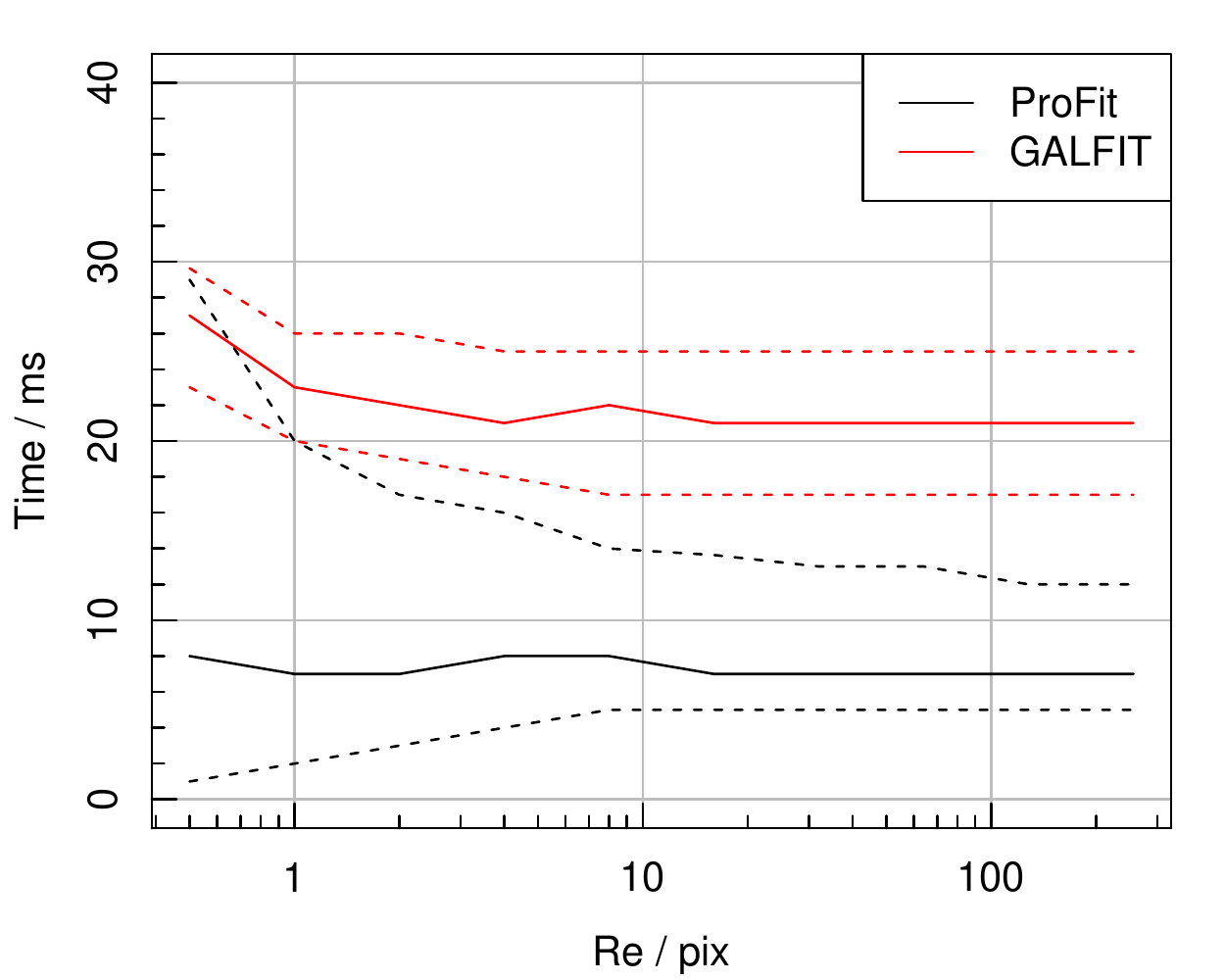}
	\includegraphics[width=\columnwidth]{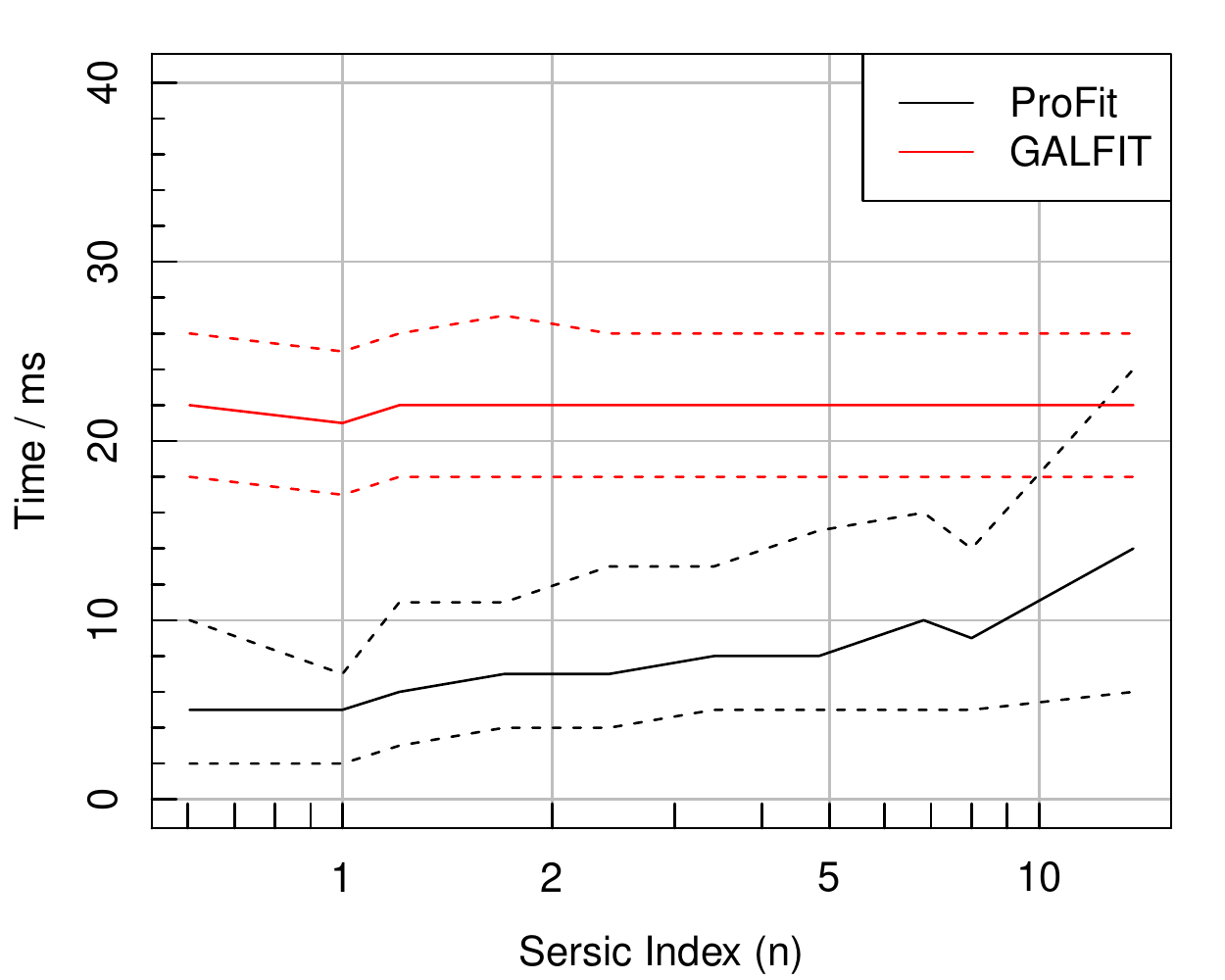}\\
	\includegraphics[width=\columnwidth]{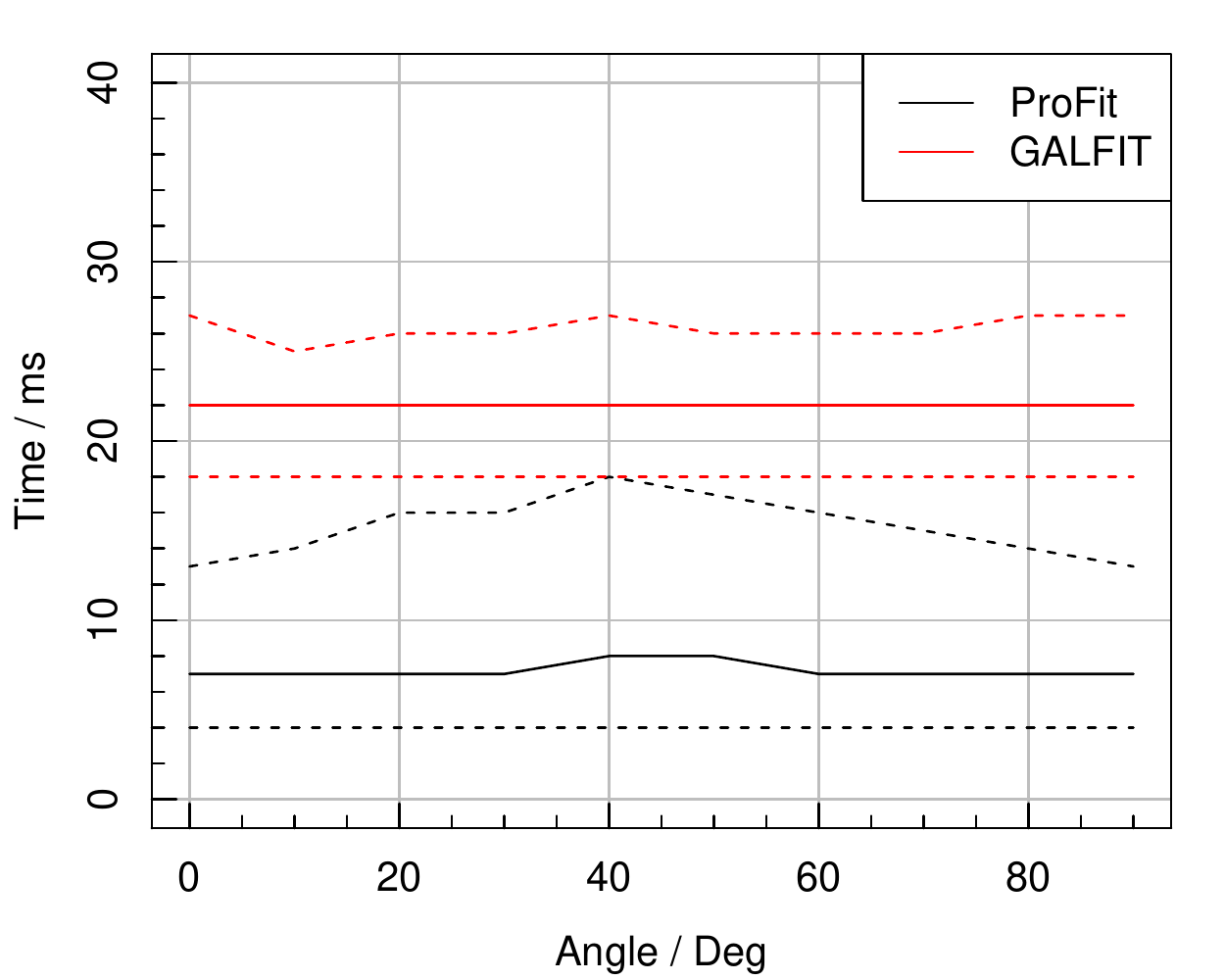}
	\includegraphics[width=\columnwidth]{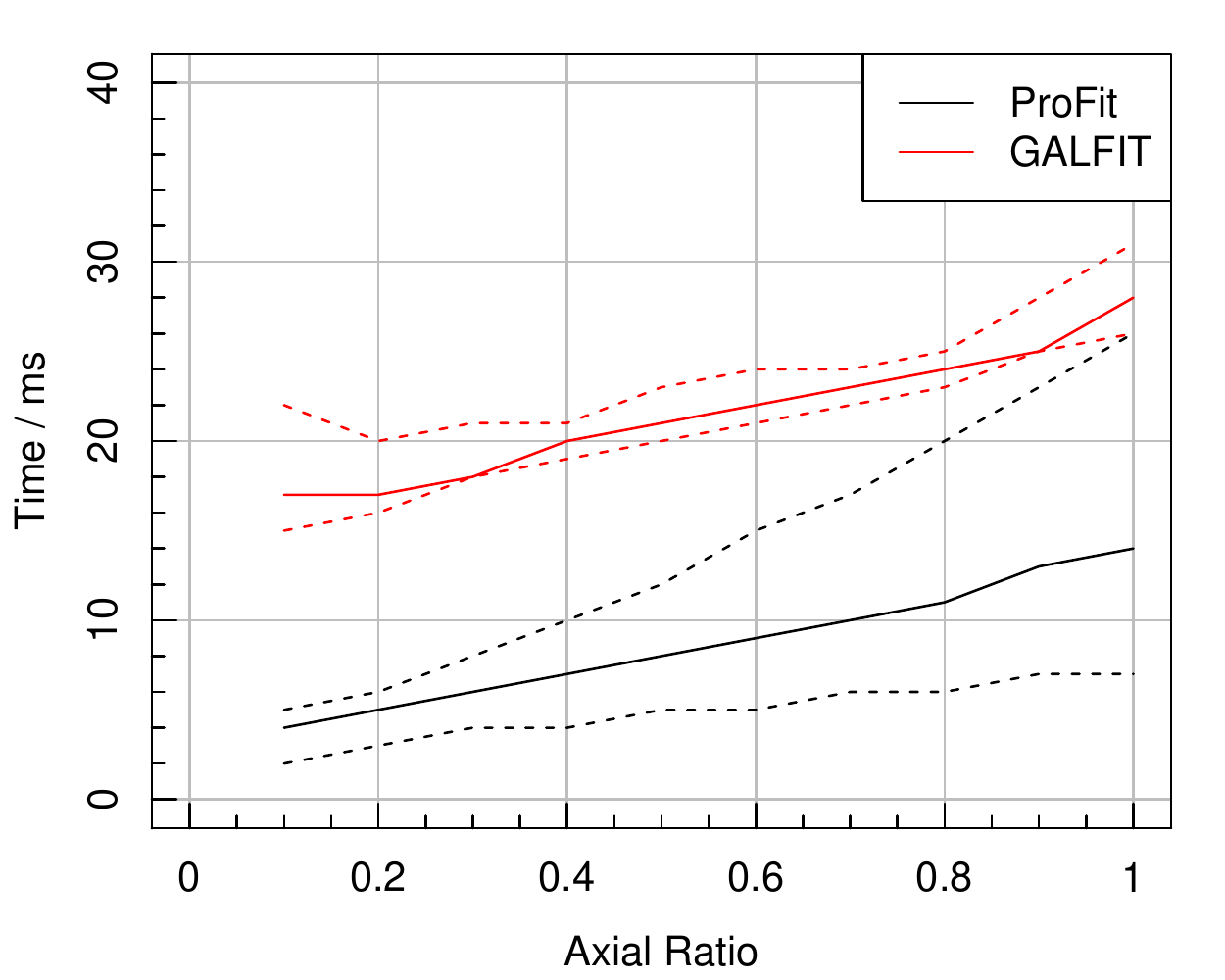}\\
	\caption{A large grid of model parameters were used to create $200\times200$ pixel images with centred single \sersic component galaxies (see Fig.~\ref{fig:TimevTime} for details). The marginalised median image generation time are shown in these plots for \libprofit{} and \galfit, with the effective 1$\sigma$ computation time spread shown as dashed lines.}
	\label{fig:paramtime}
\end{figure*}

The computation times of \libprofit{} and \galfit for \sersic profiles are investigated in more detail in Fig.~\ref{fig:TimevTime} and Figure \ref{fig:paramtime}. For all these tests the same MacBook Pro running El-Capitan with 2.6 GHz i7 processors and 16 GB of RAM was used. In Figure \ref{fig:TimevTime} we see that \libprofit{} is factors of a few faster compared to \galfit for most combinations of model parameters. In Fig.~\ref{fig:paramtime} we see there is weak dependence of computation time on $R_e$ and $n$, and a strong dependence on axial ratio (in all three cases larger means longer computation time for \libprofit{}). For reasons not immediately clear, \galfit actually takes longer to compute compact models --- perhaps because its integration scheme is not sufficiently adaptive.

In general, we find a wide range of potential computation times, spanning a factor of $\sim$10 for \libprofit{} versus $\sim$3 for \galfit even though the target image is $200 \times 200$ pixels in all cases. This makes it hard to predict in general how long a particular fit might take given the image size. In a pathological situation of the target galaxy having large $R_e$, large $n$, and being circular in projection, the computation could be a factor 10 longer than for the case of a compact galaxy with low $n$ and fairly elongated projection.

It is possible to configure \libprofit{} to exceed $\sim$0.1 per cent flux-weighted image accuracy. This default was deemed appropriate for typical (and certainly our) uses of \libprofit{} since it equates to 0.001 mag accuracy which in practice far exceeds the systematic and random measurement uncertainties for even relatively high fidelity photometric images. It also allows for rapid image fitting since small gains in accuracy begin to take factors longer to compute. However, a combination of increasing the oversampling and recursion depth allows for arbitrarily high levels of integration accuracy should this be required by the user. By including a simple interface to the Cuba library in the \R \profit package for all six radial profile types, it is possible to tune these parameters in order to achieve the desired level of accuracy and speed.

\subsection{Image Convolution}
\label{subsec:convolution}

ProFit supports both brute-force convolution and FFT-based convolution using the FFTW library. In the \R implementation, both methods are benchmarked and the optimal method for the given image and PSF size is chosen. Brute-force convolution is often significantly faster than FFT-based methods but scales directly with image \emph{and} PSF size, being $\mathcal{O}(n_{image}n_{PSF})$, whereas FFT convolution is $\mathcal{O}(4n_{image}\log{4n_{image}})$, accounting for necessary zero-padding. Testing shows that both methods give nearly identical answers, with fractional differences of less than $10^{-12}$.

One important consideration is that the accuracy of convolution depends on the resolution of the PSF relative to the pixel scale. A poorly resolved PSF with a FWHM $\le 3$ pixels may give flux errors on the order of a few percent for the innermost pixels. \profit allows for model and PSF oversampling for more accurate convolution, whereby the model pixel grid is subdivided by an integer factor (preferably odd). This is independent of the oversampling for the purposes of accurate model convolution, and requires either an analytic PSF or prior interpolation of the empirical PSF. Interpolating a noisy empirical PSF offers a limited gain in accuracy, so for this reason (and as mentioned in Section \ref{subsubsec:psf}), analytic PSFs are generally preferred. When this process is attempted empirically it is usually via stacking or drizzling of the data (e.g.\ with the use of a program like MultiDrizzle\footnote{http://stsdas.stsci.edu/multidrizzle/}). Accounting for pixel covariance when drizzling is a complex problem, and best handled by fitting the native resolution images simultaneously rather than attempting to combine them in an information-lossy process. Simultaneous image fitting of multi-exposure data and covariance likelihoods are high on the list of future extensions to \profit, but are not present in v1.0 (although experimental covariance likelihoods are already being tested).

\section{Image Fitting with \profit}
\label{sec:fit}

Having demonstrated that \libprofit{} integrates and convolves analytic profiles quickly and accurately, the remaining ingredients for model fitting are the likelihood function and optimization and/or likelihood sampling method. This logic is implemented in the \profit{} \R package. \R is becoming increasingly widely used in astronomy and other fields of advanced data analysis, and since it features a large number of statistics packages and optimization methods (maximum-likelihood and Bayesian), it is ideally suited for the purpose of Bayesian galaxy image fitting.

% The following three paragraphs ("having established", "having easy" and "in practice") were originally in the opening of section 2, 
% but after moving them down here it seems like the first one is not needed anymore

% Having established that the code would not be tied to any specific optimiser, but instead be general enough in design that almost any standard optimiser can be used, the main effort was invested in making model image that were accurate and fast. Once accurate model images have been made the next task is to then compare this model to data (in either photon, electron or ADU count space) with user defined errors and likelihood distributions. This task must be executed in a sufficiently general manner, such that the model and data comparisons can be made using third party optimisation engines.

Having easy access to a wealth of optimizers was a key design goal of \profit. A large subdivision of statistical data science and computer science has invested huge efforts into describing and implementing a multitude of optimizers that have a range of strengths and weaknesses. Depending on the specific problem at hand, one optimizer might be preferred over another for reasons that are very hard for us to intuit. Since this is such a key component of galaxy modelling, where a typical use case might involve fitting dozens of partly degenerate parameters, giving the user flexibility to experiment with optimization engines was deemed critical.

In practice an initial global parameter search might be made using a fast but biased downhill gradient optimizer, or one which is known to get easily trapped in local minima. From this point a more expensive but robust Markov Chain Mote Carlo (MCMC) optimizer could be used to accurately refine the parameter posteriors (this is the typical mode of operation recommended by the authors and described in detailed vignettes included with the \R package version of \profit). In the \R implementation alone the user has easy plug-and-play access to hundreds of optimizers accessible though the Central R Archive Network (CRAN) with hundreds more served from online repositories (e.g.\ GitHub). We do not wish to be overly prescriptive in how samplers are used to model data, but some typical applications will be discussed in the following sections.

%HERE 16:53 26/09/2016

\subsection{Required Observational Data}

A number of inputs are required to meaningfully fit a profile to a target galaxy observation. For good-quality data where the whole image is to be used for fitting, the bare minimum is strictly an image where pixels contain integer photoelectron counts and an analytical description of the PSF.

In practice, a more complete set of inputs can include the flux image ($D$, in whatever linear units are appropriate), a segmentation map (that is pixel matched to $D$, where distinct integer values represent different detected structures in the image as per SExtractor), a binary mask (that is pixel matched to $D$, where TRUE means mask out and ignore for analysis, and FALSE means do not mask), and the $\sigma$ image ($\sigma$, that is pixel matched to $D$, where values represent Normal errors in the same units as for $D$).

The PSF can be specified in two main ways: either via an analytical description using any of the available radial profiles (in practice, the Moffat function is a good profile for this task), or with an image of the PSF that is on the same pixel scale as $D$. \profit has a high-level setup function ({\tt profitSetupData}) that can take these basic inputs and create an internal object that is appropriate for fitting a \profit model.

\subsection{Image Likelihood}

There is a wealth of routes to compute meaningful image versus model likelihoods, depending on the data regime. If many (i.e.\ hundreds of) photo-electrons are registered in pixels for the observed galaxy then the data are likely to be operating in the Normal statistics regime, where \profit likelihoods can be computed with

%1/(?(2 ?) ?) e^-((x - ?)^2/(2 ?^2))
\begin{eqnarray}
x_{i,j}&=&\frac{(D_{i,j}-M_{i,j})^2}{2\sigma_{i,j}^2},\\
\ln\mathcal{L}&=& \frac{\ln{2\pi}}{2} - \sum_{i=1}^{N_i}\sum_{j=1}^{N_j} x_{i,j} -\ln{\sigma_{i,j}}.
%\ln\mathcal{L}= - \sum_{i=1}^{N_i}\left( \sum_{j=1}^{N_j} \left(\ln{\sqrt{2\pi}} -\ln{\sigma} - \frac{(D_{i,j}-M_{i,j})^2}{2\sigma^2} \right) \right).
\end{eqnarray}

\noindent Here, $D_{i,j}$ represents the observed pixel data at pixel $i, j$, $M_{i,j}$ represents a computed \profit model at pixel ${i, j}$ and $\sigma_{i,j}$ represents the estimated Normal uncertainty at pixel $i,j$.

An alternative view is that cumulative sum of squared residuals should follow a $\chi^2$ distribution, with a \profit likelihood computed by

%1 / (2^(n/2) ?(n/2)) x^(n/2-1) e^(-x/2)
\begin{eqnarray}
\chi^2&=&\sum_{i=1}^{N_i} \sum_{j=1}^{N_j} \frac{(D_{i,j}-M_{i,j})^2}{\sigma_{i,j}^2},\\
\ln\mathcal{L}&=& - \frac{k}{2} \ln{2} - \ln{\Gamma{(k/2)}} + \left(\frac{k}{2}-1 \right) \ln{\chi^2} - \frac{\chi^2}{2}.
\end{eqnarray}

\noindent Here $k$ represents the degrees of freedom of the $\chi^2$ distribution. For fitting of this type this will usually be determined by the number of pixels containing galaxy flux (here, $N_i N_j$) minus the number of parameters being used to fit the model to the data (up to eight in the case of a single component \sersic model). In the regime of the statistics being truly Normal and the model fitting well (i.e. a reduced $\chi^2$ near unity), the Normal and $\chi^2$ distribution likelihoods converge to similar results. The chief distinction is that the $\chi^2$ statistic penalizes overfitting, since extremely small residuals across all pixels are highly unlikely. Both statistics are sensitive to residuals significantly above the shot noise ($\chi^2$ somewhat more so), such as non-axisymmetric features such as bars, spiral arms, and dust lanes. Nonetheless, if the dominant source of uncertainty in the image is truly shot noise rather than systematics, and the signal to noise is high, the $\chi^2$ statistic is appropriate.

% Is this really true? Both Normal and $\chi^2$ statistics disfavour bad models; I think the above distinction in (dis)favouring overfitting is more important

% However, if there are residual structures not represented by the model (e.g.\ spiral arms) then the two can diverge to differing solutions. Typically the $\chi^2$ will be more sensitive to extreme outliers in the likelihood, and using the $\chi^2$ likelihood will alter the fit such these are better explained by the model.
 
However, if the count rates are relatively low, then the Poisson statistics must be used. In this case, \profit likelihoods are computed with

%p(x) = ?^x exp(-?)/x!
\begin{eqnarray}
\ln\mathcal{L}=\sum_{i=1}^{N_i} \sum_{j=1}^{N_j} D_{i,j} \ln{\lambda_{i,j}} -\lambda_{i,j} -D_{i,j}!.
\end{eqnarray}

\noindent Here, $\lambda_{i,j}$ is the expectation for the number of photoelectrons counted in at pixel $i,j$, and $D_{i,j}$ is the integer number of photoelectrons recorded. Dropping the $D_{i,j}!$ term (since it depends only on the data) applies a global offset to log-likelihood, which makes no difference to any posterior inference since this is all computed in a relative manner. Expressed in this form and multiplied by a factor of $-2$ we recover the `Cash' statistic $C$ \citep{cash79}, which is similar to $\chi^2$ but for Poisson statistics. In the same manner that $-2 \ln\mathcal{L} = \chi^2 + a$ for Normal statistics, $-2 \ln\mathcal{L} = C + a$ for Poisson statistics.

It is worth highlighting that returning images to true photo-electron counts is non-trivial and in our experience many published data sets do not produce enough ancillary information to return the images to true counts to better than an accuracy of a few factors. A particularly pernicious issue is that counts must be known within the CCD itself, not for photons at the top of the atmosphere. The latter is relatively trivial to compute after the fact if we know intrinsic properties of the source spectral energy distribution; however, the former requires estimates of atmospheric transmission, instrumental losses, detector gain and quantum efficiency, etc. Sometimes this information is easy to obtain at least approximately \citep[e.g.\ for the Sloan Digital Sky Survey, SDSS:][]{ahn14}, but this level of ancillary meta-data is in general rare. Added to this is the fact that optical survey data generally have copious counts per pixel: typically thousands of photoelectric counts. For these reasons, Poisson likelihoods should seldom be used for optical survey data, even if quite shallow. It is more appropriate for fitting X-ray data, where the counts are low and have the required meaning for Poisson statistics. Indeed, X-ray astronomy is the main subdivision that uses Poisson (usually in the guise of Cash) statistics.

If the data are nominally in the Normal statistics regime (i.e.\ many photoelectron counts per pixel), but in practice have a significant number of data points that are poorly represented by our attempted model (typical when trying to fit smooth two-dimensional \sersic profiles to well-resolved galaxies with asymmetries) then more robust Student-T distribution statistics might be appropriate. These distributions are approximately Normal in the core but have broad Lorentzian wings, which puts more likelihood mass at large distances from the Normal core. This behaviour is controlled by the `degrees-of-freedom', which in practice is not free and is instead estimated via maximum-likelihood from the data directly. The \profit Student-T likelihood is computed by

\begin{eqnarray}
x_{i,j}&=&\frac{(D_{i,j}-M_{i,j})^2}{\sigma_{i,j}^2},\\
\ln\mathcal{L}&=&-\ln \left( \sum_{i=1}^{N_i}\sum_{j=1}^{N_j} \frac{\Gamma{(\frac{\nu+1}{2})}}{\sqrt{\nu \pi}\Gamma{\frac{\nu}{2}}}\left(1+ \frac{x_{i,j}}{\nu} \right)^{-\frac{\nu+1}{2}} \right).
\end{eqnarray}

\profit gives the users easy access to all four types of commonly used likelihood statistics, and it is for the user to decide the most appropriate given the quality and type of data at hand. For high $S/N$ images of galaxies, the Student-T statistic behaves well over a broad regime given its robustness to outlier flux values. For this reason, it is selected as the default option when setting \profit up for fitting. In future, we will add covariance likelihoods \citep[more appropriate for data that exhibits large amounts of pixel correlation, e.g. NIR, see][]{andr14}, but these are not present in v1.0.

\subsection{Parameter Optimization}

Once the observational data have been correctly assembled and the most appropriate type of likelihood statistic has been chosen, the user is free to fit a model image. \profit requires the user to suggest an initial parameterization of the model, which need not be close to the optimal parameters with robust optimizers. The user is free to fix some components and fit others (e.g.\ they might wish to fix the disc \sersic index to 1). They are also free to provide prior distributions for each parameter (important for formal Bayesian analysis), hard limits and/or constraints on the allowed fitting region, to specify whether parameters are optimized in log or linear space and to specify additional constraints between parameters (e.g.\ pair parameters together). The detailed process for specifying these options is provided in the \profit manual and example vignettes, so we will not repeat the full description here.

Once the fitting structure has been specified within \profit ({\tt profitSetupData}, as mentioned above) the user can interface with a number of popular optimization routines. The \R{} \profit package provides examples of using the base {\tt optim} function, the open source {\textsc LaplacesDemon} package that gives access to gradient optimizers via {\tt LaplaceApproximation} and a large suite of Markov Chain Monte Carlo (MCMC) samplers via {\tt LaplacesDemon}, and the genetic algorithm {\textsc CMA-ES} package\footnote{Versions of {\tt LaplacesDemon} and {\textsc CMA-ES} implementing runtime limits for use on shared (super)computers are available at \url{https://github.com/taranu/LaplacesDemon} and \url{https://github.com/taranu/cmaeshpc}, respectively.}. These four routes to optimisation offer over a hundred distinct optimisers. The CRAN collates packages that tackle particular problems types: optimization\footnote{cran.r-project.org/web/views/Optimization.html} and Bayesian\footnote{cran.r-project.org/web/views/Bayesian.html} analysis task views suggest over a hundred additional packages that themselves potentially contain multiple algorithms.

To the uninitiated this can seem daunting, but the reality is that a core few packages and optimizers give the user a solid foundation for a diverse range of fitting problems. Sample vignettes provided in the \R \profit package goes through some example fitting problems in detail, but we will summarize the approach here. The basic strategy for preparing the inputs for galaxy fitting is

\begin{itemize}
\item REQUIRED: read in an {\bf image} of an observed galaxy (in principle any format, but e.g.\ FITS).
\item OPTIONAL: read in a {\bf mask} with the same size, pixel scale, and astrometry as the {\bf image}.
\item OPTIONAL: read in a {\bf sigma} image with the same size, pixel scale, and astrometry as the {\bf image}.
\item OPTIONAL: read in a {\bf segmentation} map with the same size, pixel scale, and astrometry as the {\bf image}.
\item OPTIONAL: read in a {\bf PSF} image with the same pixel scale as the {\bf image}.
\item REQUIRED: read in an {\bf initial} model list (as detailed above).
\item OPTIONAL: read in a list with the same structure as {\bf initial} detailing which parameters \profit will {\bf fit} (the others will be fixed at their {\bf initial} values).
\item OPTIONAL: read in a list with the same structure as {\bf initial} detailing which parameters \profit will fit in {\bf log} space (the others will be fitted in linear space).
\item OPTIONAL: read in a function detailing the calculation of the {\bf prior} log-likelihood.
\item OPTIONAL: read in a list with the same structure as {\bf initial} detailing the allowed {\bf limits} for each parameter when fitting.
\item OPTIONAL: read in a function detailing the calculation of additional {\bf constraints} to apply to parameters, e.g.\ pairing parameters together or making them scale relative to each other in a fixed manner.
\item REQUIRED: all of the above two-dimensional matrix images, list structures, and functions are provided to the high-level {\tt profitSetupData} function that organises the information into an object of class {\bf profit.data} ready for fitting with third-party optimization engines.
\end{itemize}

Once the user has created the {\bf profit.data} object, which includes some additional options for choosing the type of likelihood and setting the verbosity of the fitting process, a basic strategy for fitting might look like as follows:

\begin{itemize}
\item Initially find a maximum likelihood galaxy model solution using the base \R{} {\tt optim} function using the Broyden, Fletcher, Goldfarb, and Shanno \citep[BFGS:][]{flet70} algorithm (which is fairly robust to spurious minima).
\item Take the solution from the BFGS optimization as a starting point and use {\tt LaplacesDemon} component-wise hit-and-run metropolis (CHARM) to make MCMC samples of the likelihood space (this is very robust to local minima, and can move large distances from poor solutions; hence hit-and-run),
\item Use diagnostic tools provided with the {\textsc LaplacesDemon} package to test for the quality and convergence of the model fit, including tests for auto-correlation between parameters, the number of effective stationary samples and traditional parameter triangle plots.
\item Assuming the fit is flagged as being well converged and well behaved, the log marginal likelihood (LML) can be computed to aid comparisons between models.
\end{itemize}

The above fitting process can be carried out for models of varying complexity. If they are well-converged then the LML computed can be used to calculate Bayes factors between simpler and more complex models, offering an objective route for deciding how complex the model needs to be.

\section{\profit in Practice}
\label{sec:examples}

\subsection{Single galaxy case-study of G266033}
\label{sec:example1}

The \profit package includes 10 example data sets with all the required inputs for fitting. These examples use public Sloan Digital Sky Survey \citep[SDSS;][]{ahn14} and the Kilo Degree Survey \citep[KiDS;][]{kuij15} data for galaxies selected from the Galaxy And Mass Assembly \citep[GAMA;][]{driv11, lisk15} survey that are at reasonably low redshift and are deemed to have multiple components \citep[as found in][]{lang16}. The vignettes included with the software give examples of fitting these data using high-level \R and \Python interfaces.

\begin{figure*}
	\centering
	Bulge component of galaxy fit:\\
	\includegraphics[width=\columnwidth*2]{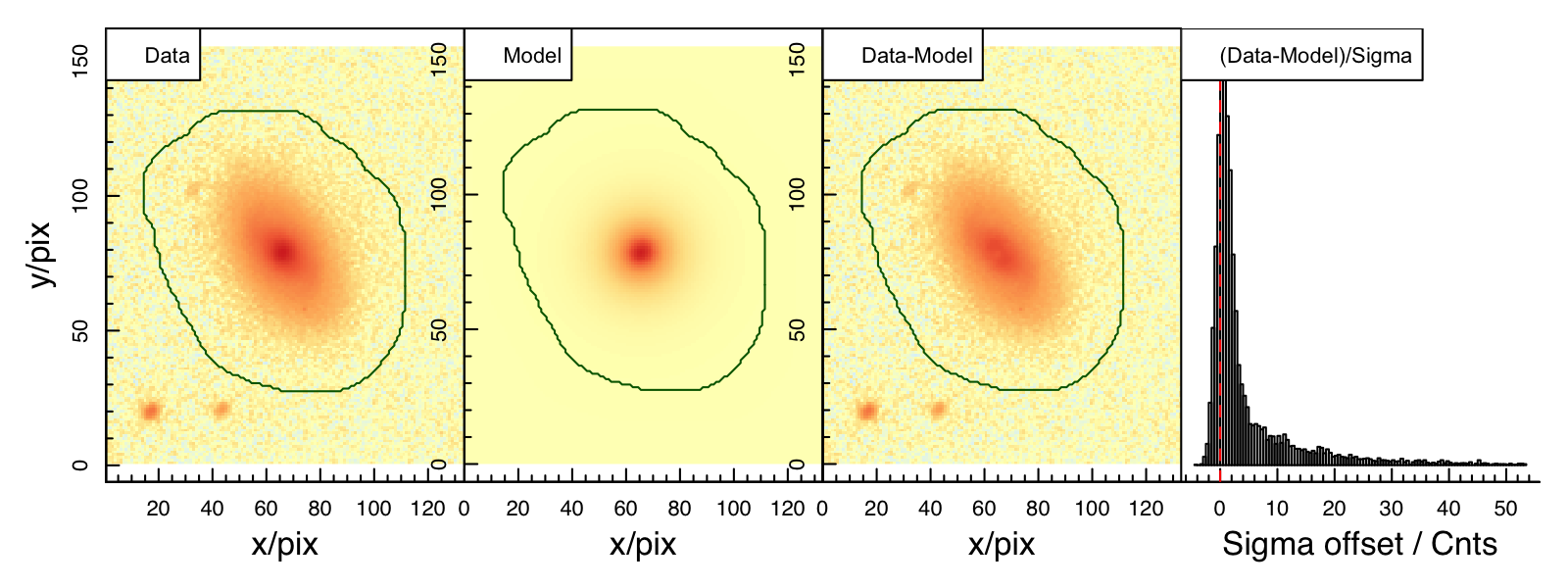}\\
	Disc component of galaxy fit:\\
	\includegraphics[width=\columnwidth*2]{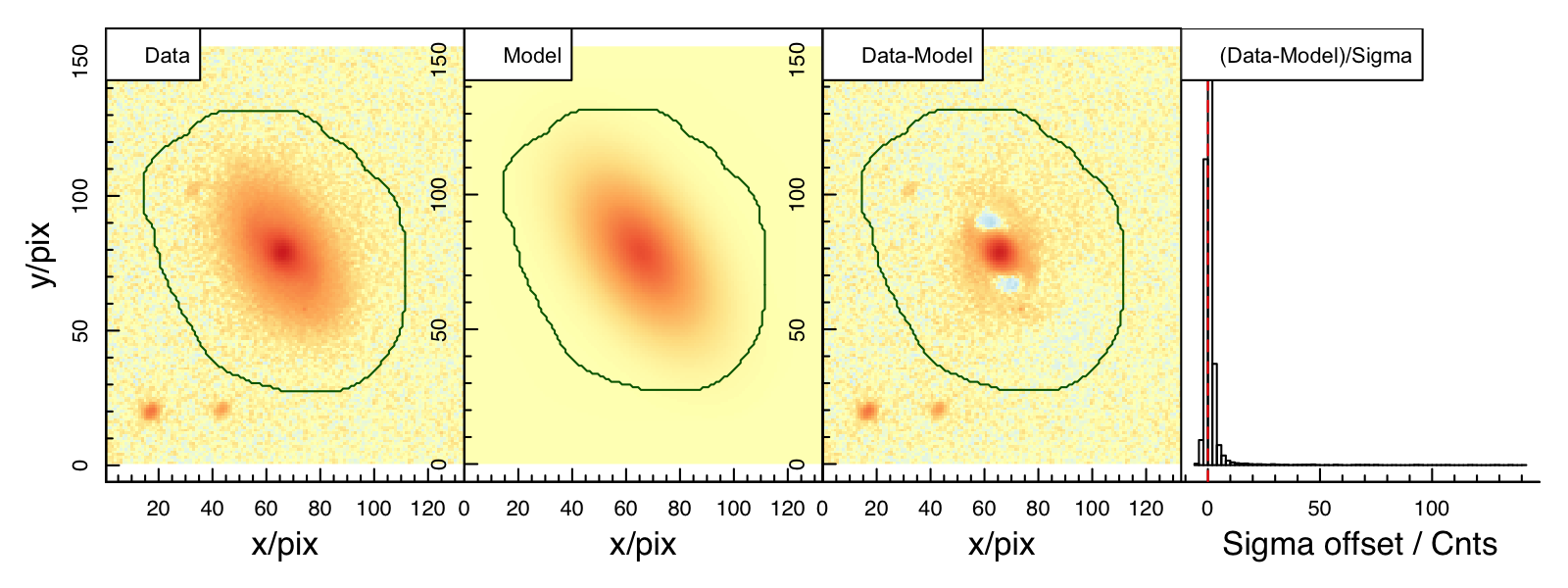}\\
	Bulge+Disc components of galaxy fit:\\
	\includegraphics[width=\columnwidth*2]{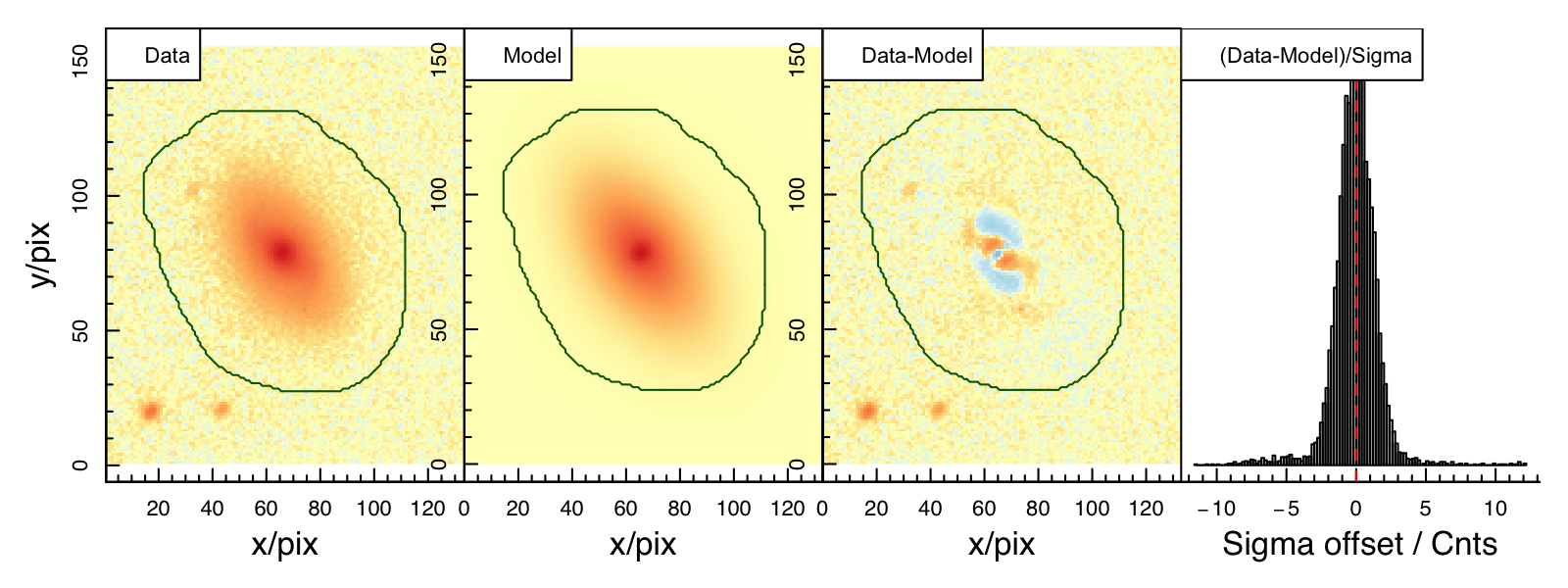}\\
	\caption{Result of the $r$-band MCMC decomposition of G266033. The top row shows the data (left), the bulge model (second left), the data-model (second right) and the histogram of residuals. The second row shows the data (left), the disc model (second left), the data-model (second right) and the histogram of residuals. The bottom row shows the data (left), the bulge-disc model (second left), the data-model (second right) and the histogram of residuals. Even with an optimal bulge-disc model there is still evidence of residual structure.}
	\label{fig:modelfit}
\end{figure*}

\begin{figure*}
	\centering
	\includegraphics[width=\columnwidth*2]{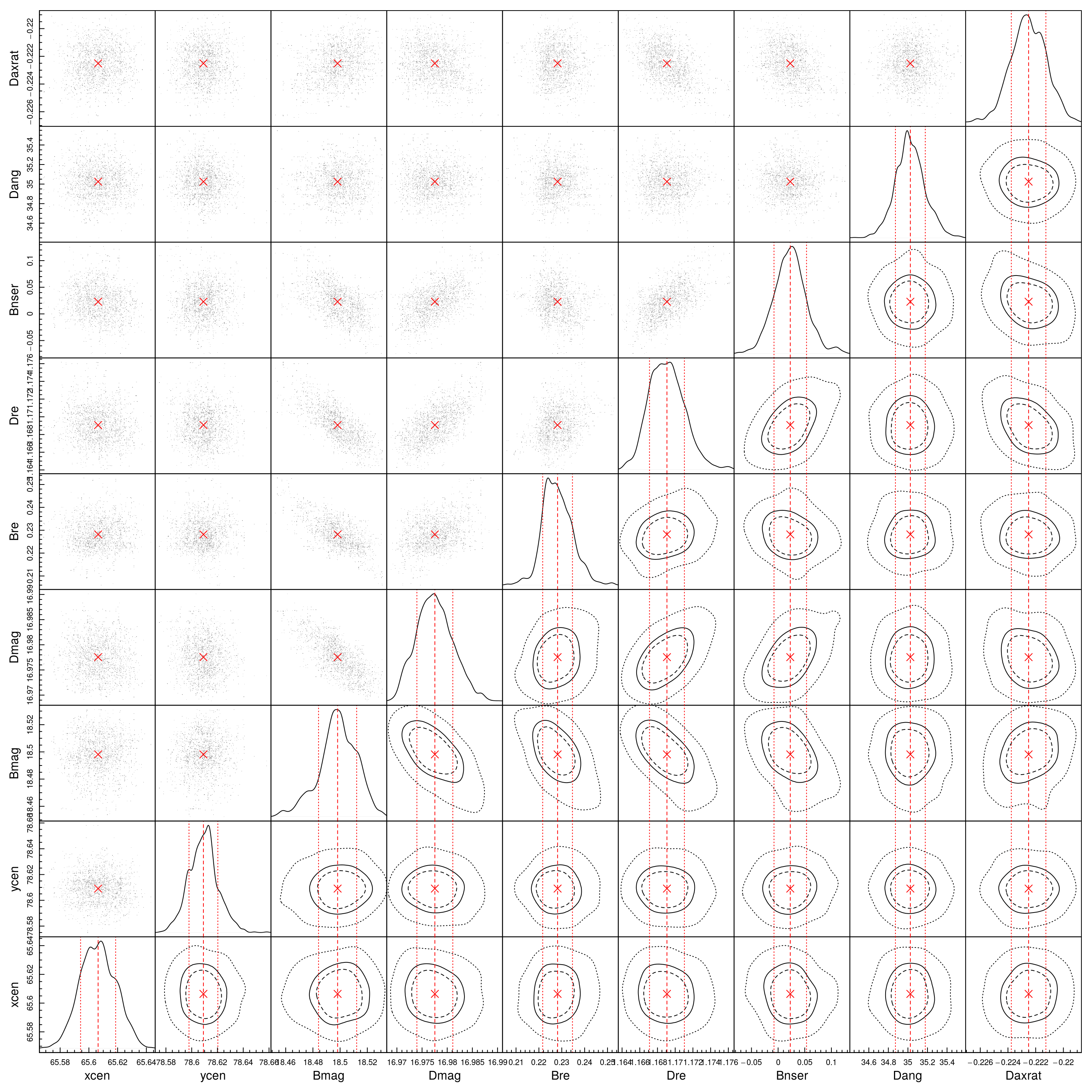}
	\caption{The triangle plot of the stationary MCMC chains for the model fit of G266033 shown in Fig.~\ref{fig:modelfit}. The top-left of the triangle shows the raw samples, the bottom right shows the contoured version of the samples, with dashed/solid/dotted lines containing 50 per cent / 68 per cent / 95 per cent of the samples. The diagonal one-dimensional density plots show the marginalised distributions for each parameter. In this case nine parameters were fitted using CHARM in the {\tt LaplacesDemon} function: $x_{cen}$ (xcen), $y_{cen}$, the bulge magnitude (Bmag), the disc magnitude (Dmag), $R_e$ of the bulge (Bre), $R_e$ of the disc (Dre), the \sersic index of the bulge (Bnser), the rotation angle $\theta$ of the disc (Dang) and the axial ratio of the disc (Daxrat). From the triangle plot it is clear that in this case the bulge magnitude (Bmag) and disc magnitude (Dmag) show the most covariance with other parameters, but in general the fit is well converged. Note the covariances seen tend to be galaxy specific, and might not be present in generic galaxy modelling cases.}
	\label{fig:modeltriangle}
\end{figure*}

As a case study, Fig.~\ref{fig:modelfit} shows the quality of MCMC fit and residuals that are possible using \profit for G266033. A qualitative assessment of the residuals shows that we are able to remove the major bulge and disc structural components, but that non-axisymmetric components remain (the spiral arms in particular). The three rows of panels show the bulge (top), disc (middle) and combined (bottom) components. The outputs of the CHARM MCMC sampling are shown in Fig.~\ref{fig:modeltriangle}. Here only stationary samples are shown (i.e.\ there is no `burn-in' of the sampling), and whilst there is evidence of some weak covariance between some parameters (e.g.\ disc magnitude (Dmag) and effective radius (Dre)), in general the posteriors look well sampled and consistent with multivariate Normal distributions. This latter characteristic is useful when estimating marginalized parameter errors.

\begin{figure*}
	\centering
	\includegraphics[width=\columnwidth*2]{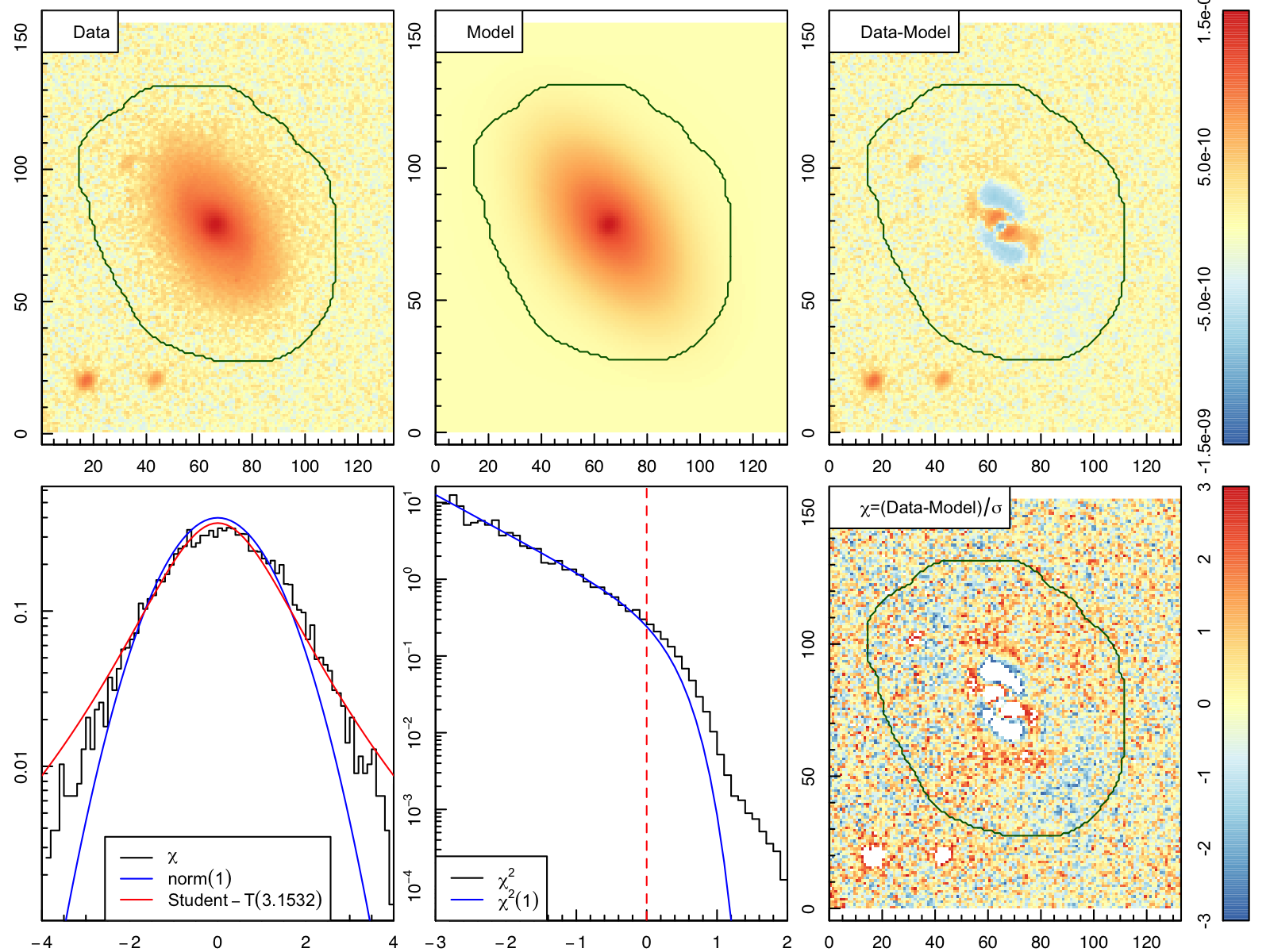}
	\caption{The output of the provided high-level {\tt profitMakePlots} function included in the \R version of \profit. The top panels are the same as Fig.~\ref{fig:modelfit} for G266033. The bottom panels show the histogram of residuals on a scale of $\sigma$ (left), the $\chi^2$ residuals compared to a $\chi^2$ distribution with one degree of freedom (middle) and the two-dimensional residuals shown in terms of $\sigma$ significance. In this case the part of the galaxy that is well approximated by the model has well behaved Normal distribution errors, but there are substantial deviations for residuals in regions containing non-smooth structure (e.g.\ spiral arms etc).}
	\label{fig:modelfitchisq}
\end{figure*}

Using the \R interface to \profit, more quantitative information can be extracted from the fit. Fig.~\ref{fig:modelfitchisq} shows the output of the {\tt profitMakePlots} function, which as well as producing the standard residuals, also outputs how they compare to a reference Normal and Student-T distribution. The $\chi^2$ excess is also demonstrated with respect to a $\chi^2$ distribution with one degree-of-freedom. Finally, a residual image scaled by the pixel error $\sigma$ is provided. This shows that there are some significant residual spiral arm structures that are highly unlikely from the point of view of the idealized generative model we are fitting.

\begin{figure}
	\centering
	\includegraphics[width=\columnwidth]{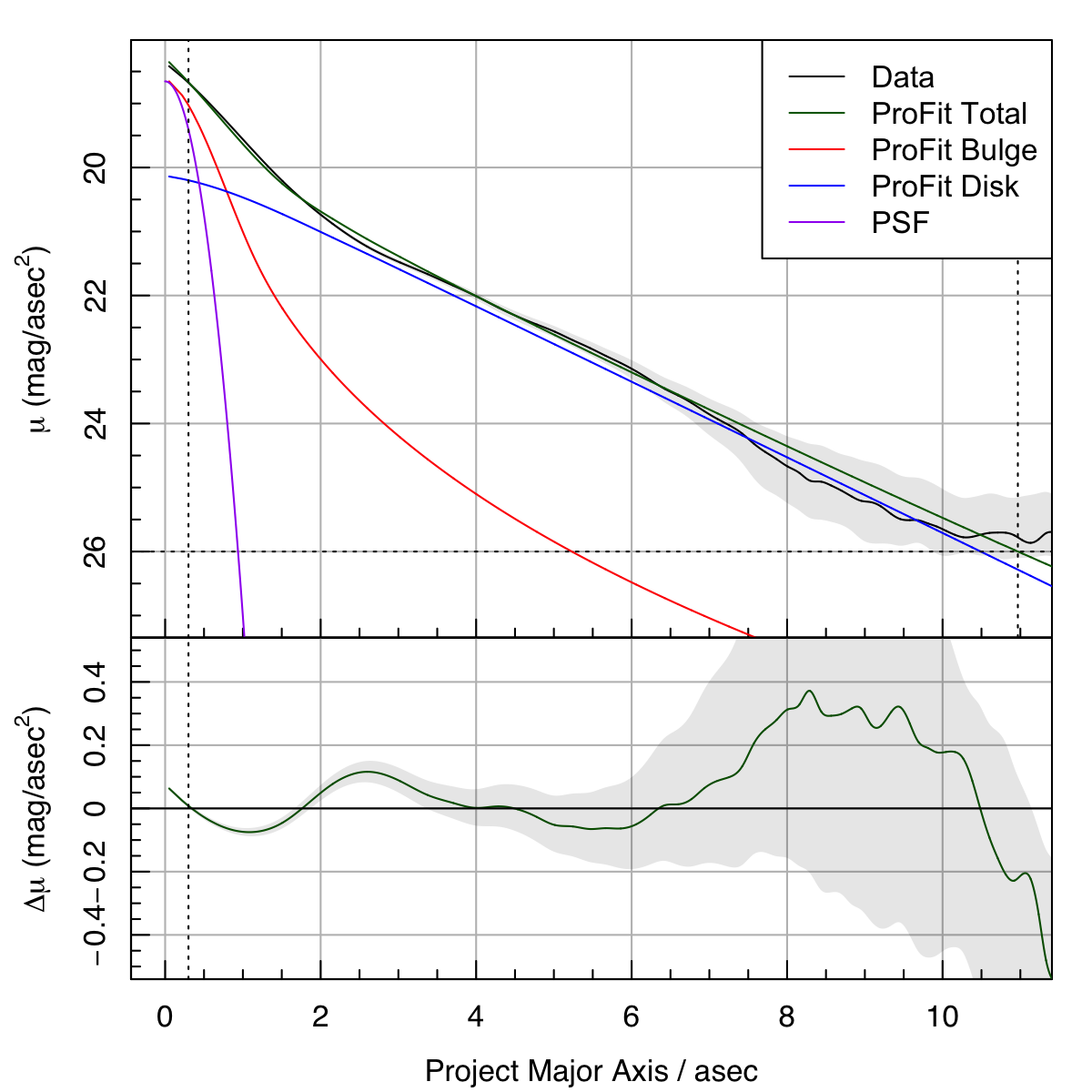}
	\caption{The approximate projected one-dimensional profile for G266033. The grey shaded region shows the 1$\sigma$ error region for the data pixels, and ideally the total model (green line) should sum to follow this distribution quite closely. At a few radii the data and model are in some tension (at the 0.1 mag level in surface brightness), but for the majority of the profile they agree very well.}
	\label{fig:modelfitoned}
\end{figure}

As discussed in the Introduction, it is common in the field of galaxy modelling to collapse two-dimensional data into a one-dimensional form. Whilst \profit does not evaluate likelihood using this one-dimensional form, it does provide functions for producing such isophotal data and the associated plots ( {\tt profitEllipse} and {\tt profitEllipsePlot}). In brief these functions use the geometric parameters of the disk and bulge components of a simple two-component model to extract isophotal annuli, allowing for the standard geometric distortions such as ellipticity and boxiness (if present). The one-dimensional profile of G266033 with overlaid model fits is shown in Fig.~\ref{fig:modelfitoned}.

\begin{figure*}
	\centering
	\includegraphics[width=\columnwidth*2]{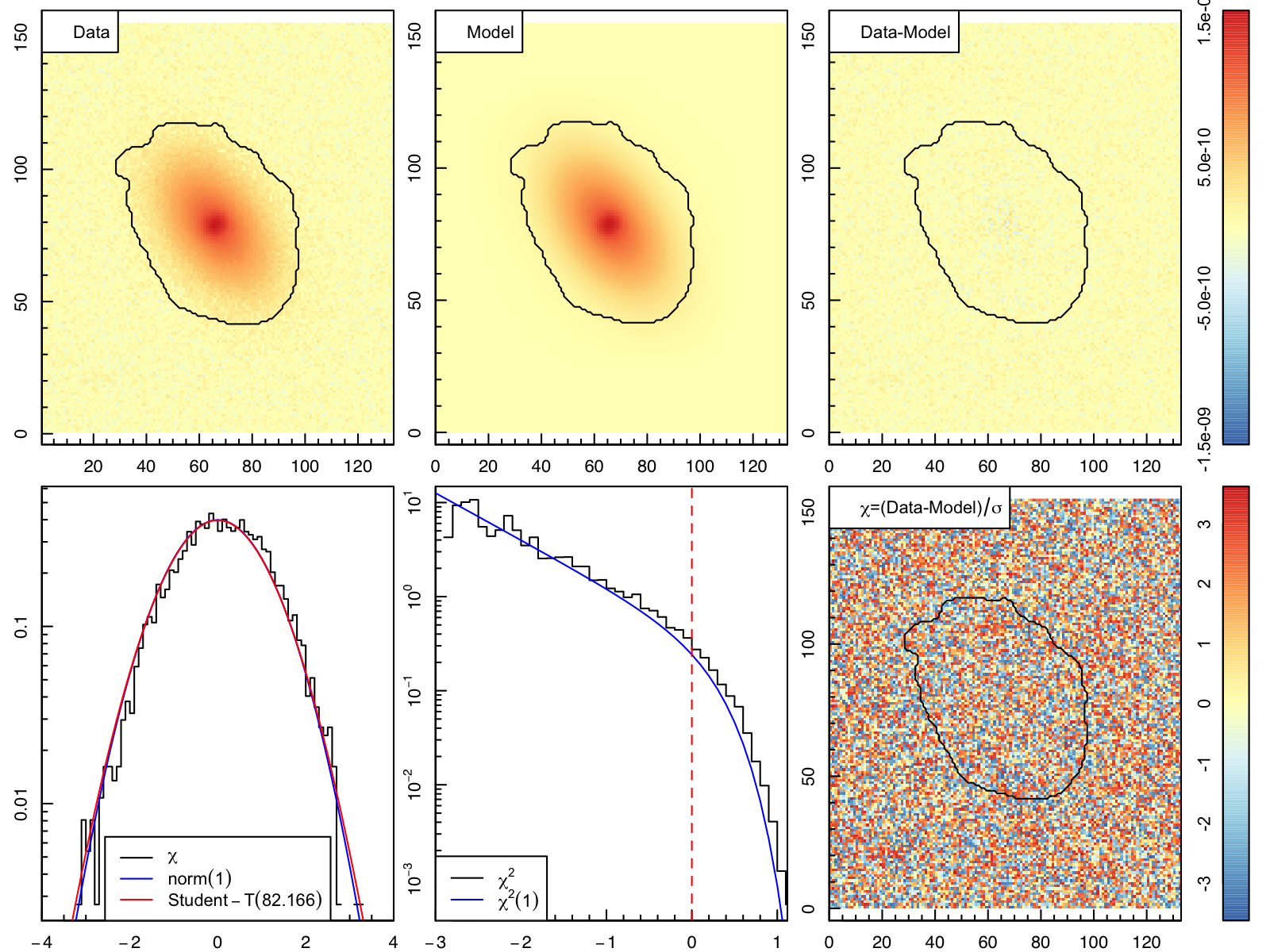}
	\caption{As per Figure \ref{fig:modelfitchisq}, but with the residuals removed. Image noise has been added back in via Poisson sampling of the convolved model image. This means an exact model can be fit to the data, hence the residuals are consistent with an idealised model fit.}
	\label{fig:modelfitchisqmod}
\end{figure*}

Given the residual flux not replicated by the model, the question remains how representative any posterior error distributions might be. To estimate the impact an imperfect model can have on the returned posterior error estimates, the residual is subtracted from the initial image. Fig.~\ref{fig:modelfitchisqmod} shows the effect of fitting such an input where the image noise has been added back via Poisson sampling of the convolved model image. The residuals almost entirely disappear, and the $\sigma$ scaled residuals show no radial structure at all. Comparing the posteriors to the original input data and the residual-removed input data, the former has typical marginalized posterior errors for each parameter that are a factor of $\sim$2 {\it larger} (spanning the range 1.3--3) with best-fitting values that agree to within 1\%. This is encouraging, since it suggests in this particular case, where \profit is correctly identifying the major structural components, the presence of non-axisymmetric components creates posteriors with larger errors. How such fitting behaves will depend on the data at hand (i.e.\ this should not be assumed to be true in general), but such a test is easy to implement, and should be part of a standard \profit workflow.

%Any more ProFit related puns for section titles add here:
%\section{ProFit Philosophy}
%\section{Capitalising with ProFit}
%\section{ProFit Margins of Error}

\subsection{Running \profit and \galfit on a small sample from SDSS and KiDS}
\label{sec:example2}

As mentioned above, the \R{} \profit package comes with 10 fairly isolated example galaxies included that have imaging from SDSS and KiDS. All of these are well resolved galaxies taken from the GAMA survey where we already have well converged bulge and disc fits from \citet{lang16}. To create this sample, 40 galaxies from \citet{lang16} were randomly sampled and these were ranked in terms of fit-ability (removing galaxies with more complex backgrounds and nearby sources). Of these, the 10 largest were chosen as example galaxies. As such this sample, whilst small, is broadly representative of how common different classes of bulge-disc systems are (e.g.\ we are dominated by lower $B/T$ systems). Consistent sky subtractions were made using LAMBDAR \citep{wrig16}, and segmentation maps and PSFs were created using an updated version of SIGMA \citep{kelv12}. We then proceeded to fit these 10 galaxies using both SDSS and KiDS inputs, and using \profit and \galfit. The free parameters used throughout were $x$ and $y$ centres, bulge mag, $R_e$ and $n$ ($A/B$ for the bulge was fixed to be 1), and disc mag, $R_e$, $A/B$ and $\theta$ ($n$ for the disc was fixed to be 1, i.e.\ exponential).

The input parameters were taken from the fits in \cite{lang16}, although it should be noted that \profit (used in full MCMC mode) is largely insensitive to the inputs used, as long as the total input magnitude is approximately correct (i.e.\ within a couple of magnitudes of the correct value). Used in pure downhill gradient mode \profit suffers from similar local minima issue to \galfit \citep[see the discussion of \galfit convergence in][]{lang16}.

The full range of comparison Figures A1--A6 are included in Appendix \ref{app:consist}. Here we compare how well these different combinations of decomposition codes and data sources affect the returned values for bulge and disc magnitudes.

\subsubsection{Comparison of bulge magnitudes}

\begin{figure*}
	\centering
	\includegraphics[width=\columnwidth*2]{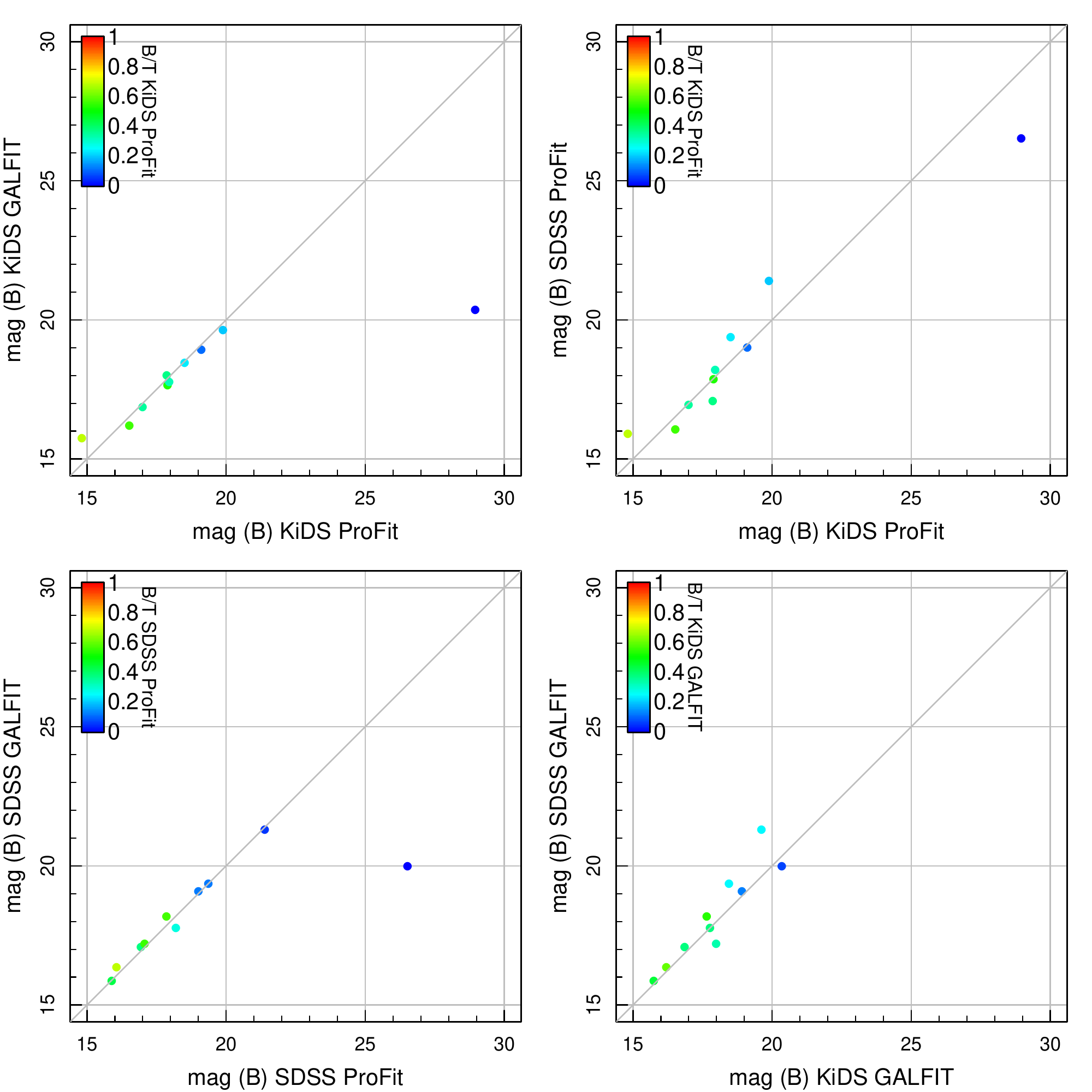}
	\caption{Panels comparing the measured bulge magnitude for the 10 well resolved galaxies included with the \R \profit package. Top-left shows the comparison of \galfit and \profit using KIDS data. Bottom-left shows the comparison of \galfit and \profit using SDSS data. Top-right shows the comparison of SDSS and KiDS using \profit. Bottom-left shows the comparison of SDSS and KiDS using \galfit. In all cases the B/T that is used to colour the data points is derived from the fit used for the x-axis. Fit errors provided by both \galfit and \profit are smaller than the points, so are not plotted here.}
	\label{fig:compmagB}
\end{figure*}

\begin{figure*}
	\centering
	\includegraphics[width=\columnwidth*2]{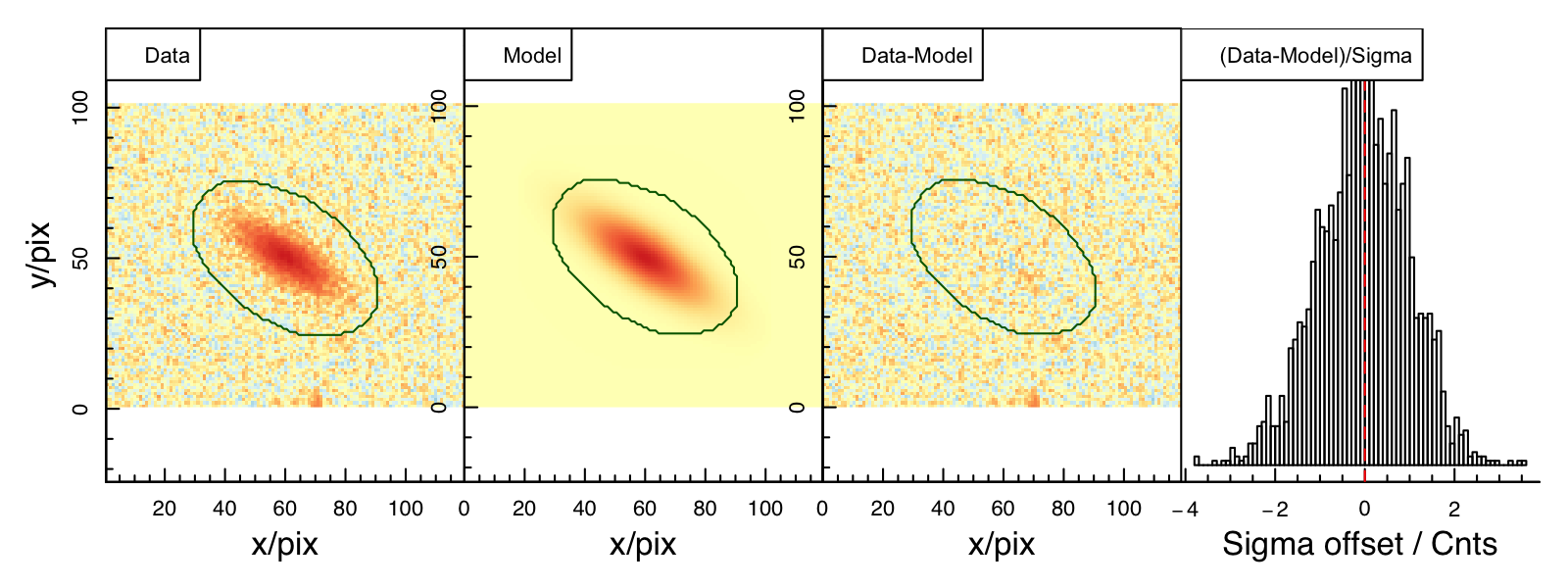}
	\caption{The \profit SDSS decomposition for G266035 which effectively has no bulge since $B/T<10^{-4}$. Judging by the almost pure noise residuals it seems reasonable to conclude that the data has little requirement for a bulge component.}
	\label{fig:sdssdecomp}
\end{figure*}

Fig.~\ref{fig:compmagB} compares the agreement between estimated bulge magnitudes. The one major disagreement between codes is the faintest bulge measured. In \profit the decomposition preferred removing the bulge entirely, often hitting our specified lower limit of 30 mag. Using \galfit, the code does not allow the solution to move huge distances from the initial estimates unless the likelihood terrain is very smooth, which occurs only when the model reproduces the data with no clear residuals. Fig.~\ref{fig:sdssdecomp} shows the fit and residuals for the \profit fit for this galaxy, suggesting there is little need for any bulge at all.

Otherwise the main conclusion that can be drawn is that the codes correlate tightly, and so do the data sets, but in all cases the scatter is much larger than the errors suggested by \galfit or \profit. Although the errors returned by \profit were consistently larger for these galaxies, they are still smaller than the points in the panels.

A general remark is that we find more consistency between codes than between data sets. Removing the outlier galaxy G266035, we measured the intrinsic scatter using the \hf code of \citet{robo15} and find it to be reduced by more than a factor of two. It is at a minimum for the SDSS data (0.14 dex versus 0.20 dex scatter orthogonal to the 1--1 line), suggesting that some aspect of our data processing is more internally consistent for the SDSS data. This might be the background subtraction or the PSF determination. 

\subsubsection{Comparison of disc magnitudes}

\begin{figure*}
	\centering
	\includegraphics[width=\columnwidth*2]{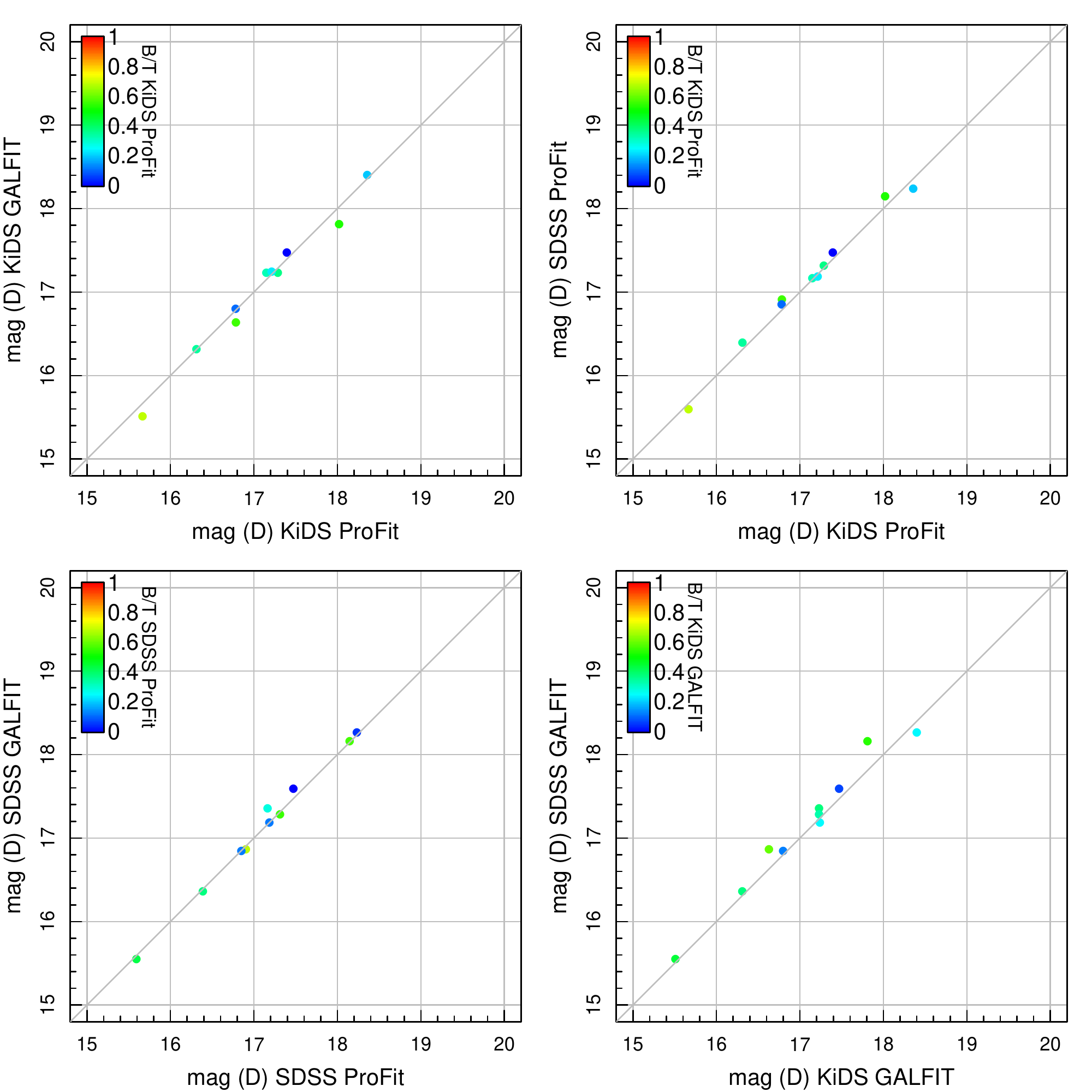}
	\caption{Panels comparing the measured disc magnitude for the 10 well resolved galaxies included with the \R \profit package. See Fig.~\ref{fig:compmagB} for details.}
	\label{fig:compmagD}
\end{figure*}

Fig.~\ref{fig:compmagD} compares the agreement between estimated disc magnitudes. In general we find more self-consistent fits for discs both between codes and between data. Whilst the smallest intrinsic scatter is again found for SDSS using \profit and \galfit, the results for \profit using SDSS and KiDS are not far behind. The intrinsic scatter for KiDS using \profit and \galfit, and \galfit using SDSS and KiDS are notably worse (the latter showing the most scatter).

For all these fits we find that the \profit MCMC solution provides a {\it more likely} fit both when forcing the \profit parameters through \galfit, and when forcing the \galfit parameters through \profit. Usually the increases in likelihood when using \profit are substantial, suggesting the difference is due to local convergence issues with the Levenberg-Marquardt \citep{marq63} algorithm built into \galfit, rather than issues to do with model accuracy that we explored in Section \ref{sec:profit-acc}.

A general conclusion is that we appear to be able to recover more self-consistent results for disc properties compared to bulge properties. Further evidence for this is shown in Appendix \ref{app:consist}, where disc $R_e$ is found to be more reliably measured than bulge $R_e$. Bulge $n$ is notably the hardest property to measure consistently, a finding in agreement with recent work by \citet{savo16}. We investigate in more detail how well we expect to be able to recover various bulge-disc properties for different quality and depth data in the next section.

%\subsection{Comparing \profit KIDS and SIGMA SDSS Results}
%
%\label{sec:example3}
%
%NOTE, MIGHT HAVE TO DROP THIS SECTION GIVEN IT USES UNRELEASED KIDS DATA. WORST CASE IT WILL HAVE TO MOVE TO WHATEVER FUTURE GAMA PAPER USES THE KIDS RESULTS FROM PROFIT.
%
%In the previous Section we investigated a small sample where we could make the four-way comparison of \profit, \galfit, SDSS and KiDS. Having established that the major structural properties of magnitude and $R_e$ are well recovered across all comparisons, we now proceed to compare a bigger sample where the SDSS results were derived from a the SIGMA wrapper software that runs \galfit in a convenient manner, and the new KIDS results for the same galaxies were obtained using \profit.
%
%In \citet{lang15} the SIGMA code of \citet{kelv12} was used to fit a sample of low redshift galaxies. Here we run \profit of the same data, ultimately using the same inputs as per SIGMA. The only major difference with the SIGMA wrapper is that we used the new sky subtraction routine developed for LAMBDAR and dicussed in \citet{wrig16}.

\section{Applying \profit to Simulated Current and Future Imaging Surveys}
\label{sec:future}

Having described and tested our new \profit decomposition code in some detail, we finally investigate how it performs on data of different depth and quality. For the purposes of these tests we create and fit bulge-disc systems in $40 \times 40$ arcsec image stamps. We investigate three sources of survey imaging that are indicative of the types of data that \profit will be used on en-masse in future: SDSS, KiDS and LSST. Since these results are purely indicative we make some simplifying assumptions for the three surveys. Table \ref{tab:surveys} presents the assumptions made when simulating and fitting the galaxies. These are not precisely the values advertised for the various surveys, but are indicative given the inevitable variations in data quality. For LSST these estimates are for the stacked 5-yr survey. Since we know the precise PSF used to convolve the model and the images themselves are precisely background subtracted, the following results are the upper-limits on the decomposition performance we can hope to expect.

The simulations themselves sample 10,000 profiles from a uniform grid of total magnitude ($15 \le m_{T} \le 22$), bulge fraction ($0 \le B/T \le 1$), logarithmic bulge effective radius ($0 \le \log_{10}(R_e) \le 1$), logarithmic disc effective radius ($0 \le \log_{10}(R_e) \le 1$), logarithmic bulge \sersic index ($0 \le \log_{10}(n) \le 1$) and disc axial ratio ($0 \le A/B \le 1$). The bulge axial ratio was assumed to have $A/B=1$ and the disc \sersic index was fixed to $n=1$. \profit never appears to struggle to find the correct disc angle $\theta$, so we only simulated galaxies with the minor axis varying on the $x$-axis of the image. Note we do not impose internal correlations between parameters, so these results span an overly generous range of structural parameter space.

\begin{table}
\caption{Assumed image quality parameters for different surveys.}
\begin{center}
\begin{tabular}{llll}
Survey	& Scale ("/pix)	& PSF FWHM (")	& $1 \sigma$ $r$-band depth	\\
\hline
SDSS	& 0.4				& 1.4				& 24	\\
KIDS	& 0.2				& 0.6				& 26	\\
LSST	& 0.2				& 0.6				& 28	\\
\end{tabular}
\end{center}
\label{tab:surveys}
\end{table}%

\begin{figure}
	\centering
	\includegraphics[width=\columnwidth]{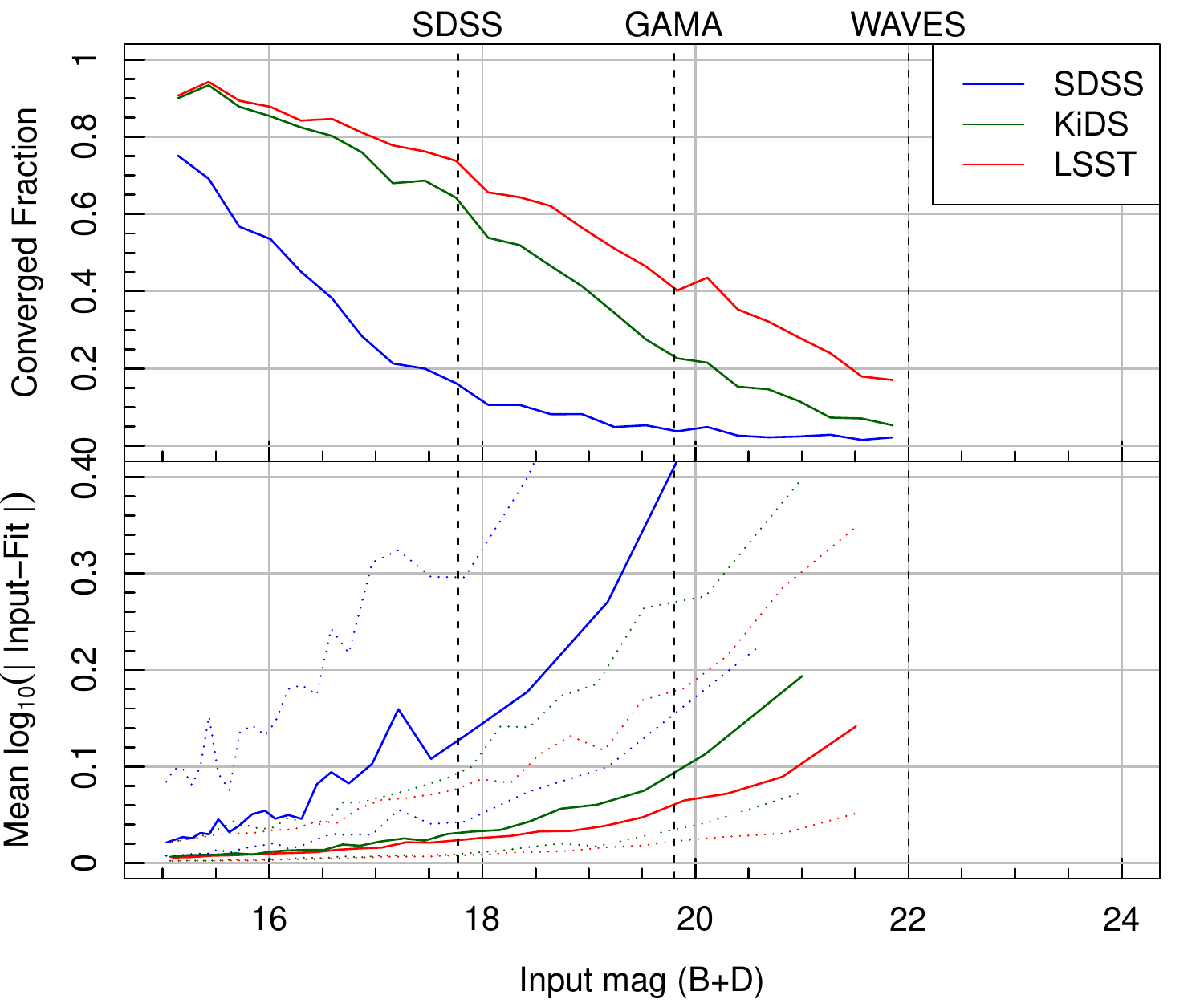}
	\caption{Comparisons of 10,000 random simulations for SDSS, KiDS and LSST quality data. Top panel shows the fraction of \profit fits that converge on a solution that is not a hard limit of the fit as a function of total magnitude. Bottom panel shows the mean Euclidian logarithmic distance between the six input parameters and the \profit fit for the galaxies that have converged fits taken from the top panel (where the distance in flux is $\log_{10}(flux)$ rather than magnitude, i.e.\ $m/2.5$). The solid lines show the median of the results, and the dotted show the $14^{th}$ and $86^{th}$ percentile ranges (akin to the $1\sigma$ range if the distributions are Normal). In simple terms, larger value are better in the top panel and lower values are better in the bottom panel.}
	\label{fig:converge_frac}
\end{figure}

Fig.~\ref{fig:converge_frac} shows the results of these simulations in a compact manner, comparing the fraction of fits that are found to converge (which means they find solutions that are not hard limits imposed within \profit) and the typical average separation between all input and output parameters in log-space. These properties are plotted in Figures B1--B7 for all major observables in Appendix \ref{app:sims}, but here we show the results versus the most observable input parameter: the total magnitude of the galaxy. Unsurprisingly, all the results are best when the galaxy is intrinsically bright and the source signal dominates over the sky background. Of perhaps more surprise is that we see a large improvement in decomposition performance when moving from SDSS to KiDS quality data, but a relatively minor improvement when we move from KiDS to LSST. This appears to be due to the major improvement being limited to surface brightness depth, the pixel scale and the typical PSF FWHM is not expected to be much (if at all) improved between KiDS and LSST (one caveat being the latter has a much larger field-of-view, so further un-simulated sky-subtraction gains could be expected).

To briefly summarize the other trends presented in Appendix \ref{app:sims}: lower bulge and disc $R_e$, lower disc $A/B$, lower bulge $n$ and a moderate $B/T\sim0.5$ all lead to improved fitting results. For $R_e$, this is simply because lower $R_e$ means high surface brightness for a given magnitude, and the components become easier to distinguish against the background. We start our simulations at an $R_e$ of 1 arcsec, so the caveat to make is that much below this ($R_e$ less than half the PSF FWHM), and the parameters become indistinguishable structurally. Even in this regime we can still expect the component magnitude to be well recovered, it is just hard to resolve structure inside the PSF scale.

It is clear from these results that only in the most favourable regimes can we expect complete {\it and} robust bulge-disc decompositions, even using the best current and next generation deep survey data. This is true even when we are operating in the regime of moderately low redshift galaxies extracted from surveys such as GAMA. A future low-redshift focused WAVES-Wide survey \citep{driv16}, being inherently lower redshift due to its photo-$z$ pre-selection (with typical redshift likely to be similar to SDSS DR10) is the best prospect for such future studies given the target of high redshift completeness, and a guarantee it will have deep good quality imaging data (since KiDS will act as the primary input source for targets).

\section{Discussion and Conclusions}
\label{sec:discuss}

In this paper we have presented our new publicly available galaxy modelling and decomposition code \profit. The core code comes in a standalone {\textsc c++} library (\libprofit) that allows for easy access to higher level languages (e.g. \R and \Python) through a simple API.

The core advances that \profit offers over currently available software are that it:

\begin{itemize}
\item is fully open-source with multiple active developers,
\item offers a standalone library (\libprofit) for accurate and fast pixel integrations when generating a model,
\item can be extended with new profiles in a simple well-documented manner,
\item allows for simple or complex priors on parameters (an important aspect of Bayesian analysis),
\item offers a range of likelihood calculations,
\item is untied to any specific optimizers but has easy access to downhill minimization, genetic algorithm, and MCMC routines,
\item can fit parameters in log or linear space,
\item allows for simple or complex additional constraints between parameters,
\item offers brute-force and FFT PSF convolution options, with automatic benchmarking to select the fastest strategy.
\end{itemize}

Initially \profit comes in three varieties: a fully featured \R package ({\tt ProFit}, discussed in detail in this paper), a basic \Python wrapper ({\tt pyprofit}) and a command line terminal interface ({\tt profit-cli}). The \R package is the most advanced in terms of features, and we have a longer term aim to bring the \Python package up to the same level of sophistication. The command line interface only exists for the easy generation of model images via calling the underlying \libprofit library, so it does not provide any built-in optimisation. By separating the model image generation (which should be an objective black-box task achieved through an API) into a separate \libprofit library, users are not tied to our high-level solution for the much more subjective problem of galaxy fitting (with many caveats over masking, convolution, likelihoods etc). We are aware of at least one community fork of \libprofit already that uses diffusive nested sampling \citep[DNEST;][]{brew11} in order to overcome sampling of highly multimodal data (Huijser et al., in preparation). Extensions to other languages using the library API are encouraged, but simpler workarounds using the command line interface are also possible for less programming-savvy users. To avoid community confusion, we request that any forks that are released append the name in some manner (e.g.\ {\textsc ProFit\_STAN}). The aim would be to pull any substantial and useful changes into the main \profit branch in the longer term.

The first versions of the \profit code stemmed from preliminary work by ASGR. Following this, DST and RT have added many new features and have heavily modified and expanded the functionality of the software. From the early stages, \profit has been designed in a highly modular manner. This is important to prevent the user becoming limited to our default choice of profile, likelihood evaluation, or optimization method. The core \libprofit library is designed to be fast and accurate at integrating and convolving target model images, where the likelihood function to compute, and optimization engine to use, are largely choices for the user via higher level interfaces. The \R package version of \profit contains a large number of examples using simple downhill gradient schemes, more complex genetic algorithms and more computationally expensive MCMC techniques. Users are encouraged to build from these examples to use more or less sophisticated engines as appropriate.

The various libraries and higher level interfaces are all available on GitHub, and future support and functionality will be added through these repositories. This paper necessarily refers to a static v1.0 of \profit, as such specifics regarding operation and options should be derived from the public repositories and documentation found there, rather than this text. The most reliable longterm location for the base \libprofit library is \url{https://github.com/ICRAR/libprofit}. The higher-level \R implementation of \profit is available at \url{https://github.com/ICRAR/ProFit} and the \Python variant is maintained at \url{https://github.com/ICRAR/pyprofit}. This paper was written using the \R implementation exclusively. 

Community use and feedback is encouraged, especially via raising {\it issues} and {\it pull} requests through GitHub. In particular, the requirements outlined in Section \ref{section_introduction} are not an exhaustive list, so users are encouraged to submit feature proposals and requests. Currently planned extensions include built-in support for Monte Carlo image generation, simultaneous fitting of multiple images (in the same band) and/or multiple bands, zero-point calibration, covariance likelihoods and PSF fitting. Indeed, the \R package already includes a vignette with PSF fitting examples, a full description of which is omitted here for brevity.

In this paper, we have investigated the application of \profit to both SDSS and KiDS data, using \galfit reference runs as a comparison. In pragmatic terms, \profit and \galfit achieve consistent results given the same input data, with at least as much scatter being produced by changes between data source (in this case, SDSS versus KiDS). This is encouraging in that it validates the broad findings of recent work that made exclusive use of \galfit, and lends credibility to our future applications of \profit to a number of different data sources. The main difference is a large degree of intervention was required when running GALFIT via running on a large grid of initial conditions and post-processing the results \citep[see][]{lang16}, whereas \profit was run by AM with just a basic wrapper using one of the many available MCMC routines in \R and converged to reasonable global solutions with little user intervention.

\section{Acknowledgements}

Much of the work presented here was made possible by the free and open \R software environment \citep{rcor16}. All figures in this paper were made using the \R{} {\tt magicaxis} package \citep{robo16}. Data from the KiDS \citep{kuij15} and SDSS \citep{ahn14} surveys were used for this work. Funding for SDSS-III has been provided by the Alfred P. Sloan Foundation, the Participating Institutions, the National Science Foundation, and the U.S. Department of Energy Office of Science. The SDSS-III web site is http://www.sdss3.org/. Parts of this research were conducted by the Australian Research Council Centre of Excellence for All-sky Astrophysics (CAASTRO), through project number CE110001020. Credit to E. Mannering for Fig.~\ref{fig:pixint}, she was inspired by the lamentable initial efforts of ASGR. Thank you to the anonymous referee, whose comments particularly assisted in clarifying complex parts of the paper.

\bibliographystyle{mn2e}
\setlength{\bibhang}{2.0em}
\setlength\labelwidth{0.0em}

%\bibliography{/Users/do/Dropbox/Publications/authored/Bibliography/astro}
%\bibliographystyle{/Users/do/Dropbox/Publications/authored/Bibliography/my_mn2e}

\appendix

\section{Comparisons of \profit Example Galaxies}
\label{app:consist}

To supplement Figs \ref{fig:compmagB} and \ref{fig:compmagD}, below are the various other \profit, \galfit, SDSS and KiDS comparisons for the 10 galaxies that are included in the standard \R \profit installation for demonstration and testing purposes.

\begin{figure}
	\centering
	\includegraphics[width=\columnwidth]{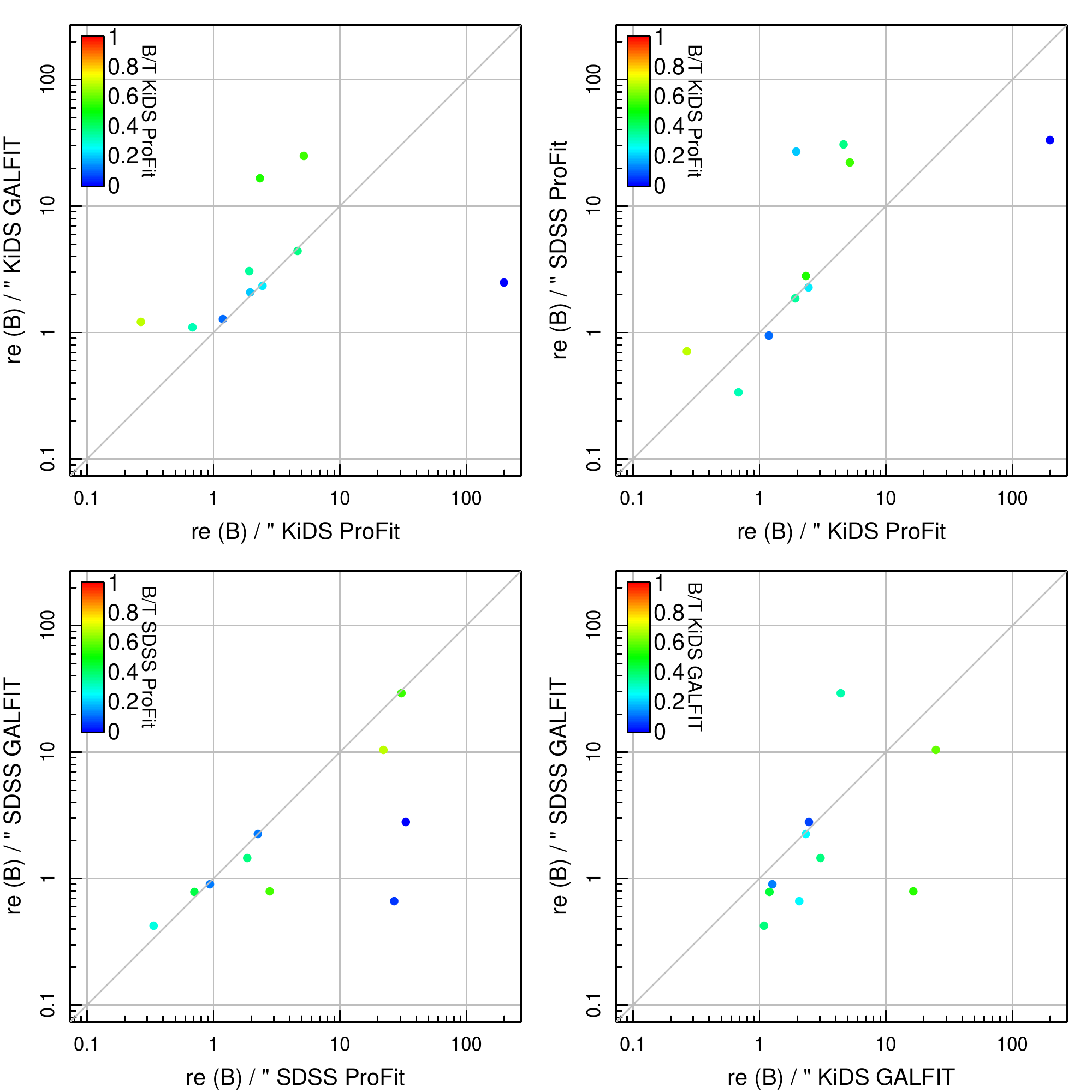}
	\caption{Panels comparing the measured bulge $R_e$ for the 10 well resolved galaxies included with the \R \profit package. See Fig.~\ref{fig:compmagB} for details.}
\end{figure}

\begin{figure}
	\centering
	\includegraphics[width=\columnwidth]{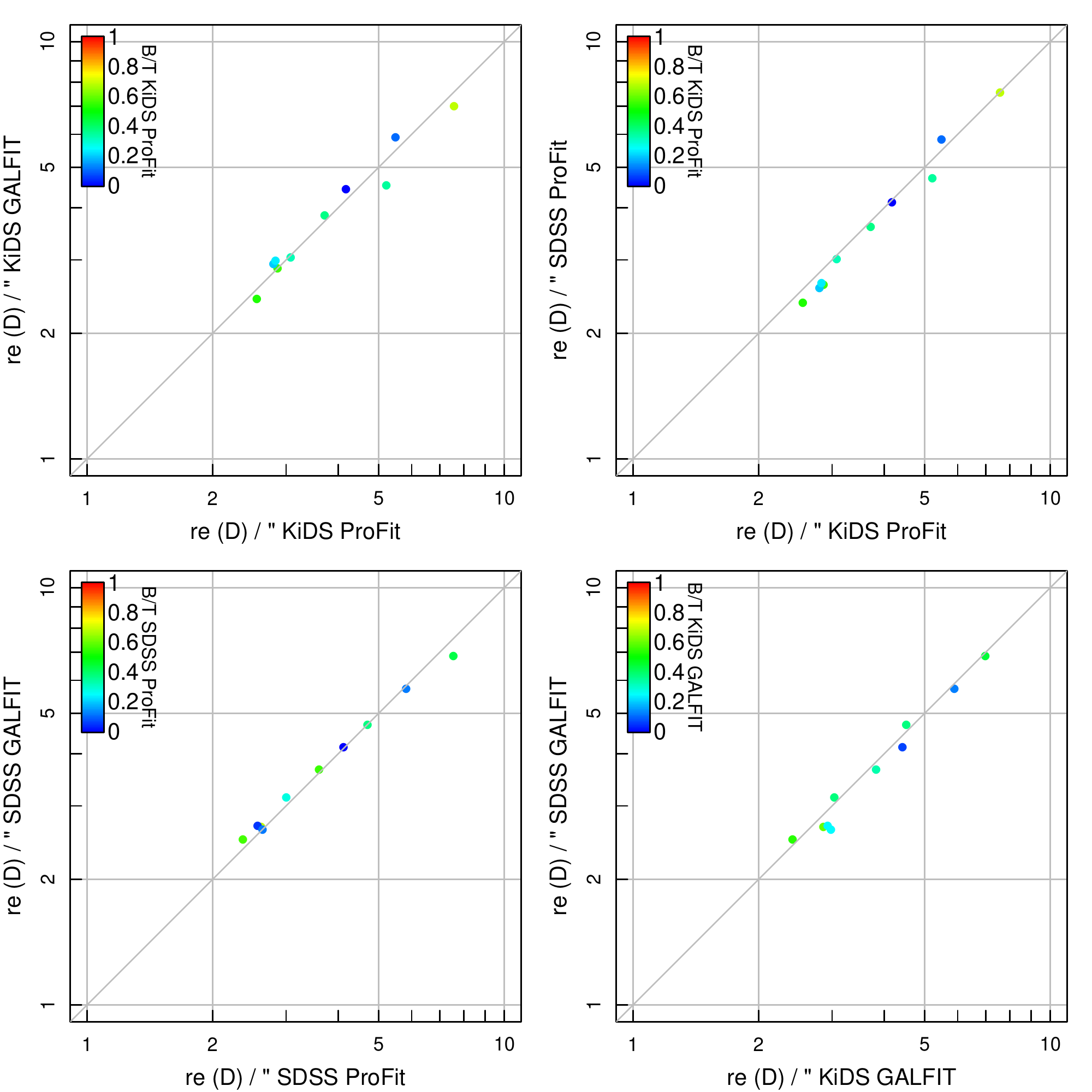}
	\caption{Panels comparing the measured disc $R_e$ for the 10 well resolved galaxies included with the \R \profit package. See Fig.~\ref{fig:compmagB} for details.}
\end{figure}

\begin{figure}
	\centering
	\includegraphics[width=\columnwidth]{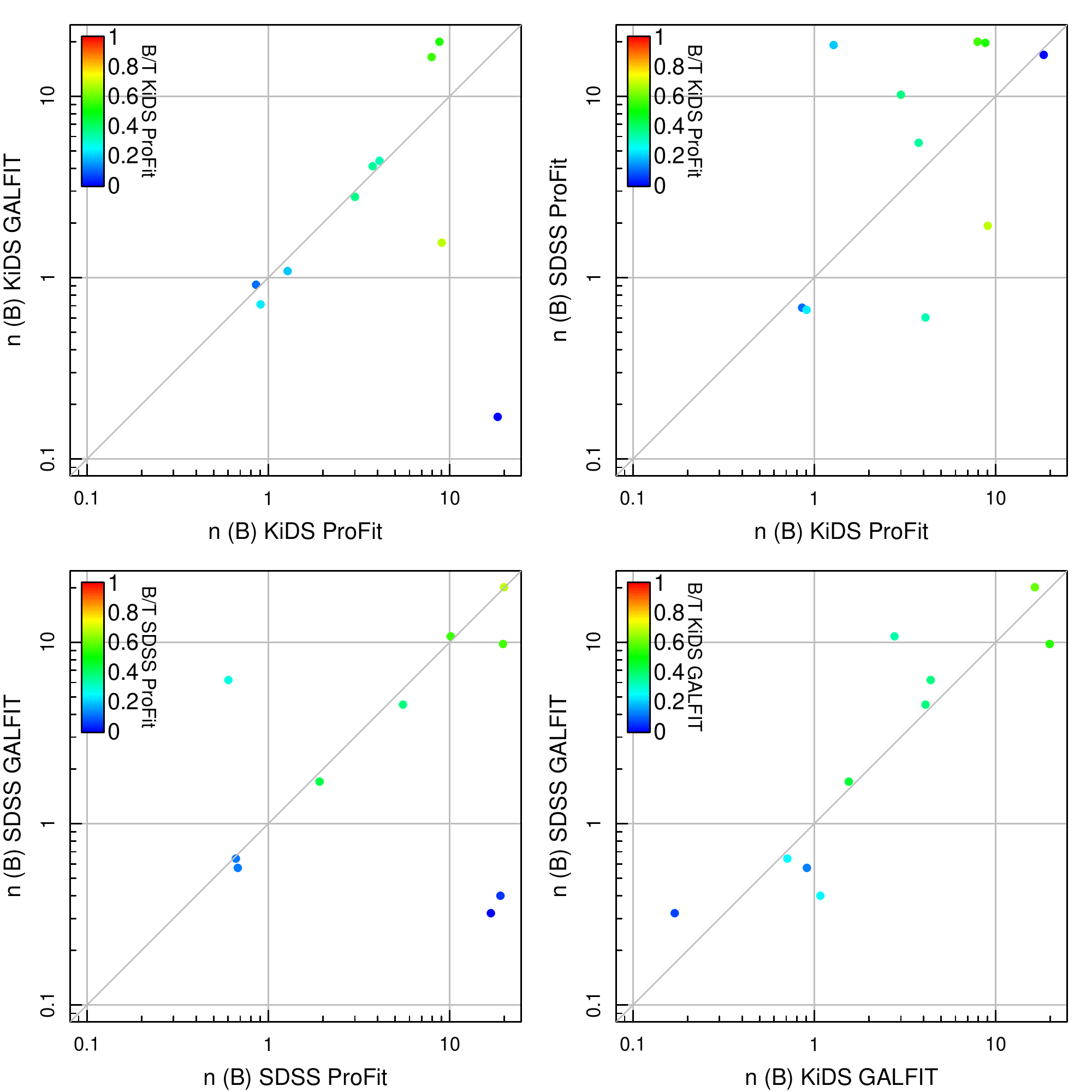}
	\caption{Figures comparing the measured bulge $n$ for the 10 well resolved galaxies included with the \R \profit package. See Fig.~\ref{fig:compmagB} for details.}
\end{figure}

\begin{figure}
	\centering
	\includegraphics[width=\columnwidth]{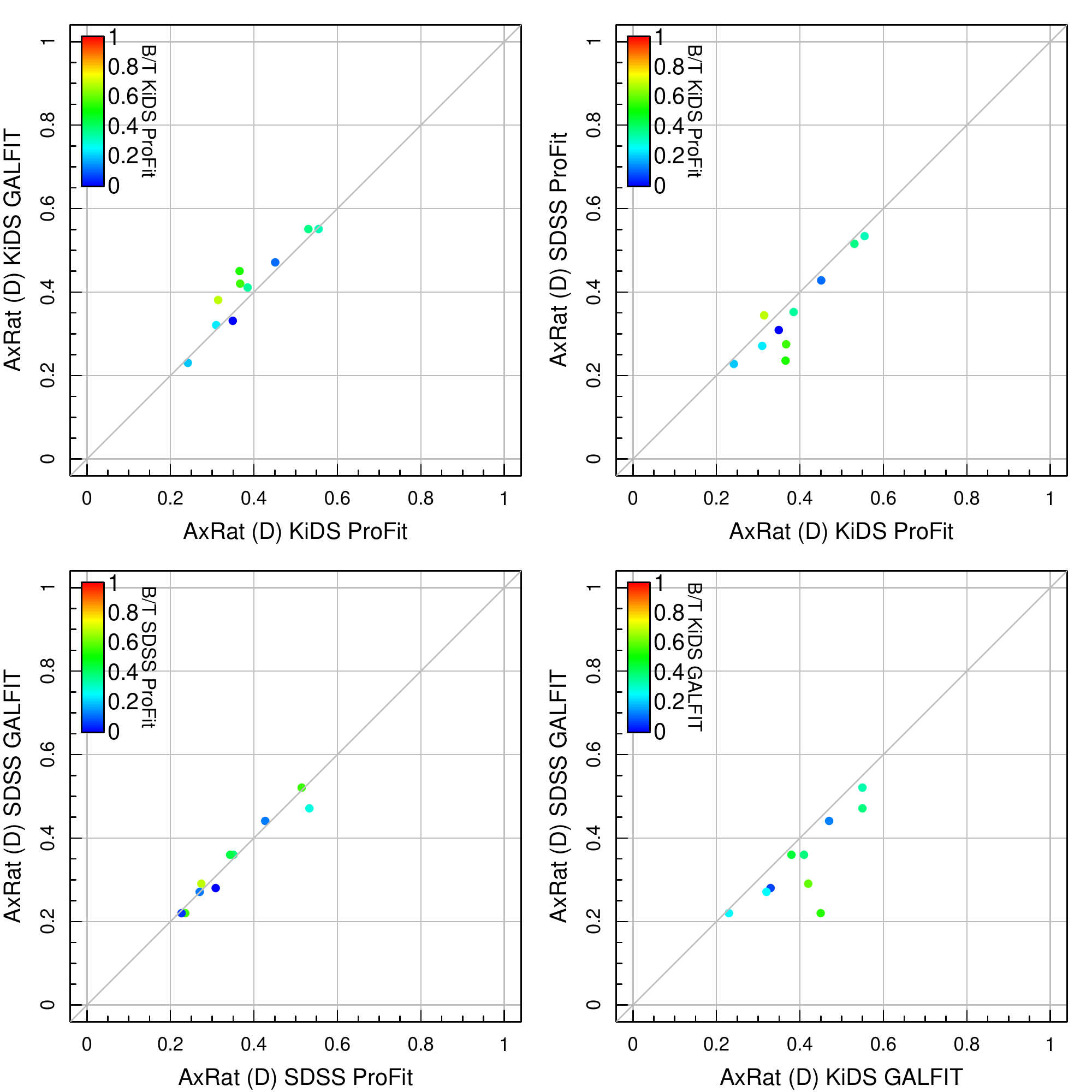}
	\caption{Figures comparing the measured disc $A/B$ for the 10 well resolved galaxies included with the \R \profit package. See Fig.~\ref{fig:compmagB} for details.}
\end{figure}

\begin{figure}
	\centering
	\includegraphics[width=\columnwidth]{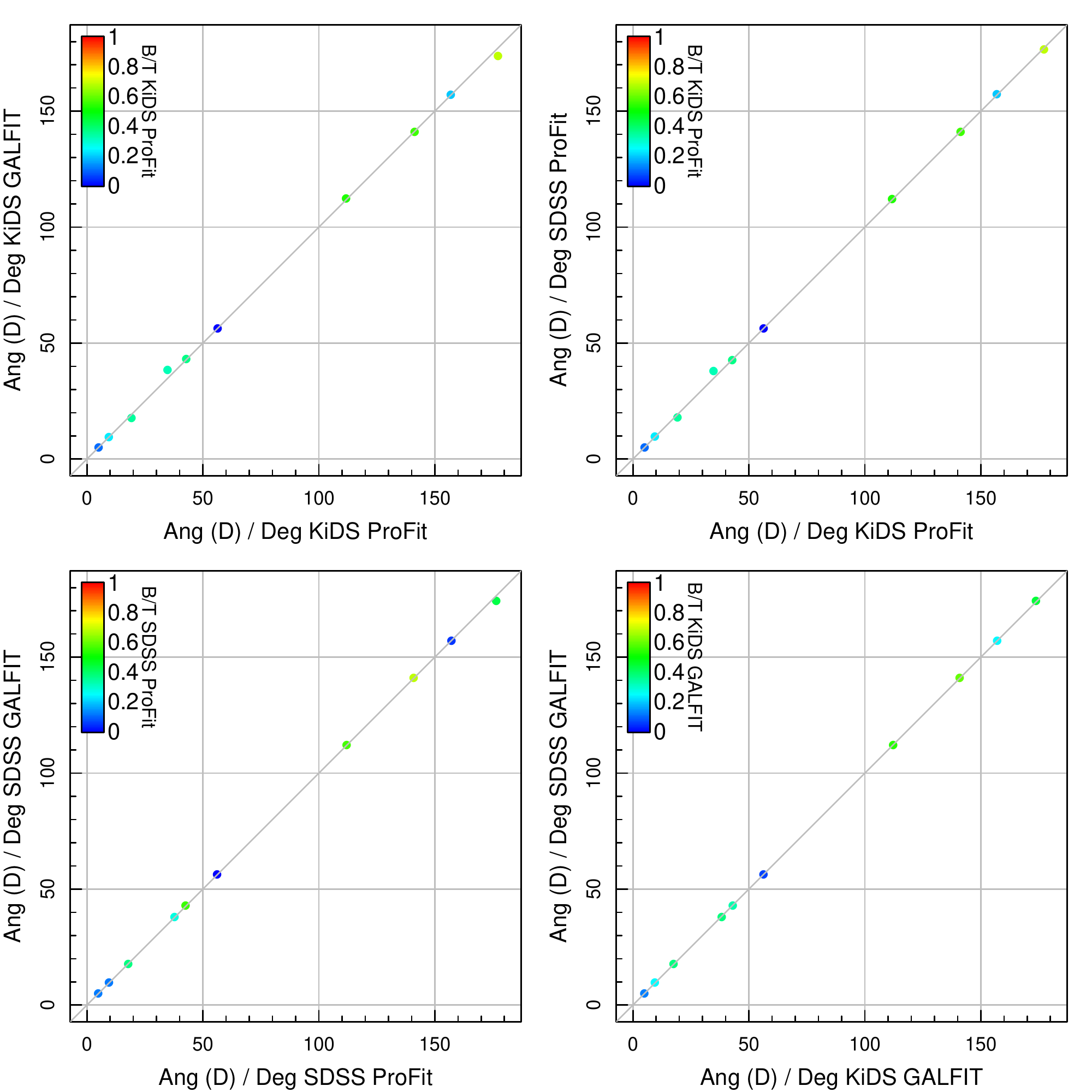}
	\caption{Figures comparing the measured disc $\theta$ for the 10 well resolved galaxies included with the \R \profit package. See Fig.~\ref{fig:compmagB} for details.}
\end{figure}

\begin{figure}
	\centering
	\includegraphics[width=\columnwidth]{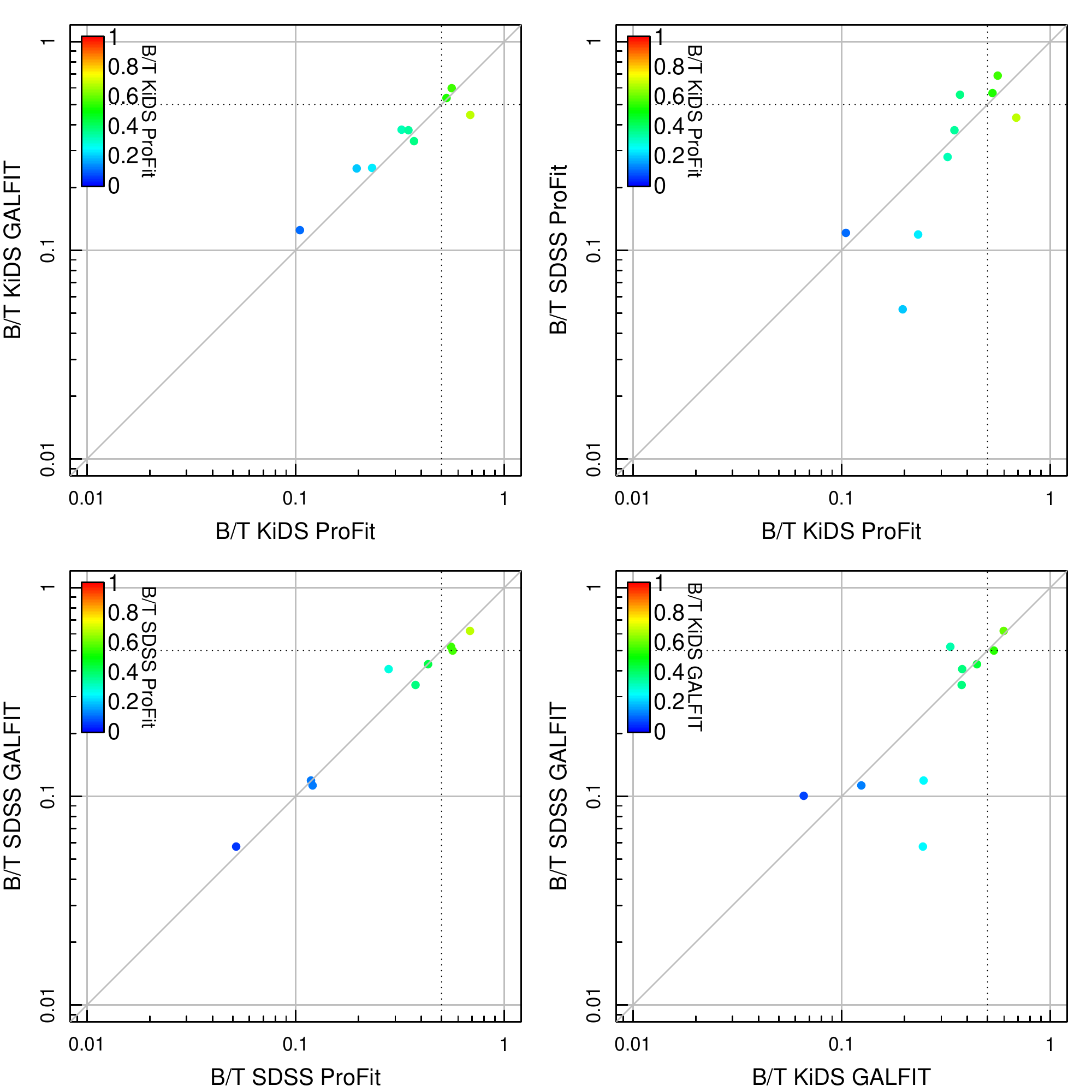}
	\caption{Figures comparing the measured $B/T$ for the 10 well resolved galaxies included with the \R \profit package. See Fig.~\ref{fig:compmagB} for details.}
\end{figure}

\section{Comparisons of \profit Simulations for SDSS, KiDS, and LSST Type Data}
\label{app:sims}

\begin{figure}
	\centering
	\includegraphics[width=\columnwidth]{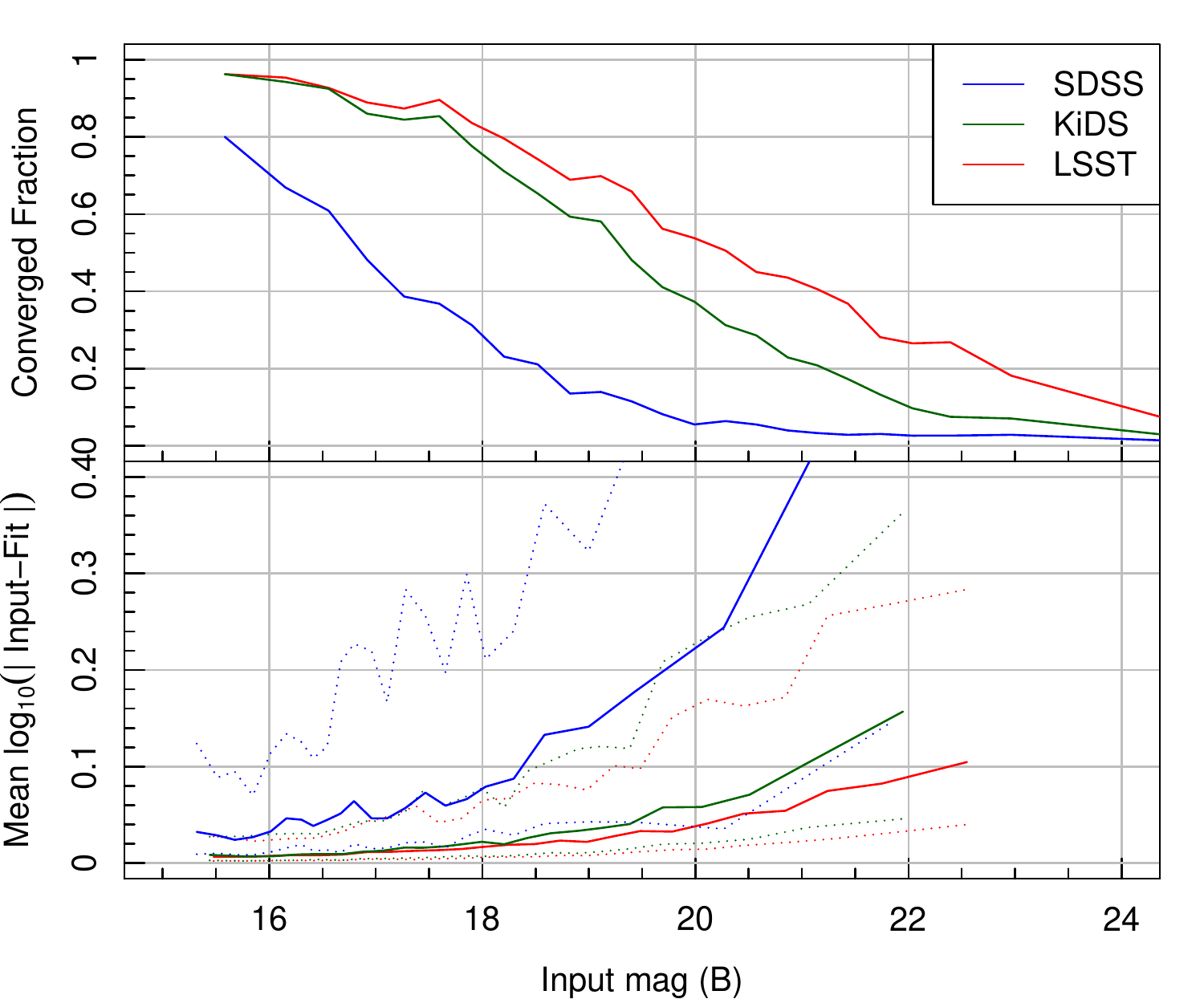}
	\caption{As per Fig.~\ref{fig:converge_frac} but for bulge magnitude on the x-axis.}
\end{figure}

\begin{figure}
	\centering
	\includegraphics[width=\columnwidth]{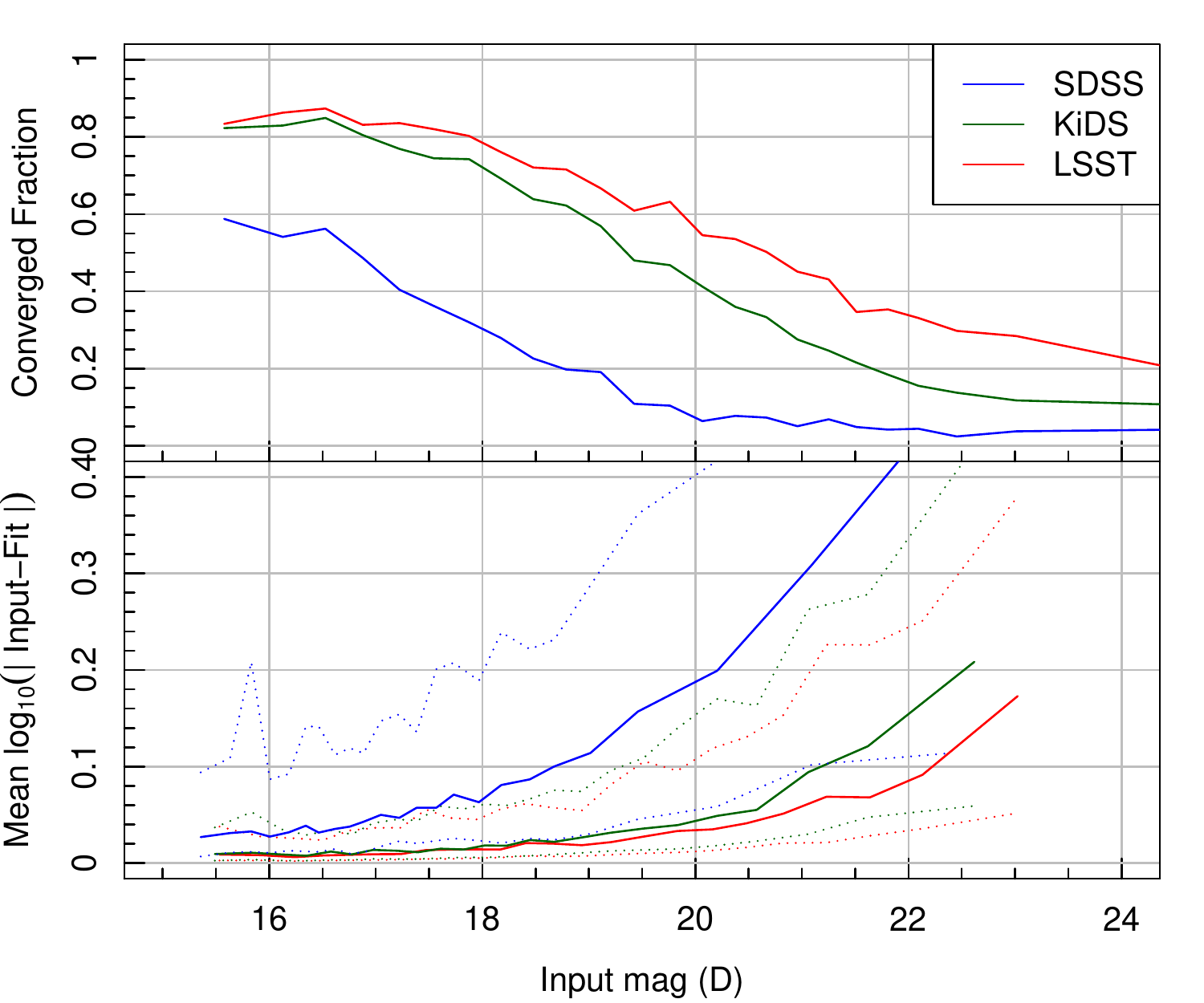}
	\caption{As per Fig.~\ref{fig:converge_frac} but for disc magnitude on the x-axis.}
\end{figure}

\begin{figure}
	\centering
	\includegraphics[width=\columnwidth]{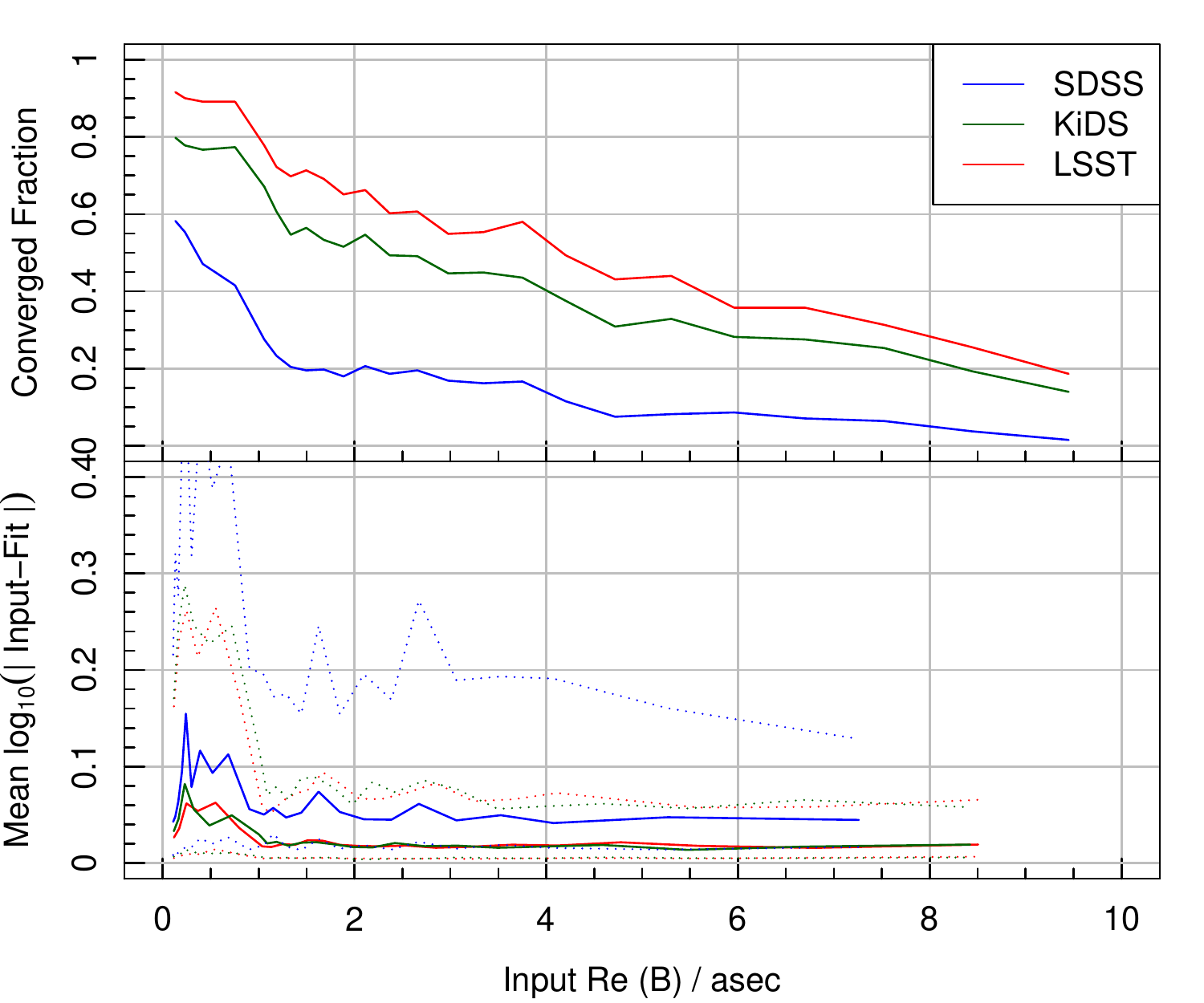}
	\caption{As per Fig.~\ref{fig:converge_frac} but for bulge $R_e$ on the x-axis.}
\end{figure}

\begin{figure}
	\centering
	\includegraphics[width=\columnwidth]{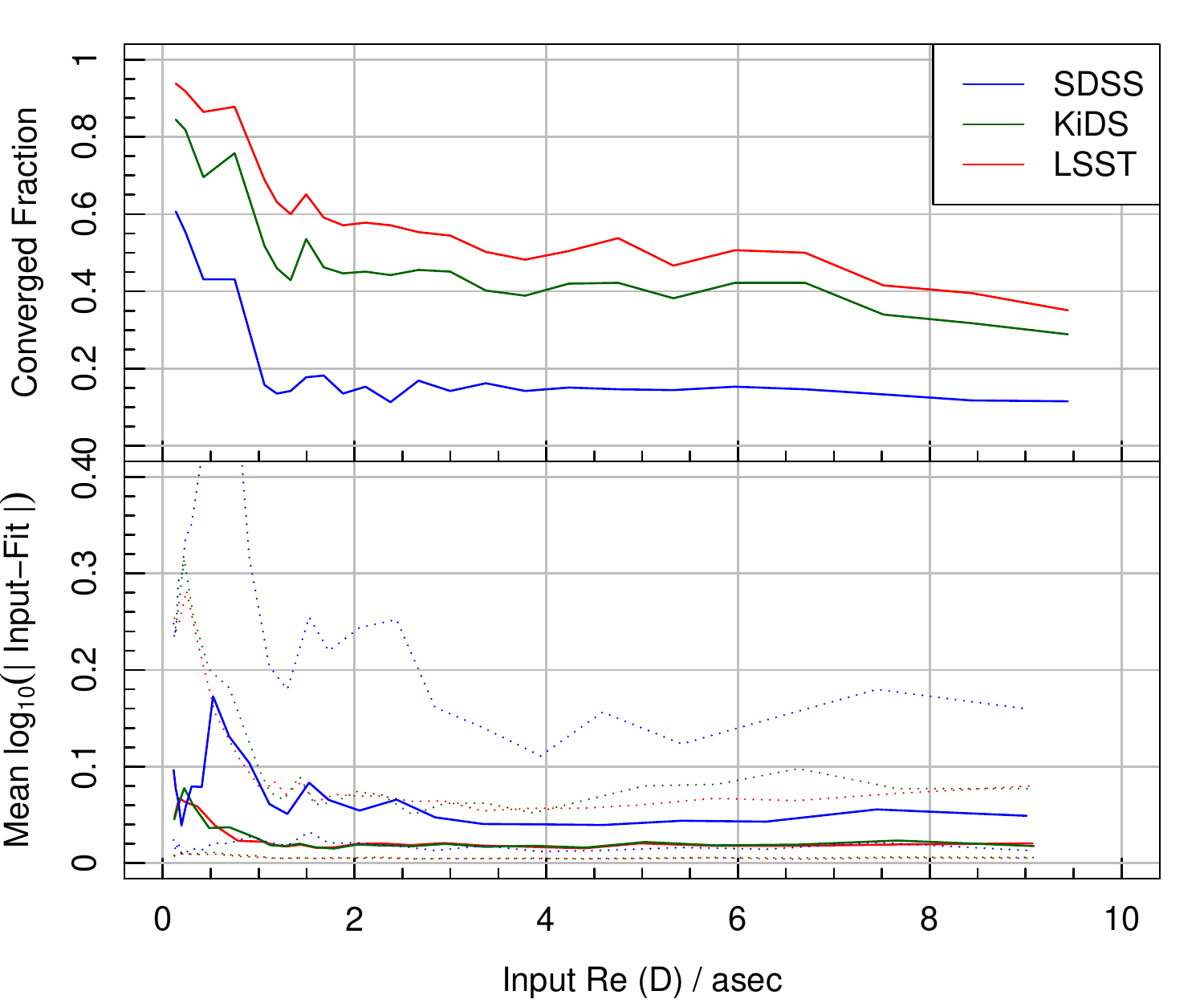}
	\caption{As per Fig.~\ref{fig:converge_frac} but for disc $R_e$ on the x-axis.}
\end{figure}

\begin{figure}
	\centering
	\includegraphics[width=\columnwidth]{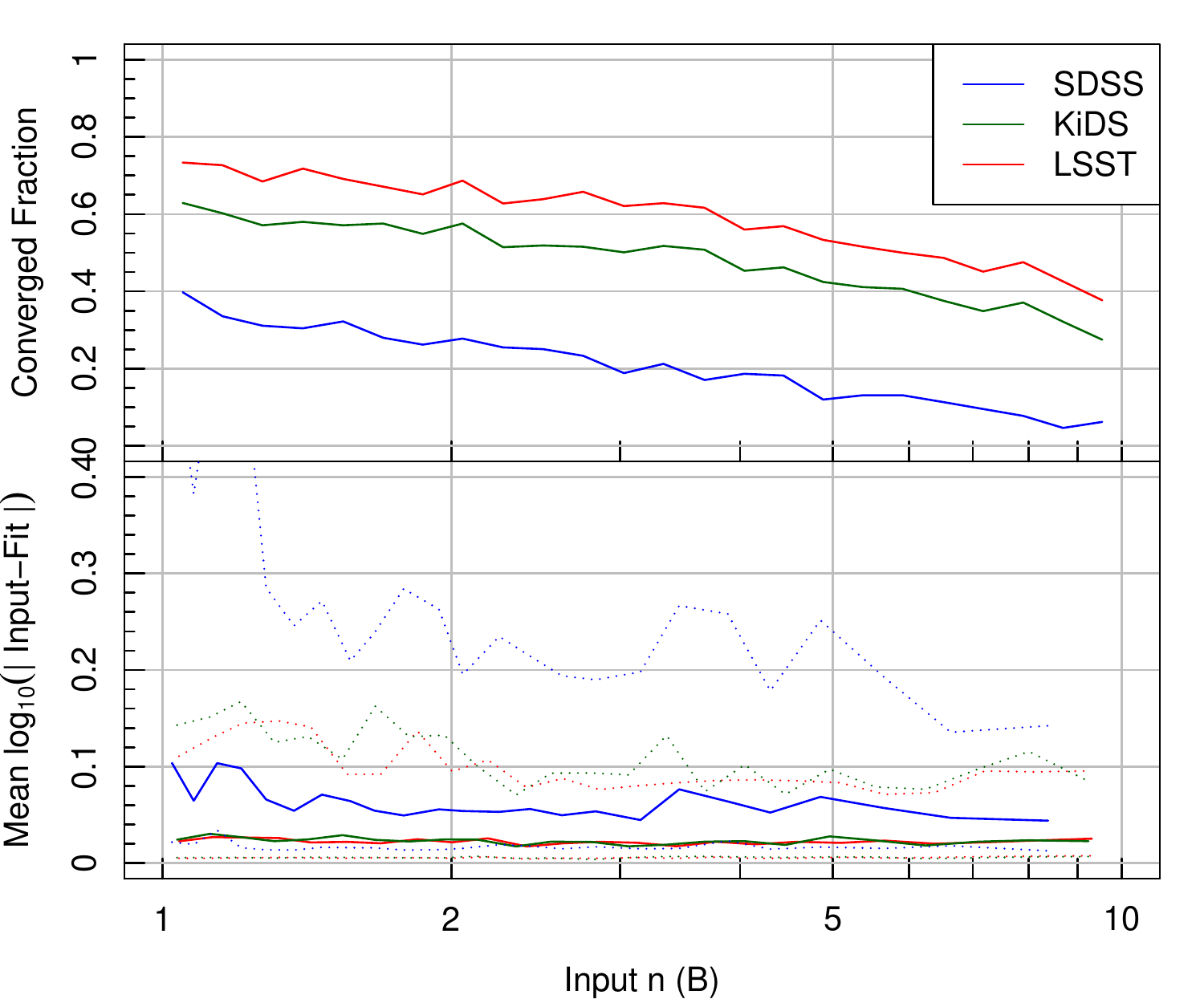}
	\caption{As per Fig.~\ref{fig:converge_frac} but for bulge $n$ on the x-axis.}
\end{figure}

\begin{figure}
	\centering
	\includegraphics[width=\columnwidth]{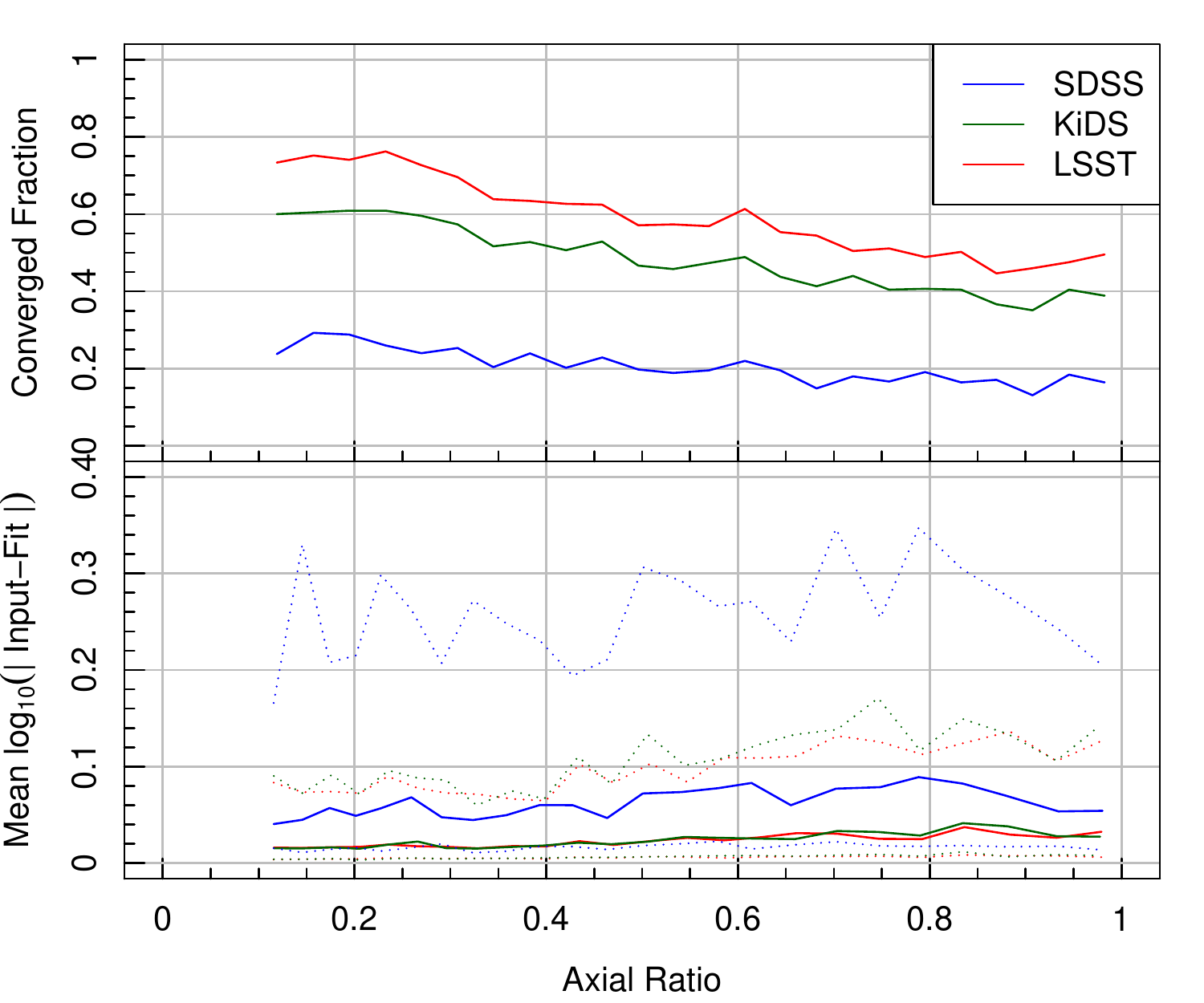}
	\caption{As per Fig.~\ref{fig:converge_frac} but for disc $A/B$ on the x-axis.}
\end{figure}

\begin{figure}
	\centering
	\includegraphics[width=\columnwidth]{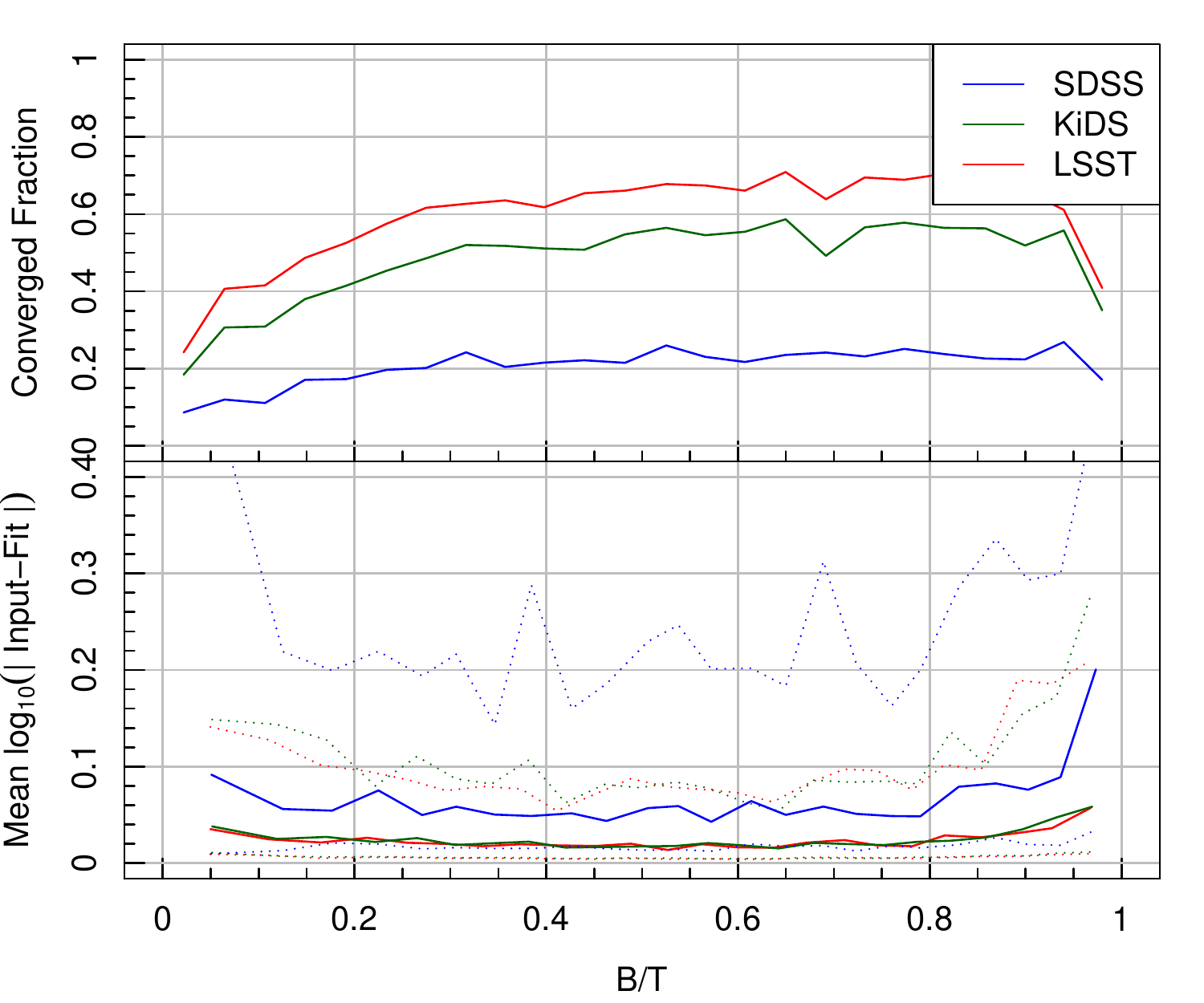}
	\caption{As per Fig.~\ref{fig:converge_frac} but for $B/T$ on the x-axis.}
\end{figure}

\end{document}